 \documentclass[aps,pre,twocolumn,float]{revtex4}
\usepackage{amsmath,bm,epsfig}

\def\Fbox#1{\vskip1ex\hbox to 8.5cm{\hfil\fboxsep0.3cm\fbox{%
  \parbox{8.0cm}{#1}}\hfil}\vskip1ex\noindent}  
\let \nn  \nonumber

\let\*\cdot
\def\<{\left\langle} \def\>{\right\rangle} \def\({\left(} \def\){\right)}
\let\p\partial \let\~\widetilde \let\^\widehat 

\newcommand{\B}[1]{{\bm{#1}}}
\newcommand{\C}[1]{{\mathcal{#1}}}    

\renewcommand{\sb}[1]{_{\text {#1}}}  
\def\Sb#1{_{\scriptscriptstyle\rm{#1}}}

\newcommand{\eq}[1]{(\ref{#1})}
\newcommand{\Eq}[1]{Eq.~(\ref{#1})}
\newcommand{\Eqs}[1]{Eqs.~(\ref{#1})}
\newcommand{\Fig}[1]{Fig.~\ref{#1}}
\newcommand{\Figs}[1]{Figs.~\ref{#1}}
\newcommand{\Sec}[1]{Sec.~\ref{#1}}
\newcommand{\Secs}[1]{Secs.~\ref{#1}}
\newcommand{\Ref}[1]{Ref.~\cite{#1}}
\newcommand{\Tab}[1]{Tab.~\ref{#1}}

\let \nn  \nonumber
\usepackage{pstricks}
\usepackage{pst-node}
\usepackage[ansinew]{inputenc}
\usepackage{amssymb,amsmath}

\let \= \equiv

\def\p{\partial}

\def\a{\alpha}

\def\g{\gamma}

\def\o{\omega}
\def\O{\Omega}

\def\<{\langle} \def\({\left(}  \def\>{\rangle} \def\){\right)}

\newtheorem{exi}{Example}

\begin{document}

\title{ Finite-Dimensional Turbulence of  Planetary Waves}
\author{Victor S. L'vov}
\email{Victor.Lvov@weizmann.ac.il}
\author{Anna Pomyalov}
\author{Itamar Procaccia}
\author{Oleksii Rudenko}
\affiliation{Department
of Chemical Physics, The Weizmann Institute of Science, Rehovot
76100, Israel}

\begin{abstract}
    \emph{Finite-dimensional wave turbulence} refers to the chaotic
    dynamics of interacting wave `clusters' consisting of finite
    number of connected wave triads with exact three-wave
    resonances. We examine this phenomenon using the example of
    atmospheric planetary (Rossby) waves.  It is shown that the
    dynamics of the clusters is determined by the types of connections
    between neighboring triads within a cluster; these correspond to
    substantially different scenarios of energy flux between different
    triads. All the possible cases of the energy cascade termination
    are classified. Free and forced chaotic dynamics in the clusters
    are investigated: due to the huge fluctuations of the energy
    exchange between resonant triads these two types of evolution have
    a lot in common. It is confirmed that \emph{finite-dimensional
    wave turbulence} in \emph{finite} wave systems is fundamentally
    different from \emph{kinetic wave turbulence} in \emph{infinite}
    systems; the latter is described by \emph{wave kinetic equations}
    that account for interactions with overlapping quasi-resonances of
    finite amplitude waves. The present results are directly
    applicable to finite-dimensional wave turbulence in any wave
    system in finite domains with 3-mode interactions as encountered
    in hydrodynamics, astronomy, plasma physics, chemistry, medicine,
    etc.
\end{abstract}


\maketitle

\tableofcontents

 \section{\label{s:intro} Introduction}
  \subsection{\label{ss:WWT} Weak-wave turbulence in finite-size systems}

 ``\emph{Wave turbulence}" refers to the chaotic dynamics of
 nonlinearly coupled oscillatory modes. The phenomenon appears in a
 variety of physical context from surface water waves, through
 atmospheric planetary waves, plasma waves, acoustic waves in solids
 and fluids etc. Depending on the strength of nonlinear interaction
 one distinguishes \emph{weak-wave turbulence} from \emph{ strong-wave
 turbulence}.  \emph{Weak-wave turbulence} is characterized by a
 smallness parameter $\zeta$ which is roughly the root mean square of
 the ratio of the nonlinear to the linear term in the equation of
 motion. For surface waves $\zeta$ is about the ratio of the wave
 amplitude to the wave-length $\lambda$, for sound in continues media
 this is the ratio of the density variations to the mean density, etc.

 The theory of weak-wave turbulence is particularly well developed in
 the limit of infinite systems where the ratio of the system size $L$
 to the characteristic wave length $\lambda$ is very large,
 $L/\lambda\to \infty$. In that limit the observed energy spectrum
 (energy distribution between modes) is well described by the so
 called ``\emph{wave kinetic equations}" that received considerable
 attention in the last half century, see
 e.g.~\cite{ZLF,NazarLvov2004,NazarLvov2005,NazarLvov2006,
 NewellZakhar2008}. We will call this regime of weak wave turbulence
 ``\emph{kinetic wave turbulence}" to distinguish it from other
 regimes, ``\emph{finite-dimensional wave turbulence}" and
 ``\emph{mesoscopic wave turbulence}" that will be introduced later.

  Notice that when the parameter $L/\lambda$ is of the order of unity
 the dynamics of waves can be very well described by low-dimensional
 chaotic models of the type studied intensively in recent decades~,
 see e.g. \cite{KL-07,KL-08}. In this paper we explore the non-linear
 dynamics of weakly interacting waves when the parameter $L/\lambda$
 is neither of order unity nor very large. This regime of parameters
 cannot be described either by kinetic equations or as low-dimensional
 chaos; it calls for new approaches and novel concepts, as partially
 demonstrated in this paper.

 To clarify the possible new regimes of weak-wave turbulence in
 various finite-domain systems we consider the general mathematical
 framework that takes the form of an energy conserving partial
 differential nonlinear equation for a field $\Psi(\B r,t)$. In a
 finite domain $S$ one expands $\Psi(\B r,t)$ in a complete set of
 eigenfunctions $\Phi_j(\B r)$ of the linearized dynamics that satisfy
 the boundary condition on the boundary $\partial S$:
 \begin{equation}
 \Psi(\B r,t) = \sum_j A_j(t) \Phi_j(\B r) \ ,
 \end{equation}
 where in general $j$ can be a multiple index and the amplitudes $A_j(t)$ are functions of time
 but not of space. Accordingly the dynamics can be represented by a set of ordinary differential
 equations for the vector of amplitudes $\B A(t)=\{A_j(t)\}$, of the form
 \begin{equation}
 \frac{dA_j(t)}{dt} = i \omega_j A_j(t) +\mbox{NL}_j (\B A)
 \end{equation}
 where $\omega_j$ is the (real) eigenfrequency of the $j$th mode; The
 symbolic term NL stands for the nonlinear contributions in this
 equation.  According to the Poincar\'e and Poincar\'e-Dulac theorems
 \cite{Arnold} the nonlinear contributions can be brought to a normal
 form by a nonlinear change of variables.  The nonlinear monomials
 that survive the change of variable are the resonant ones.  The
 $n$-tuple $(\omega_1,\dots \omega_n)$ of eigenfrequencies is said to
 be {\em resonant} if there exists a relation of the form
 \begin{equation}
 \omega_j = m_1\omega_1+m_2 \omega_2 +\dots  m_n\omega_n \ .
 \end{equation}
The order of the resonance is $\sum_k m_k $. The resonant monomials are of the form
$A_1^{m_1}\dots A_n^{m_n}$.

For the case of weak nonlinearities we invoke the smallness parameter $\zeta\ll 1$ to
discard all the higher order resonances, keeping only the lowest available order. In other words,
if there exist solutions to the equation
\begin{subequations}\label{3W} \begin{equation}
\omega_j=\omega_m+\omega_n \ , \label{3w}
\end{equation}
we keep only the resulting quadratic monomials also known as
``three-wave interactions", which satisfy the conservation law
(\ref{3w}). For example, in space-homogeneous, scale-invariant,
isotropic, infinite wave-systems, in which the dependence of the wave
frequency $\o(k)$ on the wave vector $k\=|\B k|$, $\o(k)\propto
k^\alpha$, the three wave resonances
\begin{equation}\label{3wB}
\o(k_1)+\o(k_2)=\o(|\B k_1+\B k_2|)\,,
\end{equation}
\end{subequations}
 are allowed if $\a\ge 1$~\cite{ZLF}. For $\a<1$ \Eq{3wB} has no solutions and one needs to account for  higher order resonances.

In this paper we focus on problems for which Eq. (\ref{3w}) has
solutions, determining the leading nonlinearity. We note that as the
ratio $L/\lambda$ increases, there may be more and more
eigenfrequencies that satisfy equation (\ref{3w}). In particular,
while at small values of $L/\lambda$ we can expect only isolated
resonant triads of waves, for larger values of $L/\lambda$ triads can
share a common mode and the number of coupled resonant triads
increases considerably, finally forming infinite clusters of connected
triads. Analysis of the ensuing dynamics under the influence of such
growing clusters is the main subject of the present paper. We will
focus here on the case of small enough nonlinearity parameter $\zeta$
to ensure that only waves with exact resonances are important. In this
case one can consider only clusters of connected resonant triads of
interacting waves. We will refer to the chaotic dynamics of
interacting waves in this regime as ``finite-dimensional wave
turbulence", to stress the importance of the finite number of
interacting modes with exact wave resonances. With increasing of wave
amplitudes one has to account also for quasi-resonances. This type of
wave turbulence was called ``discrete wave
turbulence"~\cite{Kar-06}. In contrast, in infinite systems, the
resonance conditions [e.g. \Eq{3wB}] has infinitely many solutions;
then usually the (\emph{kinetic}) wave turbulence can be described by
wave kinetic equations. A more detailed analysis~\cite{Naz-07} shows
that in the plane ($L/\lambda\,, \zeta$) there exists a region of
parameters where there exists weak-wave turbulence whose properties
are intermediate between finite-dimensional and kinetic regimes. Some
features of this type of turbulence, called ``mesoscopic
wave-turbulence", were observed, for example,
in~\cite{NazarLvov2006,Zakhar2006}.

To study finite-dimensional wave turbulence we focus here for concreteness on the example of the barotropic
vorticity equation on a sphere; this is an idealized model for atmospheric planetary (Rossby) waves \cite{ped}, shortly described in Sec.~\ref{APW}.  Planetary-scale motions in the ocean and
atmosphere are due to the shape and rotation of the Earth,  and play
a crucial role in weather and climate predictability
\cite{ped}. Oceanic planetary waves influence the general large-scale
ocean circulation, can intensify the currents such as the Gulf
Stream, as well as push them off their usual course. For example, a
planetary wave can push the Kuroshio Current northwards and affect the
weather in North America~\cite{Kuroshio}. Atmospheric
planetary waves detach the masses of cold or warm air that become
cyclones and anticyclones and are responsible for day-to-day weather patterns
at mid-latitudes~\cite{smth}.

Recently a new model \cite{KL-07} was developed for the intra-seasonal
oscillations in the Earth atmosphere, in terms of triads of planetary
waves whose eigen-frequencies solve Eq. (\ref{3w}). The study of the
complete cluster structure in various spectral domains shows that both
for atmospheric \cite{KK-07} and oceanic \cite{all-08} planetary waves
indeed the size of the clusters increases with the growth of the
spectral domain. In large clusters one finds both large and small
wavenumbers, meaning that the energy flux between very different
scales becomes possible, bringing with it the hallmark of turbulence.
Nevertheless, both numerical simulations \cite{zak4,T07} and
laboratory experiments \cite{DLN06} indicate that the dynamics of wave
systems with intermediate value of $L/\lambda$ do not obey the
statistical description provided by wave kinetic equations. This
domain calls for a specialized investigation which is initiated in
this paper.

  \subsection{\label{ss:sr} Structure of this paper}
 Section~\ref{APW} reviews the properties of atmospheric planetary
 waves which are important for our analyis: in \Sec{ss:bar} we
 consider barotropic vorticity \Eq{BWE} on a sphere and its dynamical
 invariants~\eq{ints1}; in \Sec{ss:proj} we project \Eqs{BWE} and
 \eq{ints1} on the spherical basis; and in \Sec{sss:nes} we analyze
 the properties of the resulting interaction coefficients.

In Sections~\ref{S:TT} and \ref{s:topology} we study the topology and
other properties of finite size clusters of resonant triads of
planetary waves that influence the dynamics of finite-dimensional wave
turbulence. In \Sec{S:TT} we begin with small clusters of resonant
triads. In \Sec{ss:TD} we overview equation of motion~\eq{triadAB} for
isolated resonant triad and its dynamical invariants~\eq{ints3},
present them in the Hamiltonian form~\eq{Ham}, and use the notion of
``active" and ``passive" modes~\cite{KL-08} denoted as \emph{ A- and
P-modes}; these notions are crucial for our theory. In the next
\Sec{ss:but} we move on to double-triad clusters, \emph{PP-, PA}- and
\emph{AA-butterflies}, their Hamiltonian equation of
motion~\eq{EMbuts}, Hamiltonian~\eq{Hambutt} and Manley-Rowe dynamical
invariants~\eq{MR2}.  Similar discussion for the triple-triad
clusters, \emph{stars}, \emph{chains} and \emph{triangles}, is given
in \Sec{ss:3-triads}. Specific for atmospheric planetary waves,
6-triad cluster, \emph{caterpillar}, is presented in \Sec{ss:cut} as
an example of a more complicated cluster structure.

In \Sec{s:topology} we study clusters of atmospheric planetary waves
in a large spectral domain $\ell, |m| \le 1000$, presenting in
\Sec{ss:met} the total number of clusters consisting of one, two,
three, etc.  triads with different topologies.  The histogram of all
cluster distributions with respect to the triad number in clusters,
ranging from 1 to 3691 is also presented. In \Sec{ss:PPred} we use a
notion of PP-irreducible clusters~\cite{KL-08} (important for the
discussion of the energy flux in finite-dimensional wave turbulence)
and show the histogram of their distribution in size, ranging from 1
to 130 triads in the PP-irreducible clusters.

Sections \ref{s:free} and \ref{s:stat} are devoted to numerical
simulation and preliminary analytical studies of finite-dimensional
wave turbulence in clusters typical for numerous physical systems
including the planetary waves described in \Secs{S:TT} and
\ref{s:topology}.  In \Sec{s:free} we begin with the analysis of free
evolution in small clusters: butterflies in \Sec{ss:butF} and
triple-triad clusters (stars and triple chains) in \Sec{ss:3-6}. The
main questions, that we discuss in this Section are: \\ --~ reasonable
choice of initial conditions, interaction coefficients and data
representation that allows to shed light on the typical features of
finite-dimensional wave turbulence, that depend on many parameters;\\
--~ how the energy flux between traids depends on the type of
connections, on the interaction coefficients, on the initial
conditions and on the cluster topology.

 In Sec.~\ref{s:stat} we study finite-dimensional wave turbulence with
 a constant energy flux in the long-chain clusters, consisting of a
 large number of triads ($N\gg 1$). For this goal we introduce pumping
 of energy into the the leading (first) triad and damping in the
 driven (last) triad.  We discuss in \Sec{sss:stas} how to mimic the
 energy pumping and energy damping in our particular problem and what
 are the necessary conditions of stationarity, \Sec{sss:ints}. In
 \Sec{ss-chain-res} we show that the distribution of mode amplitudes
 in long chains is universal in the following sense: it is
 asymptotically independent of number of triads in the chain in the
 limit of large $N$ and of interaction coefficients (in a wide region
 of their definition). Moreover, the distribution for the forced case
 practically coincides with that for free evolution from
 initial conditions, corresponding  to the forced stationary case.

In \Sec{ss:con1} we briefly summarize the main features of
finite-dimensional wave turbulence discovered in this paper and
formulate in \Sec{ss:con2} some important questions in this field that
remain unstudied. Our feeling is that the present paper presents many
more questions than answers, and all these questions (and many other
related ones) belong to a new field of study of weak-wave turbulence:
finite-dimensional and mesoscopic wave turbulence in finite-size
physical systems.

\begin{figure}
  \begin{center}
   \includegraphics[width=4 cm]{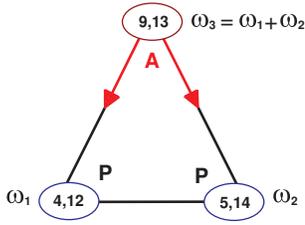} 
  \end{center}\vskip -.5cm
  \caption{\label{f:1}
Color online. Resonant triad $\Delta_1$, see
    \Tab{t:1} below. Wave numbers of modes $m\,, \ \ell$ are shown
    inside ovals. Red arrows are coming from an active mode A (with
    frequency $\o_3=\o_1+\o_2$) and show directions of the energy flux
    to the passive P-modes (with frequencies $\o_1 $ and $\o_2$).  }
\end{figure}%

\section{\label{APW}Atmospheric planetary waves}

\subsection{\label{ss:bar} Barotropic vorticity equation on a   sphere}
Planetary waves in the atmosphere pose a rich and complicated problem
which is influenced by the earth topography, the vertical temperature
profiles (varying between land and ocean), global winds etc. We do not
attempt here to take into account all this richness. The essence of
the interesting dynamics can be gleaned from simplified models.  A
very simplified model of atmospheric planetary waves (discussed,
e.g. by Silberman~\cite{54sil}) is provided by the barotropic
vorticity equation on a rotating sphere for the dimensionless stream
function $\psi(\theta, \varphi,t)$.  The variables $t, \theta$ and
$\varphi$ are the time, the latitude ($-\pi/2 \leq \theta \leq \pi/2$)
and longitude ($0 \leq \varphi \leq 2\pi$) on the sphere. The equation
reads~\cite{54sil}:
\begin{eqnarray}\label{BWE}\nn
 \Delta  \frac{\partial \psi }{\partial t}&=& \frac 1 {\sin\theta}\Big ( \frac{\p \psi }{\p \varphi}\frac \p {\p \theta} - \frac {\p \psi} {\p \theta }\, \frac \p {\p \varphi }\Big)\Big( 2\,\Omega \cos \theta + \Delta \psi \Big) \,, \\
 \Delta &=& \frac 1{\sin \theta}\Big[\frac \p {\p \theta}\Big(\sin \theta \frac{\p }{\p \theta} \Big)+ \frac 1{\sin  \theta }\frac {\p^2  }{\p \varphi ^2}\Big]\,, \end{eqnarray}
 where $\Delta$  is    the angular part of the spherical Laplacian operator.
The stream-function gives rise to the velocity $\B v=
\Omega\,  R  \, [\bf   z  \times \bm \nabla \psi] $, where $\Omega$
and $R$ are the angular velocity and radius of the  Earth and $\bf z$
is the vertical unit  vector.

Equation~\eq{BWE} conserves  the energy $E$ and the enstrophy  $H$,  which are defined by:
\begin{subequations}\label{ints1}
\begin{eqnarray}\label{intE}
E&=& \frac12 \int\limits _0^{2\pi} d\varphi \int\limits  _0^{\pi} |\nabla \psi |^2\sin \theta  d\theta  \,,  \\
\label{intH}
H&=& \frac12 \int\limits _0^{2\pi} d\varphi \int\limits  _0^{\pi}|\Delta \psi|^2 \sin \theta  d\theta \ .
\end{eqnarray}\end{subequations}

\subsection{\label{ss:proj} Projection on the spherical basis}
 The eigenfunctions of the linear part of Eq. (\ref{BWE}) are
 \begin{subequations}\label{eigen}
 \begin{equation}\label{defA}
 \Psi_j  \=\Psi_{\ell_j} ^{m_j} (\theta, \varphi,t) = Y_j (\theta, \varphi)\, \exp (\,i\, \omega_j   \, t)\,,
 \end{equation}
 where the frequencies of planetary waves are:
\begin{equation}\label{SH}
\omega_j\=\omega(\ell_j,m_j)=-  {2\, m_j\Omega} \Big / {\ell_j(\ell_j+1)} \ .
\end{equation}
From now on we use  the shorthand notation $j=(\ell_j,m_j)$ for the eigen-numbers $\ell_j, m_j$ of the  spherical harmonic $Y_{\ell_j}^{m_j}(\theta , \varphi)$
 \begin{equation}\label{defB}
   Y_j\= Y_{\ell_j} ^{m_j}(\theta , \varphi)= P_{\ell_j} ^{m_j}(\cos\theta)\, \exp (i\, m_j\, \varphi)\,,
 \end{equation}
with  the associated Legendre polynomials $P_j\=P_{\ell_j}^{m_j}(\cos\theta)$, normalized as follows:
 \begin{eqnarray}\label{defC}
 \int_0^\pi P_\ell ^m  P_{\ell'}^m \sin \theta d\theta = \delta(\ell, \ell')\ ,\quad
 P_\ell^{-m}=  P_\ell^m \ .
\end{eqnarray}
Here $\delta(\ell, \ell^\prime)$ is the Kronecker symbol (1 for
$\ell=\ell^\prime$ and zero otherwise).  The integer indices $m$ and
$(\ell-m)$ are the longitudinal and latitudinal wave-numbers of the
$\ell,m$-mode; they count the number of zeros of the spherical
function along the longitudinal and the latitudinal directions. Below
we refer to the range of $m$ and $\ell$ as the ``spectral domain". For
the approximation of two-dimensional atmosphere to hold, the
wave-length is supposed to be much smaller than the atmosphere's
hight. If we estimate the wave-length as the distance between the
appropriate zeroes of spherical function, the length of the equator at
about $40\,000$ km and the height of the atmosphere at about 40 km, we
understand that the approximation of a two-dimensional atmosphere
holds up to $\ell\lesssim 1000$.

Expanding the function $ \psi(\theta,\varphi,t)$ in the basis~\eq{defA} we get
  \begin{equation}
      \label{projA}
      \psi(\theta,\varphi,t)=\sum_j   A_j  \Psi_j \ .
  \end{equation}
\end{subequations}

Substituting in Eq. (\ref{BWE}) one obtains the governing equations for the ``slow" amplitudes $A_j\=A_{\ell_j} ^{m_j}(t)$ of the planetary waves:
\begin{subequations}\label{proj}
\begin{eqnarray}\label{projB}
  \frac{d A_j}{dt}
&=&\frac{i}{2\, N_j}\sum _{r,s}N_{r,s}\,
Z_{j|r,s}\, A_r\, A_s\\ \nn 
&&\times \exp[i(\o_r+\o_s-\o_j)t] \delta(m_j,m_r+m_s)\,,\\
N_j&\=&\ell_j(\ell_j+1)\,, \quad  N_{r,s}\=N_r-N_s\,,
\end{eqnarray}\end{subequations}
where  the three-wave interaction coefficients are
\begin{subequations}\label{ia}\begin{equation}\label{projC}
Z_{j|r,s}
\=Z_{\ell_j}^{m_j} |_{\ell_r,\,~\ell_s }^{m_r, m_s}=  \int\limits _
0^\pi  \C Z _{j|r,s}(\theta)\, d\theta\,,
\end{equation}
with the ``interaction integrand"
\begin{eqnarray}\label{projD}
\C Z_{j|r,s}(\theta) =   P_{\ell_j}^{m_j}\Big[
 m_r  P_{\ell_r}^{m_r}
 \frac{d  P_{\ell_s}^{m_s} } {d\theta }-
 m_s  P_{\ell_s}^{m_s}
 \frac{d  P_{\ell_r}^{m_r} } {d\theta }\Big] \ . ~~\end{eqnarray}\end{subequations}

 We note that Eq. (\ref{proj}) is an exact consequence of the
 barotropic vorticity equation, without any assumption about the
 existence of a small parameter. Among the interactions appearing in
 this equations there are many non-resonant ones, in which the
 exponent $ \exp[i(\o_r+\o_s-\o_j)t]$ is not unity. All these
 interactions can be removed by a change of variables, which however
 will result in new nonlinear terms (higher than quadratic), see
 e.g. Sec.~1.1.4 in \cite{ZLF}.  Assuming that the wave amplitudes are
 small enough, all these can be disregarded, bringing the final
 equations back to the same form as in Eq. (\ref{proj}), but including
 only resonant triads for which $ \exp[i(\o_r+\o_s-\o_j)t]=1$.

 \subsection{\label{sss:nes} Necessary conditions for non-vanishing interaction}
The first necessary condition that guarantees  finite  interaction amplitudes
 follows from the axial symmetry of the problem and is reflected in the Kronecker symbol in \Eq{projB}:
 \begin{subequations}\label{cond}\begin{equation}\label{condA}
 m_j=m_r+m_s\ .
 \end{equation}
Second, the spherical symmetry of the nonlinear term in \Eq{BWE} leads
to the conservation of the square of the total angular momentum of the
system. This translates to the triangle inequality for vectors $\B
\ell_j=\B \ell_r+\B \ell_s$ in each triangle with non-zero interaction
amplitude:
 \begin{equation}\label{condB}
|\ell_r-\ell_s| < \ell_j < \ell_r+\ell_s \ .
\end{equation}
Third, the explicit form  of \Eq{projC}  requires  that
 \begin{equation}\label{condC}
 \ell_j+\ell_r+\ell_s\quad \mbox{  is odd.}
 \end{equation}
\end{subequations}
Otherwise integrand~\eq{projD} is odd function of $\cos\theta$ and integral~\eq{projC} is zero.

It can be shown by integrating  \Eq{ia} by parts that whenever Eq. (\ref{condA}) is fulfilled the three interaction coefficients satisfy:
\begin{eqnarray}\label{rel1}
Z_{\ell_j}^{m_j} |_{\ell_r,\,~\ell_s }^{m_r, m_s} =Z_{\ell_s}^{m_s} |_{\ \ \ell_r,~\ell_j}^{-m_r,m_j }= Z_{\ell_r}^{m_r} |_{\ \ell_j,\ \ell_s }^{ m_j,-m_s}\ .
\end{eqnarray}
From this follows that the interaction coefficients $Z_{\dots}$ satisfy two  Jacoby identities
\begin{eqnarray}\label{Jac}
   Z_{j|r,s}+ Z_{r|s,j}+Z_{s|j,r} &=&0\,,\\ \nn
  N_jZ_{j|r,s} + N_r Z_{r|s,j}+N_s Z_{s|j,r} &=&0\ .
\end{eqnarray}
As a result,   \Eqs{proj} have two integrals of motion
\begin{subequations}\label{ints2}
\begin{eqnarray}\label{int2E}
 E&=& \frac12 \sum_j N_j |A_j|^2\,, \\ \label{int2H}
 H&=& \frac12 \sum_j N_j^2 |A_j|^2\,,
\end{eqnarray}\end{subequations}
which are nothing else but  the energy~\eq{intE} and enstrophy~\eq{intH}, presented in the basis~\eq{eigen}.

\begin{figure*}
\begin{center}
\begin{tabular}{|c|c|c|}
  \hline
 $\C A$   & $\C B$ & $\C C$ \\
  ~~\includegraphics[width=5.4cm,height=3cm]{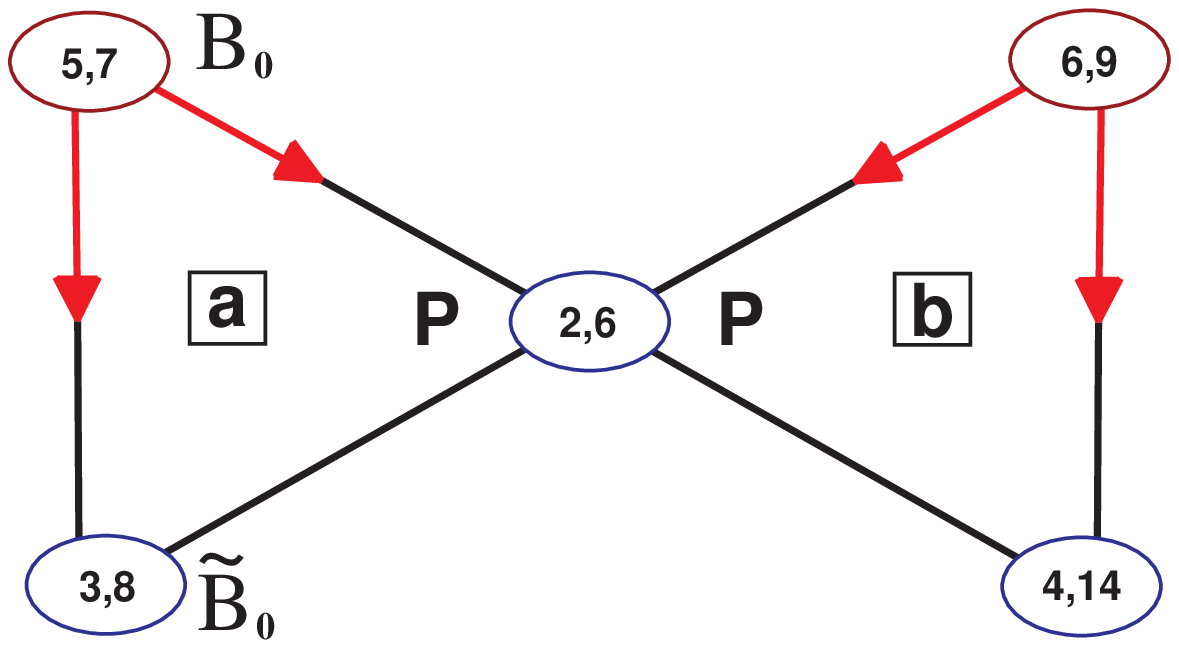}~~ &  
  ~~\includegraphics[width=5.4cm,height=3cm]{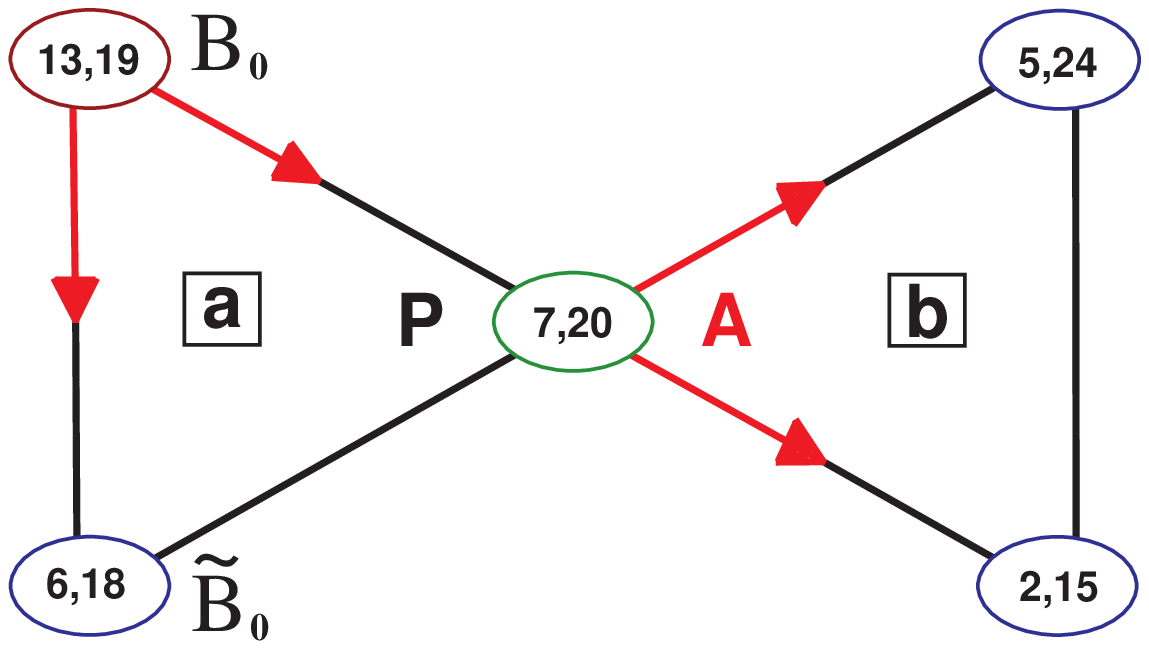}~~&   
 ~~\includegraphics[width=5.4cm,height=3cm]{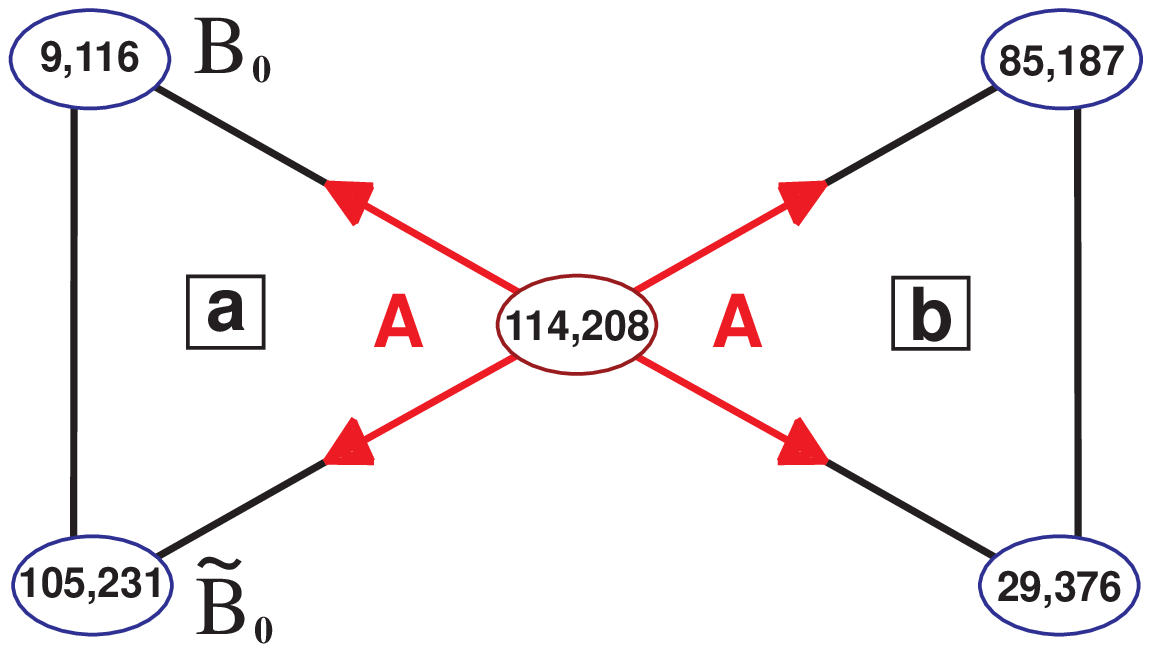}~ \\   
  PP-butterfly   & PA-butterfly &  AA-butterfly \\  \hline
\end{tabular}

\end{center}
\caption{\label{f:but} Color online. Examples of isolated
butterflies. All the notations are as in \Fig{f:1}. In particular: red
arrows are coming from the active modes A and show directions of the
energy flux to the passive P-modes.  The letters ``a" and ``b" in
square boxes denote triads and will be used as subscripts in the
corresponding evolution equations for amplitudes and for integrals of
motion.  In studies of free evolution in \Sec{s:free} the initial
energy is concentrated in the ``leading" $a$-triad in ``individual"
modes with amplitudes $B_0$ and $\~B_0$ and then goes to the ``driven"
$b$-triad.}
\end{figure*}%

\section{\label{S:TT} Small clusters of resonant triads}
In this   Section  we formulate equations of motion and motion invariants of small   clusters of resonant triads,  study their topology and other properties,  that affect on the dynamics of finite-dimensional wave turbulence. Brief analysis of these questions was given in \Ref{KL-08}.

\subsection{\label{ss:TD} Active and passive modes in a resonant triad}
In this paper we refer to a ``resonant triad" whenever we have three modes $(j,r,s)$ whose frequencies
satisfy the triad resonance condition $\o_j=\o_r+\o_s$. Accordingly, the equations for the slow amplitudes of modes in  resonant triads are written, (after relabeling  according to $r\to 1$, $s\to 2$ and $j\to 3$) as follows:
 \begin{eqnarray}\nn
 N_1\frac{dA_1}{dt}&=& i N_{3,2 } Z A_2^* A_3\,, \quad Z\=Z_{3|1,2} \,,  \\ \label {triadAB}
N_2\frac{dA_2}{dt}&=& i   N_{1,3} Z  A_1^* A_3\,,  \\ \nn
N_3\frac{dA_3^*}{dt}&=& i N_{2,1} Z  A_1^* A_2^*\ .
\end{eqnarray}
 In this case the conservation laws~\eq{ints2} take the form:
 \begin{subequations}\label{ints3}
\begin{eqnarray}\label{int23}
E&=&   \frac 12 \left(N_1 |A_1|^2+ N_2 |A_2|^2+ N_3 |A_3|^2\right) \,, \\ \label{int23}
H&=&   \frac 12 \left(N_1^2 |A_1|^2+ N_2 ^2|A_2|^2+ N_3 ^2|A_3|^2 \right)\ .
\end{eqnarray}\end{subequations}
Taking for concreteness an example for which
\begin{equation}
\ell_1>\ell_3>\ell_2 \ , \label{example}
\end{equation}
we make  in \Eqs{triadAB} a  linear change of
variables $B_i=\a_i A_i$ such that
\begin{eqnarray}\nn
\a_1&=&-i\sqrt{N_{1,2}N_{1,3}}/ \sqrt{N_2N_3}\,, \\ \label{a}
\a_2&=&i\sqrt{N_{1,2}N_{3,2}}/ \sqrt{N_1N_3}\,, \\
\a_3&=&i\sqrt{N_{1,3}N_{3,2}}/ \sqrt{N_1N_2}\ . \nn
\end{eqnarray}
This results in equations with real coefficients that involve only one interaction amplitude $Z$:
 \begin{eqnarray}\nn
 \frac{dB_1}{dt}&=&     Z B_2^* B_3\,,    \\ \label {3}
\frac{dB_2}{dt}&=&     Z  B_1^* B_3\,,  \\ \nn
\frac{dB_3}{dt}&=& - Z  B_1 B_2 \ . \nn
\end{eqnarray}
 This is a dynamical system corresponding to the  simplest possible resonant cluster.
 Equations \eq{3} are symmetric with respect to replacing two low-frequency modes $1\Leftrightarrow 2$.
The mode with highest frequency (which in this paper will be always denoted by subscript ``~$_3~$") is special. When Eq. (\ref{example}) does not hold one can find a similar change of variable
for any relations between the magnitudes of the three indices $\ell_j$.

The system~\eq{3}  has two independent conservation laws (known as Manley-Rowe integrals)
 \begin{eqnarray}\label{MR}
 \begin{cases}
 I_{23}=|B_2 |^2 + |B_3|^2 =( E\, N_1-H ){N_{23} }/{N_1 N_2
 N_3}\,,\\
 I_{13}= |B_1 |^2 + |B_3|^2 =( E\, N_2-H ) {N_{13 }}/{N_1 N_2
 N_3}\,,\\
 I_{12}=I_{13}-I_{23}= |B_1 |^2 -|B_2 |^2\,,
 \end{cases}
\end{eqnarray}
which are linear combinations of the energy $E$ and enstrophy $ H$
defined by \Eqs{ints3}.

Obviously, \Eq{3} can be written in the Hamiltonian form:
\begin{subequations}\label{Ham}
\begin{equation}\label{HamA}
i\, \frac{dB_j}{dt}= \frac{d H\sb{int}}{d B_j^*}\,,
\end{equation}
with the interaction Hamiltonian
\begin{equation}\label{HamB} H\sb{int}=i Z (B_1 B_2 B_3^*- B_1^* B_2^* B_3)\,,
\end{equation}
which is an additional integral of motion. In terms of old variables $A_j$
\begin{equation}\label{HamC} H\sb{int}= iZ \, \frac{N_{1,2}N_{1,3}N_{3,2}}{N_1 N_2 N_3}(A_1 A_2 A_3^*- A_1^* A_2^* A_3)\ .
\end{equation}
By direct calculation it is easy to check that  \Eq{HamC} is an integral of motion of the dynamical system~\Eq{triadAB}.
\end{subequations}

On the face of it Eqs. (\ref{3}) involves six dynamical variables: i.e. $B$'s and their complex conjugates.
In fact, using the standard  representation of the complex amplitudes $B_j$ in terms of real amplitudes $C_j$ and phases $\theta_j$:
\begin{subequations}\label{ap}\begin{equation}\label{apA}
B_j= C_j\exp(i \theta_j)\,,
\end{equation}
one recognizes that the right-hand-side (RHS) of \Eq{3} depends only on a single combination of phases in the triad which affects the dynamics. We refer to this combination as the \emph{triad phase}:
\begin{equation}\label{dc}
 \varphi\= \theta_1+\theta_2-\theta_3\ .
 \end{equation}\end{subequations}
The triad phase appears  in \Eq{HamB} as follows:
 \begin{equation}\label{HamD} H\sb{int}= - 2\, Z  | B_1 B_2 B_3| \sin \varphi \ .
\end{equation}
Thus we have a four
dimensional phase space with three integrals of motion, resulting in a simple periodic trajectory for
almost all conditions. For more details see ~\cite{KL-07}. Nevertheless even this simple dynamics
offers the first opportunity to discuss the energy flow within a cluster of interacting modes.

To this aim we discuss the evolution of the triad of amplitudes
with special initial conditions, when only one mode is appreciably
excited at zero time. If $B_1(t=0)\gg B_2(t=0)$ and $B_1(t=0)\gg B_3(t=0),$ then
$I_{23}(t=0)\ll I_{13}(t=0)$. The integrals of motion are independent of
time, therefore $I_{13} \gg I_{23}$ at all later times, and hence
$|B_1(t)|^2\gg |B_2(t)|^2$. Moreover, $|B_1(t)|^2\gg |B_3(t)|^2$ at
all times. Indeed, the assumption $|B_1(t)|^2 \lesssim |B_3(t)|^2$
yields $I_{13} \simeq I_{23}$, which is not tenable. This means
that  the $\omega_1$-mode, being the only essentially exited one at $t=0$
cannot redistribute its energy to   the other two modes in the triad.
The same is true for the $\omega_2$-mode. For this reason
we refer to the  lower
 frequency modes with frequencies $\omega_1<\omega_3$  and $\omega_2<\omega_3$ ``passive modes", or \emph{P-modes}.

On the other hand, the conservation laws~\eq{MR} cannot restrict the growing of P-modes
from initial conditions when only $\omega_3$-mode is appreciably
excited. In this case the P-mode amplitudes will grow exponentially~\cite{Has-67}:
$|B_1(t)|\,,\ |B_2(t)|\propto \exp [\, |Z B_3(t=0)| t]$ until all
the modes will have comparable magnitudes of their amplitudes.
Therefore we refer to the  $\omega_3$-mode as an ``active mode", or \emph{A-mode}.
An A-mode, being initially excited, is capable of shearing  its energy with
two P-modes within the triad.

\subsection{\label{ss:but} Double-triad clusters: Butterflies}
An arbitrary cluster in our wave system is a set of connected
triads. Examples of the simplest clusters, consisting of two triads
connected via one common mode are shown in \Fig{f:but}. They will be
referred to as \emph{butterflies}. Note that in principle one can have
two triads connected by {\em two} modes, but such a cluster does not
satisfy all the conditions stated above.

  The dynamics of a cluster depends on the type of the mode which is
 common for the neighboring triads.  Correspondingly we can
 distinguish three types of butterflies: PP-, AP- and AA-butterflies,
 In this Section we consider the equations of motion, the invariants
 and the restrictions on dynamical behavior, that follow from the
 existence of invariants for relatively small clusters consisting of
 two triads.  In the following Secs. we consider these questions for
 clusters consisting of three and six triads.

 Butterflies, as shown in \Fig{f:but},  consist  of two   triads  $\ a\ $ and $\ b,\ $
with wave amplitudes  $\ B_{j|a}, \ B_{j|b},\ $ $j=1,2,3$,  connected {\it via} one common mode.   For PP-butterfly the common mode is passive in both triads, say
\begin{subequations}\label{buts}
\begin{equation}\label{butsA}\ B_{1|a}=B_{1|b}\qquad\mbox{PP-butterfly}\,;
\end{equation}
for an PA-butterfly the common mode is passive in $a$-triad and active in the second, $b$-triad:
\begin{equation}\label{butsB}\ B_{1|a}=B_{3|b}\qquad\mbox{PA-butterfly}\,;
\end{equation}
while  for  an AA-butterfly the common mode is active  in both triads:
\begin{equation}\label{butsC}\ B_{3|a}=B_{3|b}\qquad\mbox{AA-butterfly}\ .
\end{equation}
\end{subequations}

The equations of motion for these systems follow from the \Eqs{proj}
under the condition of small nonlinearity and from the resonance
conditions in both triads:
 \begin{equation}\label{but-res}
 \omega_{1|a}+\omega_{2|a}=\omega_{3|a}\,, \quad \omega_{1|b}+\omega_{2|b}=\omega_{3|b}\,,
 \end{equation}
 with the obvious requirement that the frequencies of the common modes
 are the same.  After a change of variables similar to Eqs. (\ref{a}) and elimination of one  common mode,
 the resulting equations for PP-butterfly ($B_{1|a}=B_{1|b}$) are:
\begin{subequations}\label{EMbuts} \begin{eqnarray}\label{PP}
\begin{cases}
\dot{B}_{1|a}=  Z_a B_{2|a}^*B_{3|a} +   Z_b B_{2|b}^*B_{3|b}, \,, \\
\dot{B}_{2|a}=  Z_a B_{1|a}^* B_{3|a}\,,\quad    \dot{B}_{2|b}=  Z_b B_{1|a}^* B_{3|b}\,, \\
\dot{B}_{3|a}=  - Z_{a} B_{1|a} B_{2|a} \,,\    \dot{B}_{3|b}=  - Z_{b} B_{1|a} B_{2|b}\ . \\
\end{cases}
 \end{eqnarray}%
For PA-butterfly (with $B_{1|a}=B_{3|b}$) they are
\begin{equation}\label{AP}
 \begin{cases}
 \dot{B}_{1|b}=  Z_b B_{2|b}^*B_{3|b}\,, \quad  \dot{B}_{3|a}=  - Z_a B_{3|b} B_{2|a}\,, \\
\dot{B}_{2|b}=  Z_b B_{1|b}^* B_{3|b}\,,\quad    \dot{B}_{2|a}=  Z_a B_{3|b}^* B_{3|a}\,, \\
\dot{B}_{3|b}=  - Z_b B_{1|b} B_{2|b}  + Z_a B_{2|a}^* B_{3|a}\,,
 \end{cases}
 \end{equation}
  and for AA-butterfly  (with $B_{3|a}=B_{3|b}$):
\begin{eqnarray}\label{AA}
\begin{cases}
\dot{B}_{1|a}=  Z_{a} B_{2|a}^* B_{3|a}\,,\    \dot{B}_{1|b}=  + Z_{b}  B_{2|b}^*B_{3|a}\,, \\
\dot{B}_{2|a}=  Z_a B_{1|a}^* B_{3|a}\,,\quad    \dot{B}_{2|b}=  Z_b B_{1|b}^* B_{3|a}\,, \\
\dot{B}_{3|a}= - Z_a B_{1|a}B_{2|a} -   Z_b B_{1|b}B_{2|b} \ . \\
\end{cases}
 \end{eqnarray}%

\end{subequations}

All these equations can be obtained from the canonical equations of motion (\ref{HamA}) using the Hamiltonian
\begin{equation}
H_{\rm int} =2{\rm Im} \left\{Z_aB_{1|a}^*B_{2|a}^*B_{3|a} +Z_b B_{1|b}^*B_{2|b}^*B_{3|b}\right\} \ , \label{Hambutt}
\end{equation}
in which the conditions (\ref{buts}) have to be fulfilled for each particular butterfly.

 \begin{figure*}
\begin{center}
\begin{tabular}{|c||c|}
  \hline   $ \C A$ & $\C B$ \\
 ~~\includegraphics[width=6.5cm,height=4.4cm]{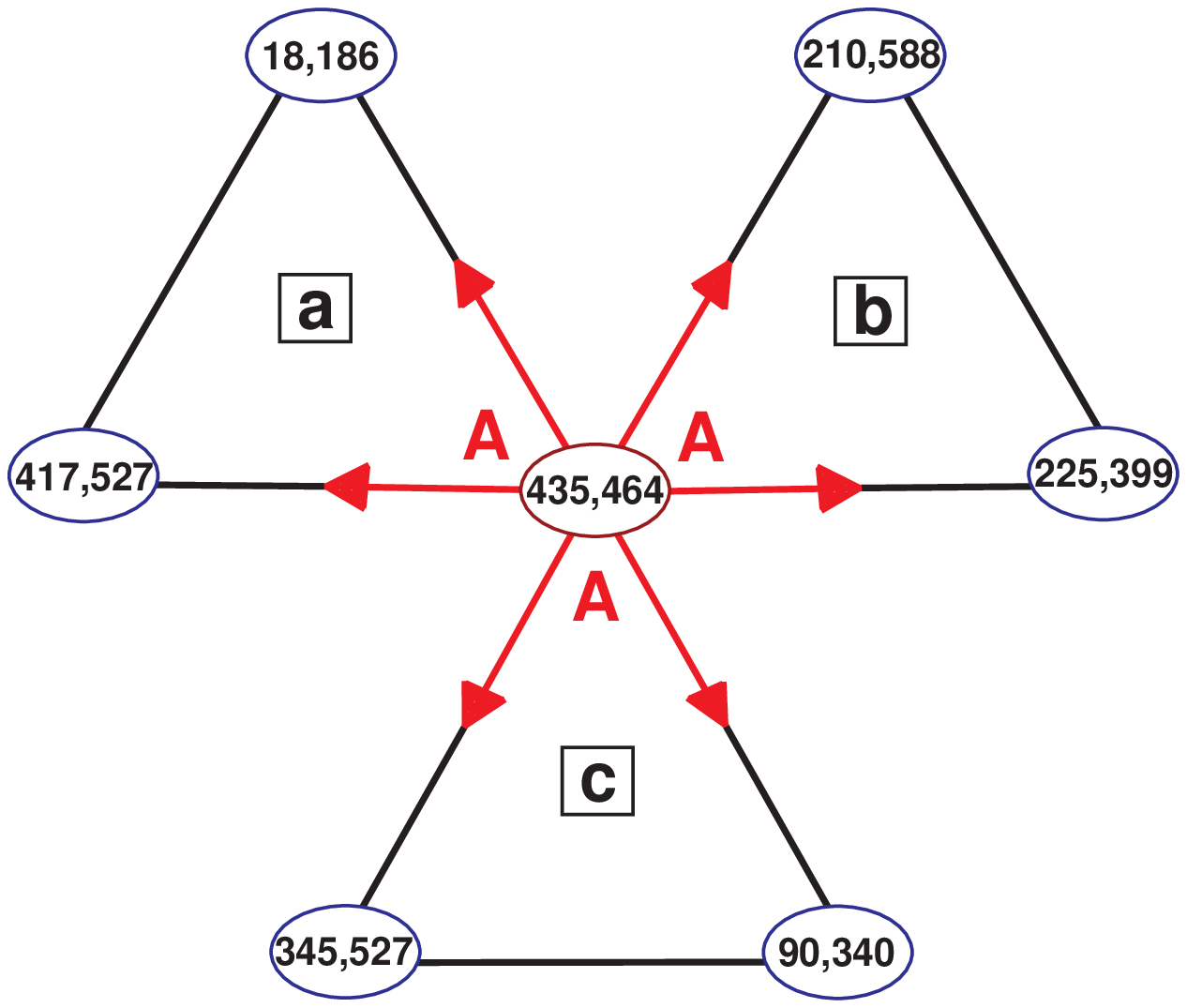}~~& 
~~\includegraphics[width=6.5cm,height=4.4cm]{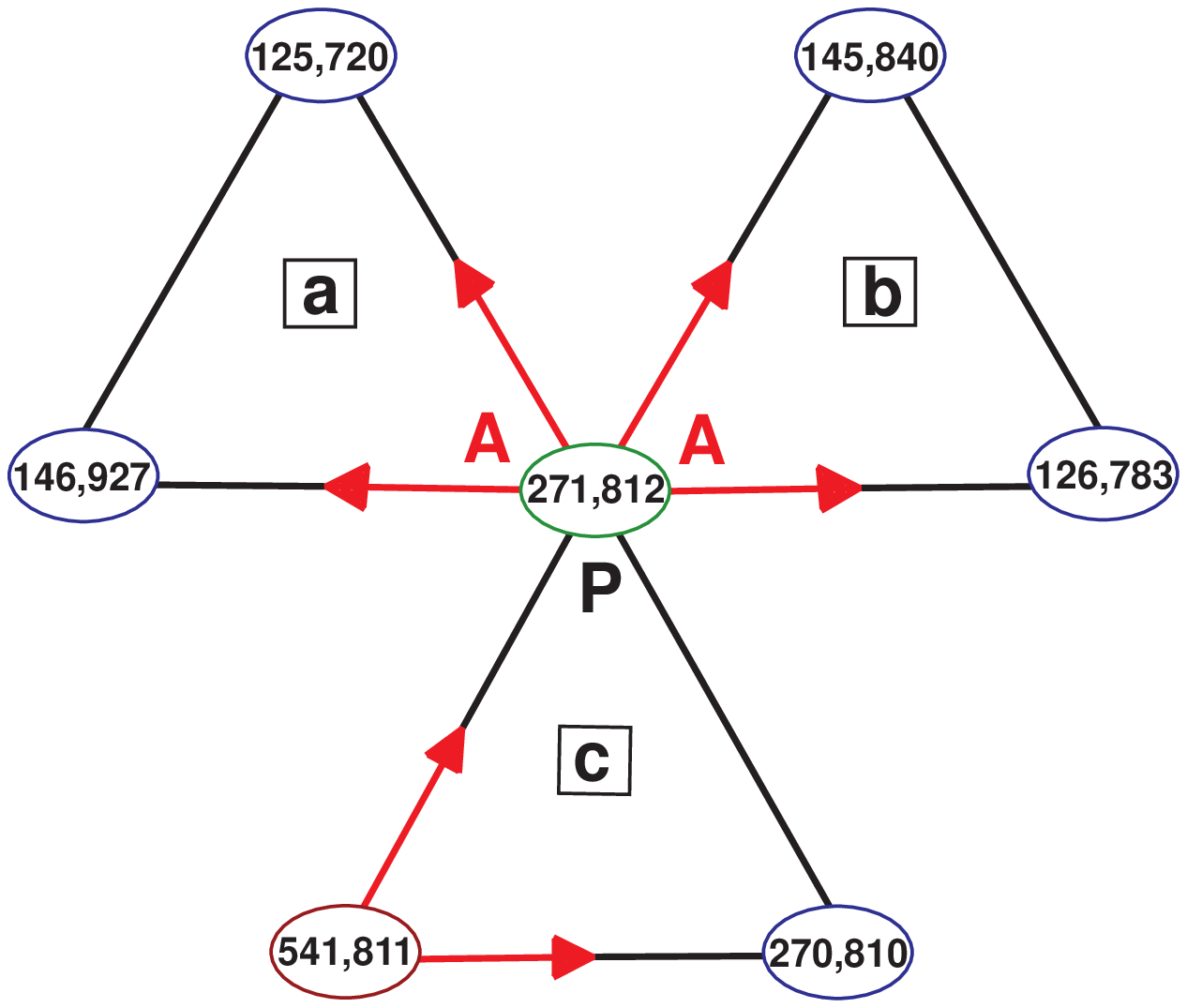}~~\\ 
One of tree AAA-stars &  One of five AAP-stars\\ \hline\hline
 $\C C$& $\C D$ \\
  \includegraphics[width=6.5cm,height=4.4cm]{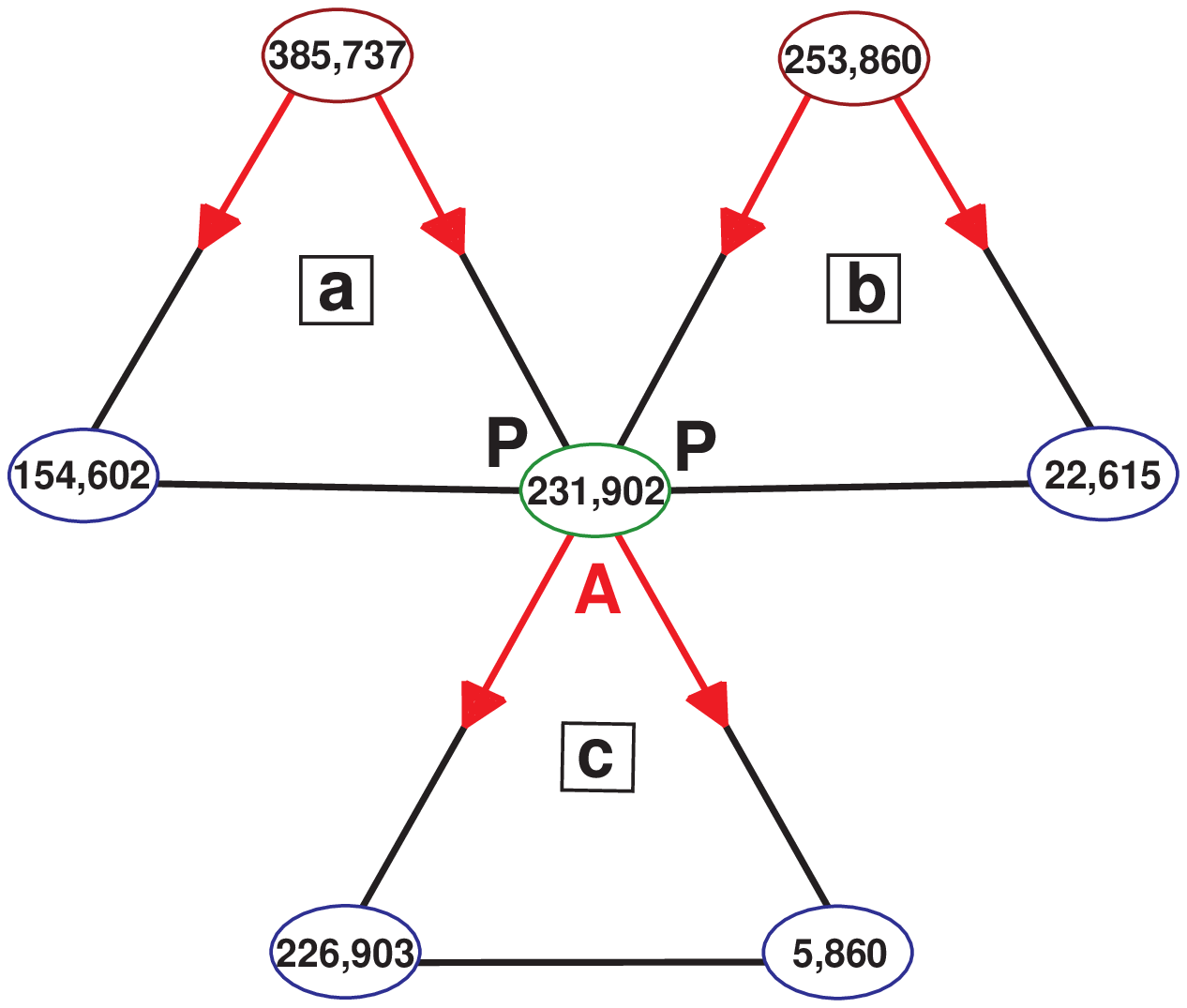}~ &  
 \includegraphics[width=6.5cm,height=4.4cm]{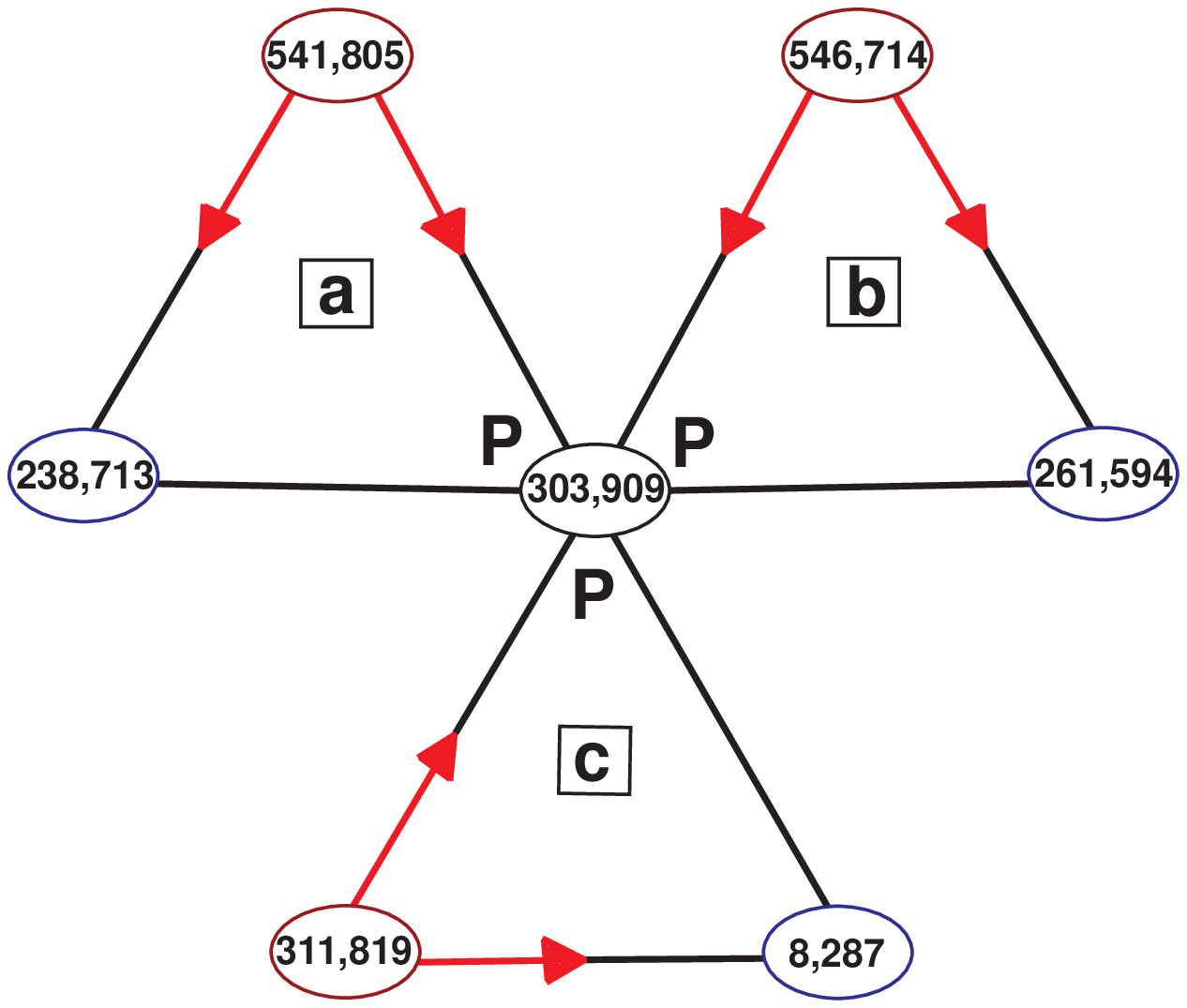}~~ \\ 
One of six APP-stars &  One of eleven  PPP-stars \\  \hline
\end{tabular}

 \end{center}
\caption{\label{f:stars} Color online. Examples of isolated triple stars in the  spectral domain $m\le \ell\le 1000$.  The letters ``a", ``b"  and ``c" in square boxes denote triads and will be used as subscripts in the corresponding evolution equations for amplitudes and for integrals of motion. All other notations as in \Fig{f:1}  and \ref{f:but}.}
\end{figure*}%

\begin{figure*}
\begin{center}
\begin{tabular}{  |c||c||c| }
  \hline

    $\C A$~~~~~~~~~~~~~~     & $\C B$~~~~~~~~~~~~~~&  $\C C$~~~~~~~~~~~~~~\\
  ~~\includegraphics[width=5.5cm,height=2.5 cm]{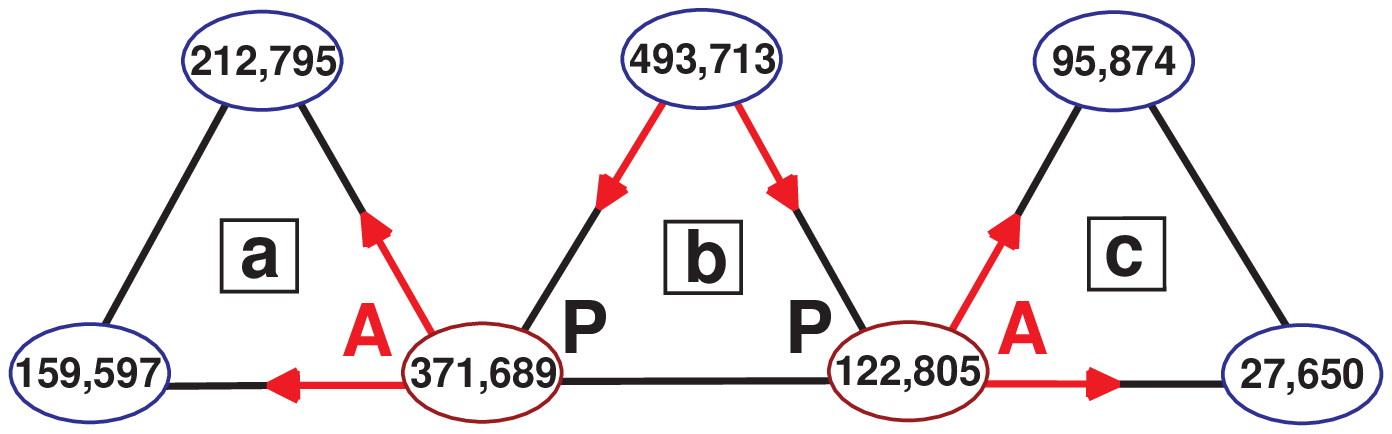}~~& 
  ~\includegraphics[width=5.5cm,height=2.5 cm]{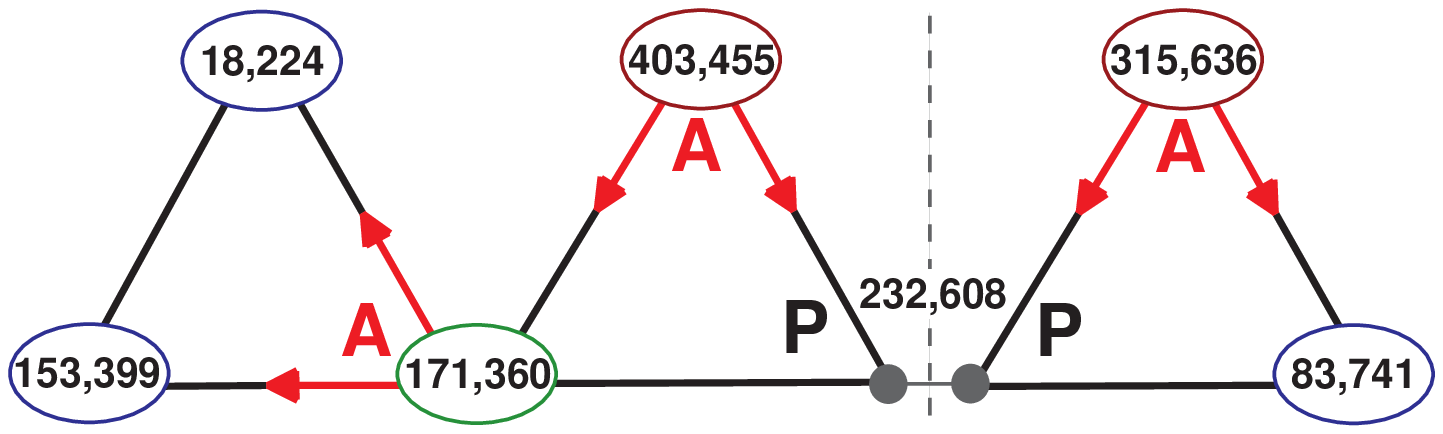}~ & 
  ~\includegraphics[width=5.5cm,height=2.5 cm]{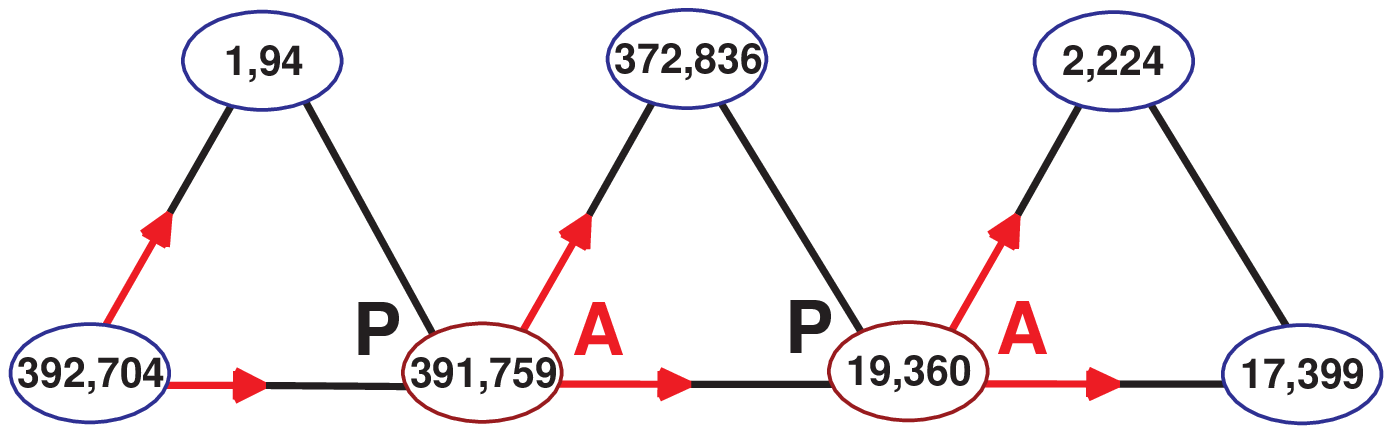}~~\\ 
  AP-PA triple chain cluster   &   One of five AP-PP  chains  &   One of six PA-PA chains  \\ \hline\hline
   $\C D$~~~~~~~~~~~~~~     & $\C E$~~~~~~~~~~~~~~&  $\C F$~~~~~~~~~~~~~~\\
 ~\includegraphics[width=5.5cm,height=2.5 cm]{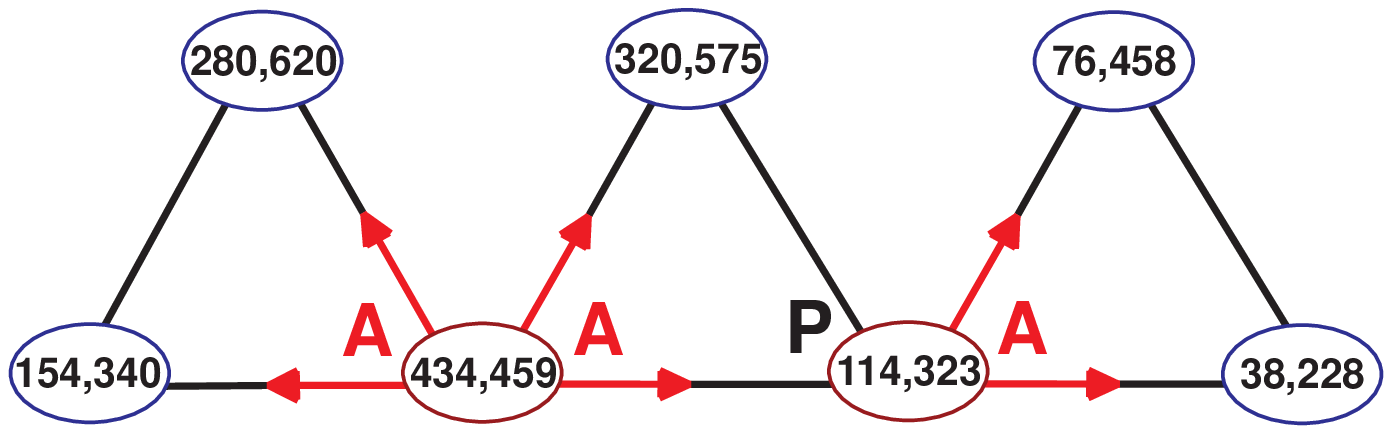}~ & 
~\includegraphics[width=5.5cm,height=2.5 cm]{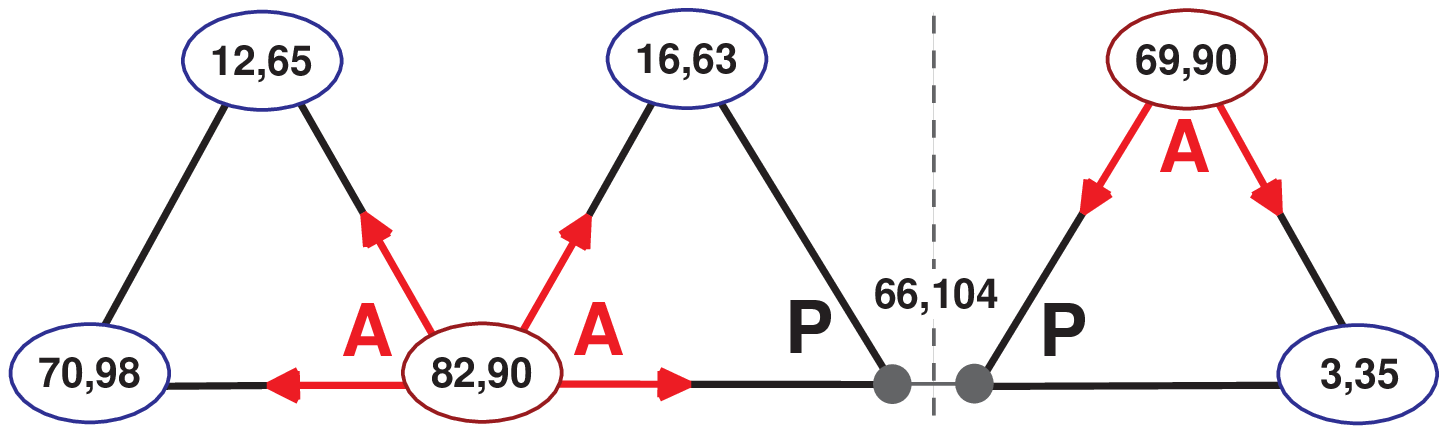}~& 
 ~\includegraphics[width=5.5cm,height=2.5 cm]{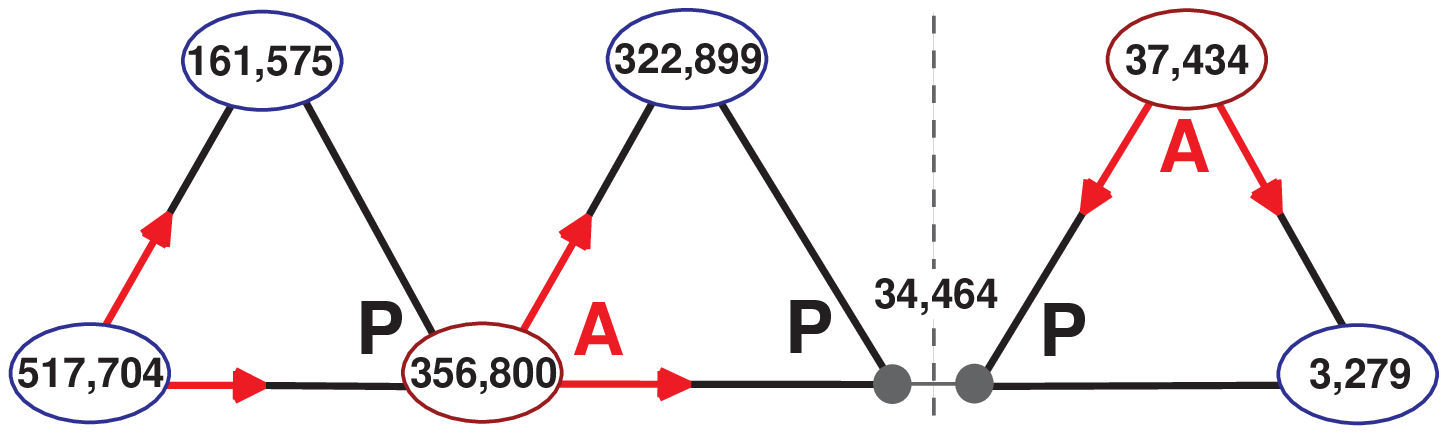}~~\\ 
   One of six AA-PA chains &   One of fourteen  AA-PP chains &
    One of sixteen  PA-PP chains\\ \hline\hline
      $\C G$~~~~~~~~~~~~~~     & $\C H$~~~~~~~~~~~~~~&  $\C I$~~~~~~~~~~~~~~\\
 ~\includegraphics[width=5.5cm,height=2.5 cm]{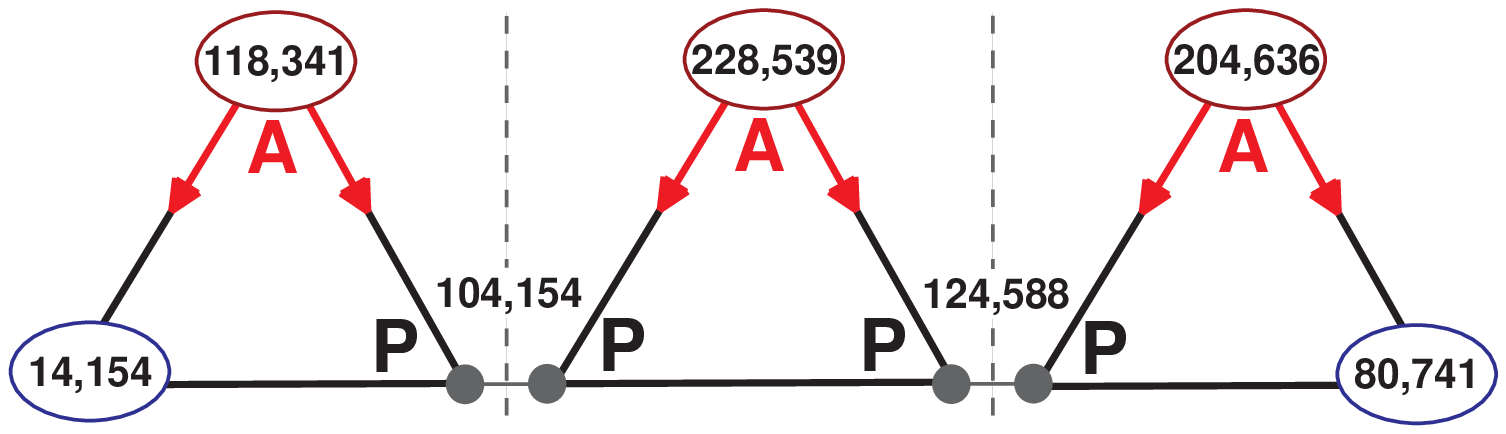}~ & 
  ~\includegraphics[width=4.2cm,height=3.8cm]{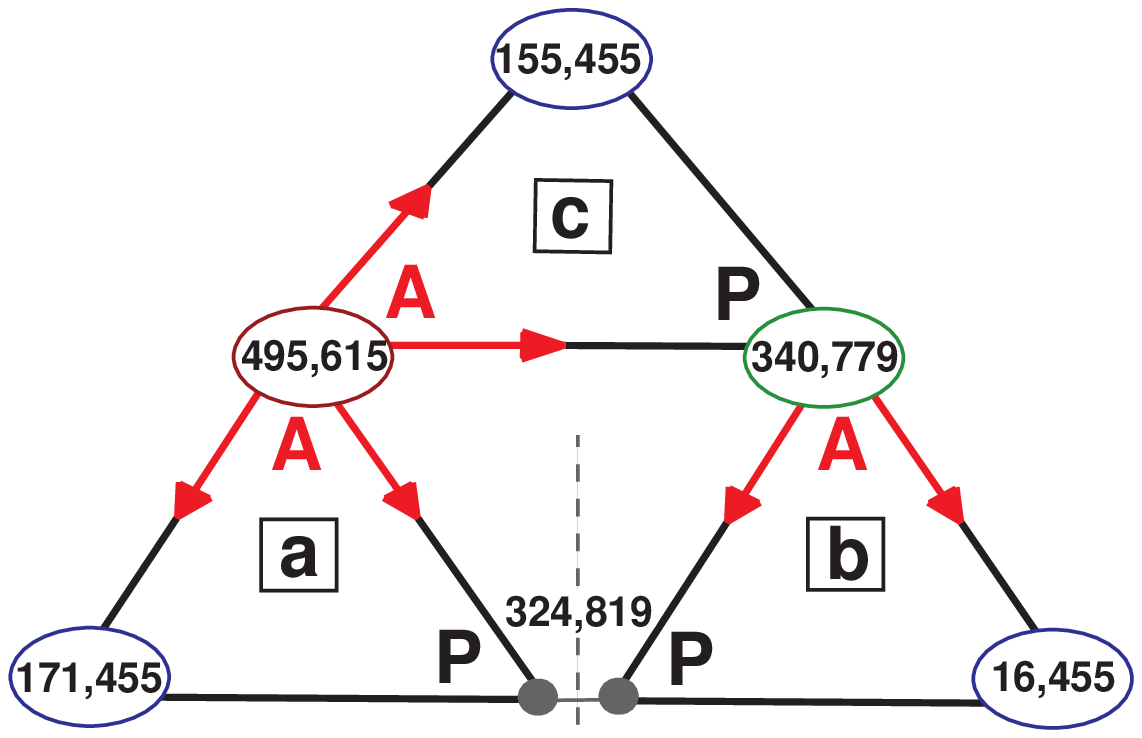}~& 
  ~\includegraphics[width=4.2cm,height=3.8cm]{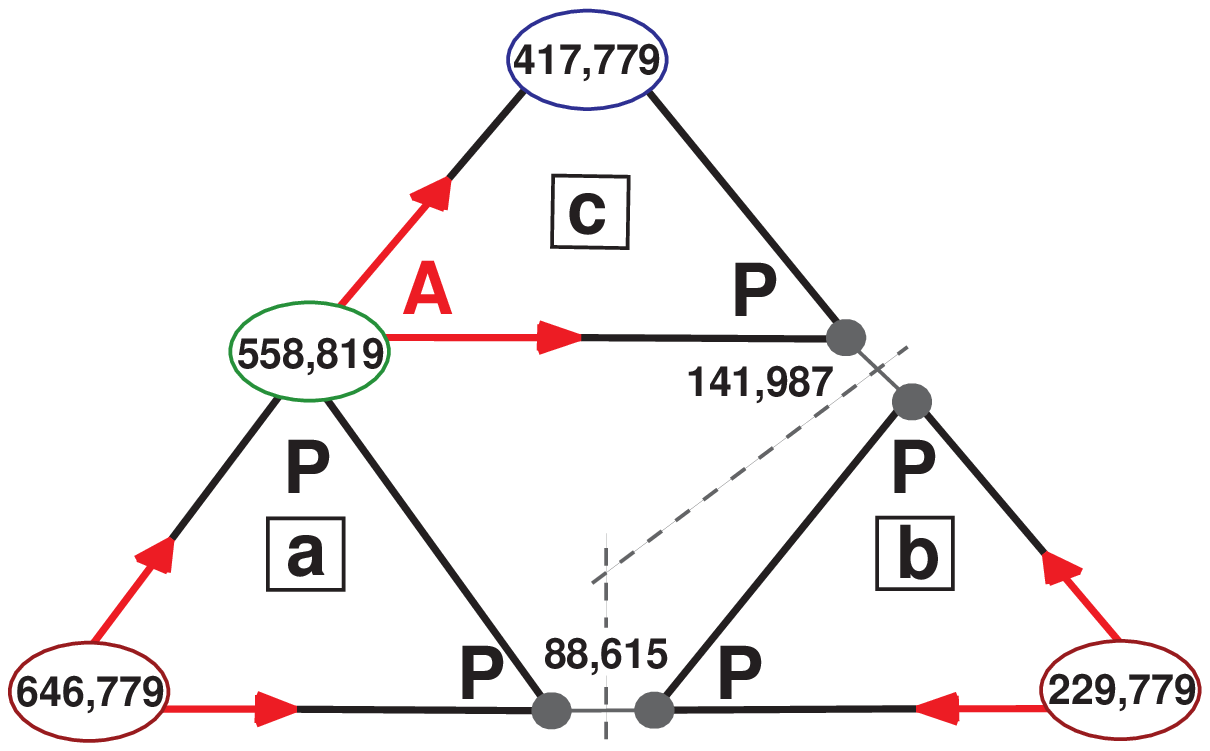} ~~\\ 
  one of eighteen  PP-PP chains  &   One of two AA-PA-PP triangles &  One of two PA-PP-PP triangles \\ \hline
\end{tabular}
 \end{center}
\caption{\label{f:3-lin}Color online. In the  spectral domain $m\le \ell\le 1000$  there are 66 isolated triple-chain clusters of seven types (examples are shown in  panels $\C A- \C F$) and  four triangle clusters with (AA-PA-PP)- and (PA-PP-PP)-connections (examples are shown in  panels $\C H\,,  \C I$)).        Common  PP-modes are split, denoting  difficulty in the energy exchange between corresponding triads. Dashed lines separate PP-irreducible sub-clusters, discussed in \Sec{ss:PPred}. }
\end{figure*}

In addition to the Hamiltonian (\ref{Hambutt}) we have three more invariants of the Manley-Rowe type for each butterfly. For the PP-, PA- and AA-butterflies they respectively  are
\begin{subequations}
\label{MR2}
\begin{eqnarray}\nn
  I_{2,3|a}&=&|B_{2|a} |^2 + |B_{3|a}|^2 \,, \quad
 I_{2,3|b}= |B_{2|b} |^2 + |B_{3|b}|^2 \,, \\ \label{int}
 I_{|a,b}&=& |B_{1|a}|^2 + |B_{3|a} |^2 + |B_{3|b}|^2 \ , \ {\rm PP;}
\\ \nn
  I_{1,2|b}&=&|B_{1|b} |^2 - |B_{2|b}|^2 \,, \quad
 I_{2,3|a}= |B_{2|a} |^2 + |B_{3|a}|^2 \,, \\ \label{APintC}
   I_{|a,b}&=&|B_{1|b}|^2+|B_{3|b} |^2 + |B_{3|a}|^2\ , \  {\rm PA;}
 \\ \nn
 I_{1,2|a}&=&|B_{1|a} |^2 - |B_{2|a}|^2 \,, \quad
 I_{1,2|b}= |B_{1|b} |^2 - |B_{2|b}|^2 \,, \\ \label{AAint}
   I_{|a,b}&=&|B_{1|a}|^2+ |B_{1|b} |^2 + |B_{3|a}|^2\ , \quad {\rm AA.}
\end{eqnarray}
  \end{subequations}
  The first two invariants for the PP butterfly,   $I_{2,3|a}$ and $I_{2,3|b}$, do not involve the common mode
$ B_{1|a}=B_{1|b}$, and are similar to the invariant $I_{23}$,  \Eq{MR},  for an isolated triad.
We can make the following observation: If
at $t=0$  the amplitudes in one triad exceed
substantially the two remaining amplitudes of the butterfly, that is
$|B_{1|a}|, |B_{2|a}|, |B_{3|a}|\gg |B_{2|b}|, |B_{3|b}|$, this relation
persists. In other words, in PP-butterfly when any of the two triads, $a$ or
$b$, has initially very small amplitudes, it is unable to absorb the
energy from  the second triad during the nonlinear evolution.

 The  invariants $I_{1,2|b}$ and $I_{2,3|a}$ for the PA butterfly do not involve the
  common mode $\ B_{3|b}=B_{1|a}$\,; they are similar to the corresponding integrals $I_{12}$ and $I_{23}$, \Eqs{MR}, for an isolated triad.
If at $t=0$ the $b$-triad is excited much
 more than the $a$-triad (and thus $\ I_{1,2|b} \gg I_{2,3|a}$) the smallness
 of the positively defined invariant $I_{2,3|a}$ prevents the $a$-triad from
 absorbing energy from $b$-triad during the time evolution. The situation is different
  when the  $a$-triad is initially excited and $\ I_{2,3|a} \gg I_{1,2|b}$. In this case the  initial
  energy of $a$-triad can be easily shared with $b$-triad. The smallness of $I_{1,2|b}$ only
  requires that during evolution $|B_{1|b}|\approx |B_{2|b}|$. Under this type of the initial conditions we will call the $a$-triad ``leading" triad, while the $b$-triad will be referred to as ''driven" triad.

Finally, the invariants for the AA butterfly,  $I_{1,2|a}$ and $I_{1,2|b}$,  do not involve the common mode $\ B_{3|a}=B_{1|b} $
 and are similar to  $I_{12}$, \Eqs{MR}, for isolated triad.
Simple analysis of these integrals of motion shows that energy,
initially held in one of the triads will be shared between both
triads dynamically.

The conclusion that we can draw from these examples is general:  any triad which is
connected to any given cluster of any size whatsoever where the connection occurs
via its passive mode cannot absorb the energy from the cluster, if
initially  the triad is not excited. In contrast,  a triad connected
to a cluster of any given size via an active mode can freely adsorb energy from the
cluster during the nonlinear evolution.

\subsection{\label{ss:3-triads} Triple-triad clusters: stars, chains and triangles}
Triple triad clusters consist of   three triads, denoted as $a$-, $b$- and $c$-triads with the mode amplitudes denoted as $B_{j|a}$,  $B_{j|b}$,   $B_{j|c}$,  $j=1,2,3$. There are three topologically different types of triple-triad clusters: with one common mode -- stars, shown in \Figs{f:stars}; with two common modes -- chains,  and with three common modes -- triangles, these clusters are shown   in   \Fig{f:3-lin}. Having in mind different types of common modes one distinguishes  13 types of triple-triad clusters, including four stars (AAA-, AAP-, APP-,  PPP-stars,  \Fig{f:stars}), 7 types of 3-chain-,  and two types of triangle-clusters,   in  \Fig{f:3-lin}. All motion equations can be written in the  canonical form~\eq{HamA}
with the Hamiltonian 
\begin{eqnarray}\nn
H_{\rm int} &=&2{\rm Im} \Big\{Z_aB_{1|a}^*B_{2|a}^*B_{3|a} +Z_b B_{1|b}^*B_{2|b}^*B_{3|b} \\
&& ~~~~~~+Z_c B_{1|c}^*B_{2|c}^*B_{3|c}\Big\} \ , \label{Ham3}
\end{eqnarray}
in which one has to equate amplitudes of common modes.

\paragraph{Stars} have one common mode in three triads.  For example, taking
$B_{3|a}=B_{3|b}=B_{3|c}$ in \Eq{Ham3} one gets from \Eq{Ham}  equations of motion for AAA-stars:
\begin{subequations}\label{star-eqs}  \begin{eqnarray}\label{AAA-eqs}
\begin{cases}
\dot B_{1|a}=Z_a B_{2|a}B_{3|a}\,, \quad \dot B_{2|a}=Z_a B_{1|a}B_{3|a}\,, \\
\dot B_{1|b}=Z_b B_{2|b}B_{3|a}\,, \quad  ~ \dot B_{2|b}=Z_a B_{1|b}B_{3|a}\,, \\
\dot B_{1|c}=Z_c B_{2|c}B_{3|a}\,, \quad ~\,  \dot B_{2|c}=Z_c B_{1|c}B_{3|a}\,, \\
\dot B_{3|a}=-Z_a B_{1|a}B_{2|a} -Z_b B_{1|b}B_{2|b} -Z_c B_{1|c}B_{2|c}\ .
\end{cases}
\end{eqnarray}
Taking $B_{3|a}=B_{3|b}=B_{1|c}$ one gets  for AAP-stars:
\begin{eqnarray}\label{AAP-eqs}
\begin{cases}
\dot B_{1|a}=Z_a B_{2|a}B_{3|a}\,, \quad \,\,\dot B_{2|a}=Z_a B_{1|a}B_{3|a}\,, \\
\dot B_{1|b}=Z_b B_{2|b}B_{3|a}\,, \quad~\,\, \dot B_{2|b}=Z_a B_{1|b}B_{3|a}\,, \\
 \dot B_{3|c}=-Z_c B_{3|a}B_{2|c}\,,  \quad  \dot B_{2|c}=Z_c B_{3|a}B_{3|c}\,,\\
\dot B_{3|a}=-Z_a B_{1|a}B_{2|a} -Z_b B_{1|b}B_{2|b} + Z_c B_{2|c}B_{3|c}\ .
\end{cases}
\end{eqnarray}
Similarly one gets motion equations for:
 \begin{eqnarray}\label{APP}\nn
&& \mbox{PPA-star:}\qquad B_{1|a}=B_{1|b}=B_{3|c}\,, \\
\label{PPA-eqs}
&& \begin{cases}
\dot B_{3|a}=-Z_a B_{1|a}B_{2|a}\,, \quad \dot B_{2|a}=Z_a B_{1|a}B_{3|a}\,, \\
\dot B_{3|b}=-Z_b B_{1|a}B_{2|b}\,, \quad\  \dot B_{2|b}=Z_a B_{1|a}B_{3|b}\,, \\
\dot B_{1|c}=Z_c B_{2|c}^*B_{1|a}\,,\qquad \dot B_{2|c}=Z_c B_{1|a}B_{1|c}\,,  \\
\dot B_{1|a}=Z_a B_{2|a}^*B_{3|a} + Z_b B_{2|b}^*B_{3|b} - Z_c B_{1|c}B_{2|c}\ .
\end{cases}
\\ \label{PPP}\nn
&& \mbox{PPP-star:}\qquad B_{1|a}=B_{1|b}=B_{1|c}\,, \,,    \\
\label{PPP-eqs}
&& \begin{cases}
\dot B_{3|a}=-Z_a B_{1|a}B_{2|a}\,, \quad \dot B_{2|a}=Z_a B_{1|a}B_{3|a}\,, \\
\dot B_{3|b}=-Z_b B_{1|a}B_{2|b}\,, \quad\  \dot B_{2|b}=Z_a B_{1|a}B_{3|b}\,, \\
\dot B_{3|c}=-Z_c B_{1|a}B_{2|c}\,,\quad\  \dot B_{2|c}=Z_c B_{1|a}B_{3|c}\,,  \\
\dot B_{1|a}=Z_a B_{2|a}^*B_{3|a} + Z_b B_{2|b}^*B_{3|b} + Z_c B_{2|c}B_{3|c}\ .
\end{cases}
\end{eqnarray}
\end{subequations}
In addition to  Hamiltonian, all triple-star clusters have four invariants of the Manley-Rowe type, three of them does not involve the common mode. For example, for
\begin{eqnarray}\label{APPint}
\begin{cases}\mbox{PPA-star: }\qquad
I_{1,2|a}=|B_{1|a}|^2- |B_{2|a}|^2\,, \\ I_{2,3|b}= |B_{2|b}|^2 + |B_{3|b}|^2\,, \quad I_{2,3|c}= |B_{2|c}|^2 + |B_{3|c}|^2\,, \\
 I_{|a,b,c}= |B_{1|a}|^2 +|B_{3|a}|^2+ |B_{3|b}|^2+ |B_{3|c}|^2\ .
\end{cases}
\end{eqnarray}
Integrals
$I_{2,3|b}$ and $I_{2,3|c}$ prevent $b$- and $c$-triads (connected via the
P-mode)  from adopting energy of initially excited $a$-triad. In
cases when $b$- and/or  $c$-triad are initially  exited, $a$-triad
can freely share  their energy via connecting A-mode.

\paragraph{Triple-chains} have two common modes in two triads. As we mentioned, there are 7 types of  triple-chains, that differ in type of connections, see \Fig{f:3-lin}.
Similarly to the triple-star clusters, one gets   equation of motion for triple-chains from    the  canonical ~\Eq{HamA}
with the Hamiltonian~\eq{Ham3},
in which one has to equate two pairs of  amplitudes of common modes. For example, for PA-PA chain,
\begin{subequations}\label{PA-PA}
 \begin{eqnarray}\label{PA-PA-eqs}
&& \mbox{PA-PA-chain:}\quad  B_{1|a}= B_{3|b} \,,
\quad  B_{1|b}=B_{3|c}\,,      \\
&& \begin{cases}
 \dot B_{1|a}= Z_a B_{2|a}^* B_{3|a}-Z_b B_{1|b}B_{2|b} \,,  \\
\label{PPP-eqs}
\dot B_{2|a}=Z_a B_{1|a}^*B_{3|a}\,,\quad \dot B_{3|a}=-Z_a B_{1|a}B_{2|a}\,,\\
\dot B_{2|b}=Z_b B_{1|b}^*B_{3|b}\,, \\
\dot B_{1|b}= Z_b B_{2|b}^* B_{3|b} -Z_a B_{1|c}B_{2|c}  \,,  \\
 \dot B_{1|c}=Z_c B_{2|c}^*B_{1|b}\,, \quad \dot B_{2|c}=Z_c B_{1|c}^*B_{1|b}
 \ .
\end{cases}
\end{eqnarray}
Again, besides Hamiltonian, all triple-chain clusters have four invariants of the Manley-Rowe type, two of them do not involve the common mode. In particular, PA-PA chain, governed by \Eq{PA-PA-eqs} has the following invariants:
\begin{eqnarray}\label{PA-PA-int}
\begin{cases}\mbox{PA-PA chain: }\qquad I_{|a,b}= |B_{1|a}|^2 -|B_{2|a}|^2+|B_{2|b}|^2\,,\\
~\,I_{|b,c}= |B_{1|b}|^2 -|B_{2|b}|^2+|B_{2|c}|^2\,, \\
I_{2,3|a}= |B_{2|a}|^2 + |B_{3|a}|^2\,, \quad
I_{1,2|c}=|B_{1|c}|^2- |B_{2|c}|^2   \ .
\end{cases}
\end{eqnarray}
\end{subequations}

\paragraph{Triple-triangles} have three common modes in three triads, see \Fig{f:3-lin}$\,\C H$ and $\C I$. Correspondingly, equations of motion for AA-PA-PP and PA-PP-PP triangles   one gets from    the  canonical ~\Eq{HamA}
with the Hamiltonian~\eq{Ham3},  in which one has to equate three pairs of  amplitudes of common modes. In particular, for AA-PA-PP
triangle one has:
\begin{subequations}\label{AA-PA-PP}
 \begin{eqnarray}\label{AA-PA-PP-con}
&&  B_{1|a}=B_{1|b}\,, \  B_{3|a}=B_{3|c}\,,  \ B_{3|b}=B_{1|c}\,,    \\
\label{AA-PA-PP-eqs}
&& \begin{cases}
\dot B_{1|a}=Z_a B_{2|a}^*B_{3|a}+Z_b B_{2|b}^*B_{3|b} \,,\\
\dot B_{2|a}=Z_a B_{1|a}^*B_{3|a}\,,\\
\dot B_{3|a}=-Z_a B_{1|a}B_{2|a}-  Z_c B_{2|c} B_{3|b}\,,  \\
\dot B_{2|b}=Z_b B_{1|a}^*B_{3|b}\,, \\
\dot B_{3|b}=-Z_b B_{1|a}B_{2|b}+  Z_c B_{2|c}^* B_{3|a}\,,  \\
 \dot B_{2|c}=Z_c B_{3|b}^*B_{3|a}
 \ .
\end{cases}
\end{eqnarray}
 Triangle clusters have three Manley-Rowe invariants. In particular, AA-PA-PP triangle, governed by \Eq{AA-PA-PP-eqs} has the following invariants:
\begin{eqnarray}\label{AP-AP-int}
\begin{cases}\mbox{AA-PA-PP triangle: }\qquad I_{|a,b}= |B_{2|b}|^2 -|B_{1|a}|^2 \,,\\
I_{|a,c}= |B_{2|a}|^2 +|B_{3|a}|^2+|B_{2|c}|^2\,, \\
I_{|b,c}=|B_{2|b}|^2- |B_{3|b}|^2 - |B_{2|c}|^2 \ .
\end{cases}
\end{eqnarray}\end{subequations}

Notice, that triple-stars and triple-chains have 10 real variables
(seven amplitudes and three triad phases) and five invariants
(Hamiltonian and four Manley-Rowes'), while triple-triangles have only 9
real variables (six amplitudes and three triad phases) and four
invariants (Hamiltonian and three Manley-Rowes'). Therefore, all
triple-triad clusters have five-dimensional effective phase space.
Recall, that butterflies  have three-dimensional phase space. Moreover, one can prove
that any $n$-triad cluster has $(2n-1)$-dimensional effective phase
space.

\subsection{\label{ss:cut} Caterpillar: 6-triad cluster}

\begin{figure*}
  \begin{center}
    \includegraphics[width= 15cm ]{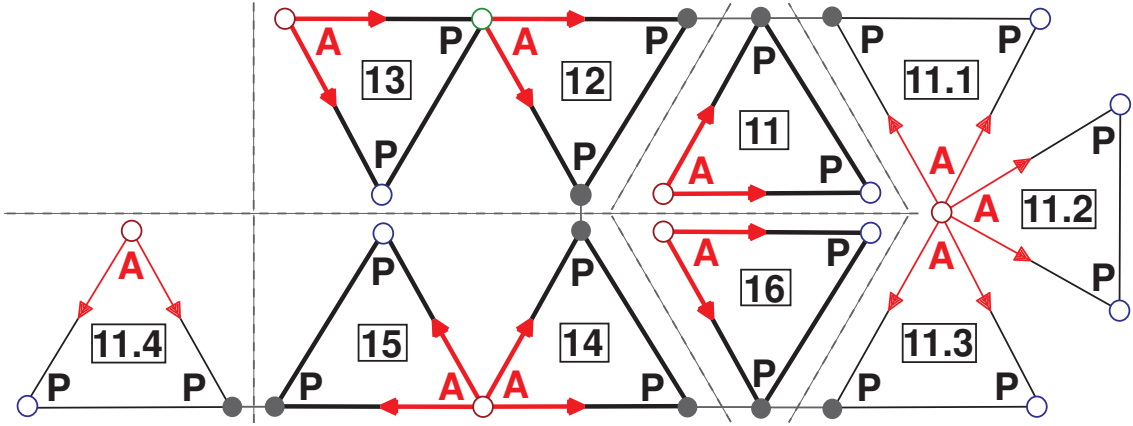} 
  \end{center}
\caption{\label{f:ext-cat}
    Color online. Triads belonging to caterpillar are drawn by bold   lines.  New, connected to them  triads, appearing in spectral domain $m,\ell \le 1000$ are drawn by thin lines. The PP-reduction of the extended caterpillar   to three triads ($\Delta_{11}$, $\Delta_{16}$ and $\Delta_{11.4}$, according to Tab.~\ref{t:1}), PA-butterfly $\bowtie_{12,13}$, AA--butterfly $\bowtie_{14,15}$, and AAA-star is shown by dashed lines. The rest of notations as in \Figs{f:1} and \ref{f:3-lin}.
}
\end{figure*}

The largest  cluster found in the domain $\ell<21$ consists of six resonant triads
$\Delta_{11}\dots\Delta_{16}$ with three PP-, one AP- and one AA-connection, see \Fig{f:ext-cat}. The equation of motion for  this cluster, (called ``caterpillar") can be obtained from Hamiltonian, similar to \Eq{Ham3}, but consisting of six terms:
\begin{subequations}\label{cat}
\begin{equation}\label{Cat-Ham}
H\sb{cat}=2\,  \mbox{Im}\sum_{n=11}^{16} Z_{n}B_{1|n}^*B_{2|n}^*B_{3|n} \,,
\end{equation}
in which we have to equate amplitudes of common modes:
\begin{eqnarray} \label{Cat-com}
\begin{cases}
B_{1|11}=B_{1|12}\,, \quad B_{3|12}=B_{1|13}\,, \quad B_{2|12}=B_{1|14}\,, \\
B_{3|14}=B_{3|15}\,, \quad B_{1|14}=B_{1|16}\ .
\end{cases}
\end{eqnarray}
Due to these five connections one has only $3\times 6-5=13$ complex equations for remaining amplitudes:
\begin{eqnarray} \label{Cut-eqs}
\begin{cases}
\dot{B}_{1|11}=  Z_{11} B_{2|11}^*B_{3|11} +   Z_{12} B_{2|12}^*B_{3|12} \,, \\
\dot{B}_{2|11}=  Z_{11} B_{1|11}^* B_{3|11}\,, \quad
\dot{B}_{3|11}=  - Z_{11} B_{1|11} B_{2|11} \,, \\
\dot{B}_{2|12}=  Z_{12} B_{1|11}^* B_{3|12}+ Z_{14} B_{1|14}^* B_{3|14} \,, \\
\dot{B}_{3|12}= - Z_{12} B_{1|12}  B_{2|12}+ Z_{13} B_{2|13}^* B_{3|13} \\
 \dot{B}_{1|14}=  Z_{14} B_{2|12}^*B_{3|14} +   Z_{16} B_{2|16}^*B_{3|16}\,, \\
\dot{B}_{3|14}=   Z_{14} B_{2|12}^*B_{1|14} +   Z_{16} B_{2|16}^*B_{3|16}\,, \\
\dot{B}_{1|15}=  Z_{15} B_{2|15}^* B_{3|14}\,, \quad \dot{B}_{2|15}=  Z_{15} B_{1|15}^* B_{3|14}\,, \\ \dot{B}_{2|16}=  Z_{16} B_{1|14}^* B_{3|16}\,, \quad \dot{B}_{3|16}= - Z_{16} B_{1|14}  B_{2|16}\ .
\end{cases}
\end{eqnarray}
Caterpillar has six  Manley-Rowe invariants:
\begin{eqnarray}\label{AP-AP-int}
\begin{cases} I_{2,3|11}= |B_{2|11}|^2 +|B_{3|11}|^2 \,,\\ I_{2,3|13}= |B_{2|13}|^2 +|B_{3|13}|^2\,, \\
 I_{1,2|15}= |B_{1|15}|^2 -|B_{2|15}|^2 \,,\\ I_{2,3|16}= |B_{2|16}|^2 +|B_{3|16}|^2\,, \\
I_{|11,12,13}= |B_{1|11}|^2 +|B_{3|11}|^2+|B_{3|12}|^2+|B_{3|13}|^2\,, \\
I_{|14,15,16}= |B_{1|14}|^2 +|B_{1|15}|^2+|B_{3|15}|^2+|B_{3|16}|^2\ .
 \end{cases}
\end{eqnarray}\end{subequations}

\begin{table}
\begin{tabular}{|c|c|c|c|c| c||c|c|}
  \hline 
  Clust. &   $N_1$ & Modes $[m,\ell]$ & $N_2$ & Connecting triads \\ \hline

\hline
$\Delta_1$& $1$ & [4,12] [5,14] [9,13]& $-$  & $-$\\
\hline
 $\Delta_2$ & $2$ & [3,14] [1,20] [4,15] & 2.1 & [4,15] [10,24] [14,20]\\

   &  &  & 2.2 & [1,20] [14,29] [15,28]\\
   &  &  & 2.3 & [1,20] [15,75] [16,56]\\
\hline
 $\Delta_3$& $3$ & [6,18] [7,20] [13,19] &3.1 & [2,15] [5,24] [7,20] \\
\hline
 $\Delta_4$& $4$ & [1,14][11,21][12,20]& 4.1 & [1,14] [9,27] [10,24] \\
\hline \hline
$\bowtie_{5,6}$  & $ 5$ & [2,6] [3,8] [5,7] &  5.1& [4,14] [9,27] [13,20]    \\
    & $ 6$& [2,6] [4,14] [6,9]&  & \\
\hline
  &$ 7$ & [6,14] [2,20] [8,15] & 7.1& [2,20] [11,44] [13,35] \\
    $\bowtie_{7,8}$ & $ 8$ & [3,6] [6,14] [9,9] & 7.2 & [2,20] [30,75] [32,56]\\
   &  &  & 7.3 & [32,56] [26,114] [58,69]\\
\hline  $\bowtie_{9,10 } $  & $ 9$ & [3,10] [5,21] [8,14] & $-$ & $-$ \\
   &$ {10}$  & [8,11] [5,21] [13,13] & & \\

\hline \hline
  & $ {11} $ & {  [2,14]}[17,20][19,19]&11.1& [2,14] [18,27] [20,24]\\
   & $ {12}$ & {  [1,6]}  {  [2,14] [3,9]}&11.2& [6,44] [14,21] [20,24] \\

 & $ {13}$& {  [3,9] } [8,20] [11,14] &11.3 & [9,35] [11,20] [20,24] \\

  $\boxtimes_{11\!-\!16} $  & $ {14}$ & {  [1,6] [11,20] [12,15]} & 11.4& [3,20] [45,75] [48,56]\\

  & $ {15}$ & [9,14] [3,20]{   [12,15] }&  & \\

& $ {16}$ & [2,7] {  [11,20]} [13,14]&  & \\

\hline
\end{tabular}
\caption{\label{t:1} The first 3 columns  provide the following data
in the domain $m, \ell \le 21 $ : the cluster's form, the triad
numbers and the modes within a cluster; in the last two -- the numbers
of additional connecting triads and their modes  that enlarge
the corresponding cluster when the spectral domain $|m|, \ \ell \le 1000 $ is
regarded.}
\end{table}%

\section{\label{s:topology} How Clusters are Organized}

This Section is devoted to the analysis of the structure of clusters of resonant triads of atmospheric planetary waves (based on the data-set  of the exact solutions of resonance conditions $\o_j=\o_r+\o_s$ with restrictions~\eq{cond}, provided by E. Kartashova~\cite{Kar-pr}). This analysis is important for the study of finite-dimensional wave turbulence throughout the present paper. A preliminary study of this issue can be found in \Ref{KL-08}.

\subsection{\label{ss:met} ``Meteorologically   significant" clusters and their extension in large spectral domain}

\begin{figure}
  \begin{tabular}{|c|}
    \hline
    $\C A$\\
    \includegraphics[width=8.5cm,height=5cm]{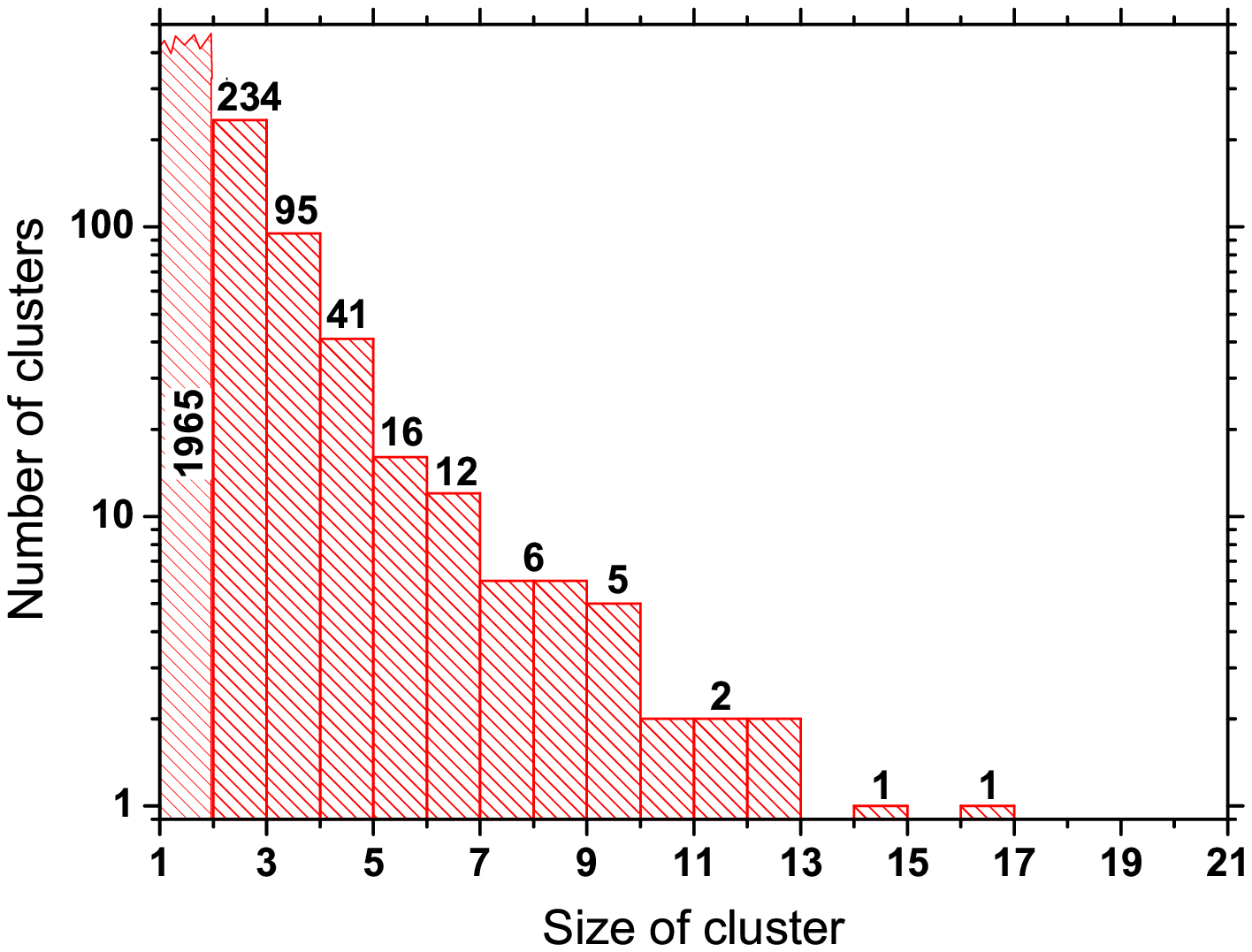}\\ 
    The 3691-triad cluster (monster) is not shown \\
    \hline\hline
    $\C B$\\
    \includegraphics[width=8.5cm,height=5cm]{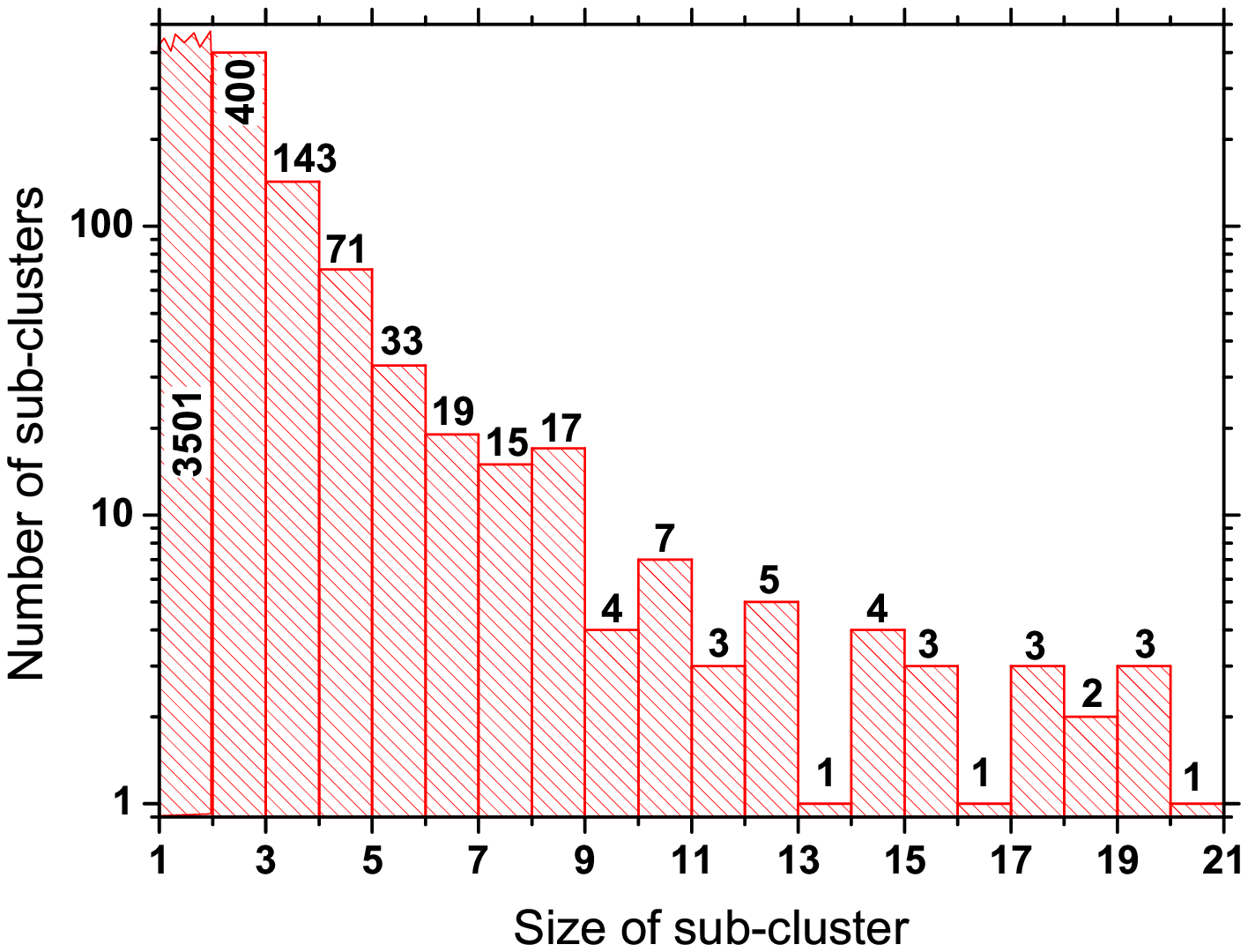}\\ 
    \begin{tabular}{|c|c|c|c|c|c|c|c|c|c|c|c|c|c|}
      \hline
     ~Sub-cl. size~& 21 & 24 & 28 & 30 & 32 & 33 & 35 & 37 & 38 & 44 & 121 & 127 &130  \\
      Amount &2 & 3 & 1 & 1 & 1 & 2 & 1 & 1 & 1 & 1 & 1 & 1& 2 \\
      \hline
    \end{tabular} \\ Amount of sub-cluster of size $\ge 21$   shown in the table \\  \hline
  \end{tabular}

  \caption{\label{f:histogram}
   Color online.  Horizontal axes denote  the number of triads in the cluster while vertical axes show  the number of corresponding clusters (panel $\C A$) and PP-irreducible subclusters (panel $\C B$).}.
\end{figure}

Dealing with Atmospheric waves one learns that the "meteorologically significant" wave numbers are believed
to be limited to $\ell<21$ \cite{05ghil}. Nevertheless, as explained above, the spectral domain for the approximate
two-dimensional atmosphere extends to $\ell\lesssim 1000$.
Counting explicitly   how many clusters we have in this spectral domain we find that there
exists altogether 1965 isolated triads and 424 clusters consisting
from 2  to 3691 connected triads. Among  them there are 234 butterflies, 95
triple-triad clusters, etc. (cf. the histogram in
\Fig{f:histogram}$\C A$).   For clarity of presentation we did not display in this histogram the largest 3691-cluster, which we refer to as the \emph{monster}.

It can be seen that about $82.2\%$ of all clusters are presented by
 isolated triads and their dynamics   has been investigated in \cite{KL-07} in all details;
the main findings are that the energy  oscillates between the three modes in the triads,
with a period of
 oscillation that is much larger then the wave period. This period is inversely proportional
 to the root-mean-square of the wave amplitude.

 234 clusters in the spectral domain ($\simeq 10.5 \%$) are the butterflies discussed above; these are further
 analyzed below. Among them there are 131 PP-, 69 AP-  and 35 AA-butterflies.
 The 95 triple-triad clusters include 25  "triple-star" clusters with one triple connection, see \Fig{f:stars}. This set  includes 3 AAA-, 5 AAP-, 6 APP- and 11 PPP-stars. There are also 66 chain clusters with two pair connections and 7 combinations of the connection types, shown in \Fig{f:3-lin}. We found also four (two pairs) triple-triad clusters with three pair connections, belonging to two different types, see \Fig{f:3-lin}.

Similar classification  can be performed for
all the other clusters.  For example, the monster includes one mode
(218,545), participating in 10 triads, three modes, participating in
9 triads,  5 modes -- in 8 triads,  23 -- in 7, 50 -- in 6, 90 in 5,
236 -- in 4, 550 -- in 3 and 1428 modes -- in 2 triads
(butterflies). The analysis of their dynamical behavior depends
crucially on the connection type as shown above and detailed below.

\subsection{\label{ss:PPred}PP-reduction of larger clusters}
\begin{figure*}
    \begin{center}
      \includegraphics[width=16cm ]{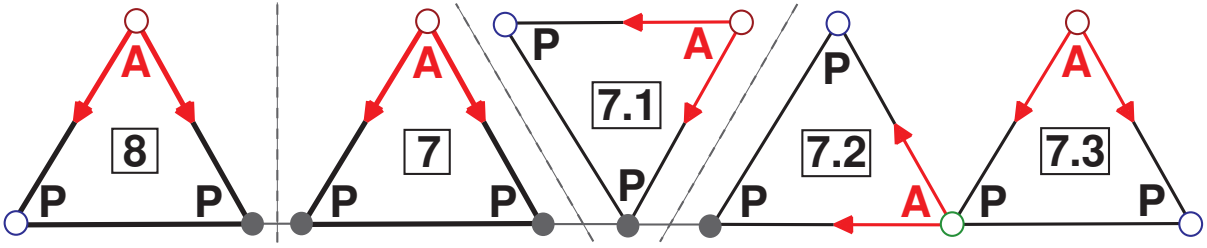} 
    \end{center}
  \caption{\label{f:PPP-reduction} 
    Color online. Triads $\Delta_7$ and $\Delta_8$, that  belong to PP-butterfly $\bowtie_{7,8}$  are drawn by bold  lines.  New, connected to them  triads $\Delta_{7.1}$,  $\Delta_{7.2}$ and $\Delta_{7.3}$, appearing in  spectral domain $m,\ell \le 1000$ (for numeration of triads see \Tab{t:1}),  are drawn by thin lines. The PP-reduction of this cluster to three triads $\Delta_7$, $\Delta_8$ and $\Delta_{7.1}$ (according to the notation in Tab.~\ref{t:1}) and one AP-butterfly is shown by dashed lines. The rest of notations   as in \Fig{f:1}.
  }
\end{figure*}

\begin{figure*}
  \begin{center}
    \begin{tabular}{|c|| c|}
      \hline    $\C A$~~~~~~   & $\C B$~~~~~~   \\
        ~~~~~~\includegraphics[width=5cm,height=3.5cm]{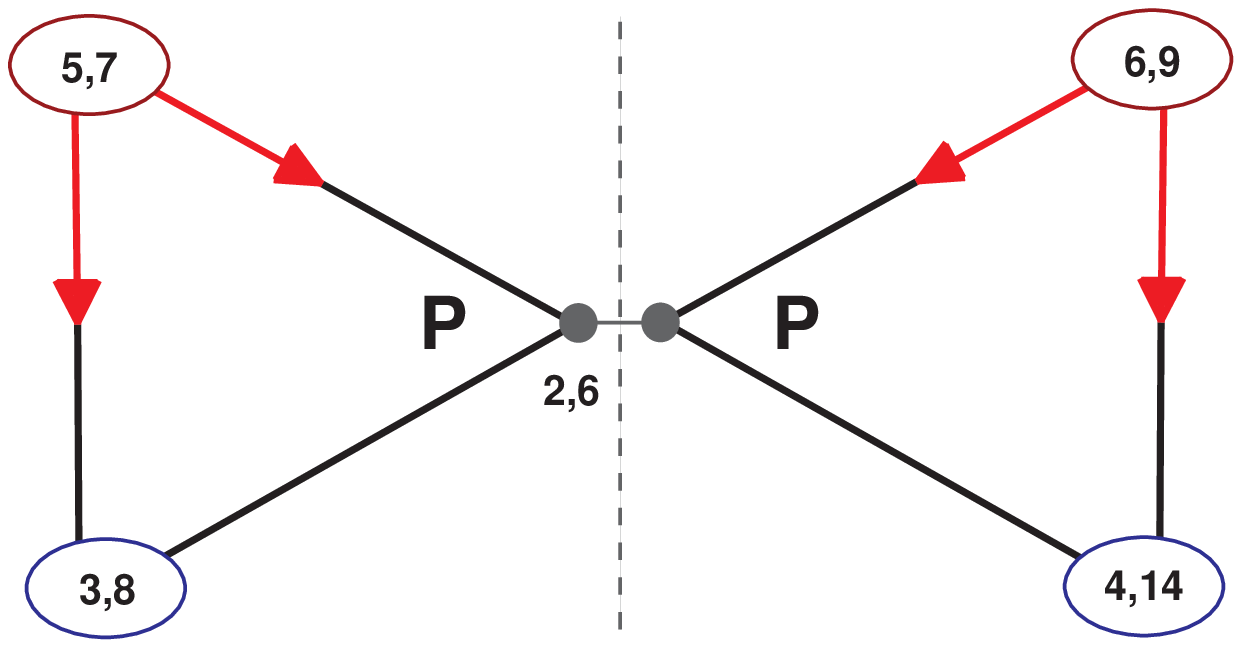}~~~~~~ & 
        ~~~~~~\includegraphics[width=7cm,height=5cm]{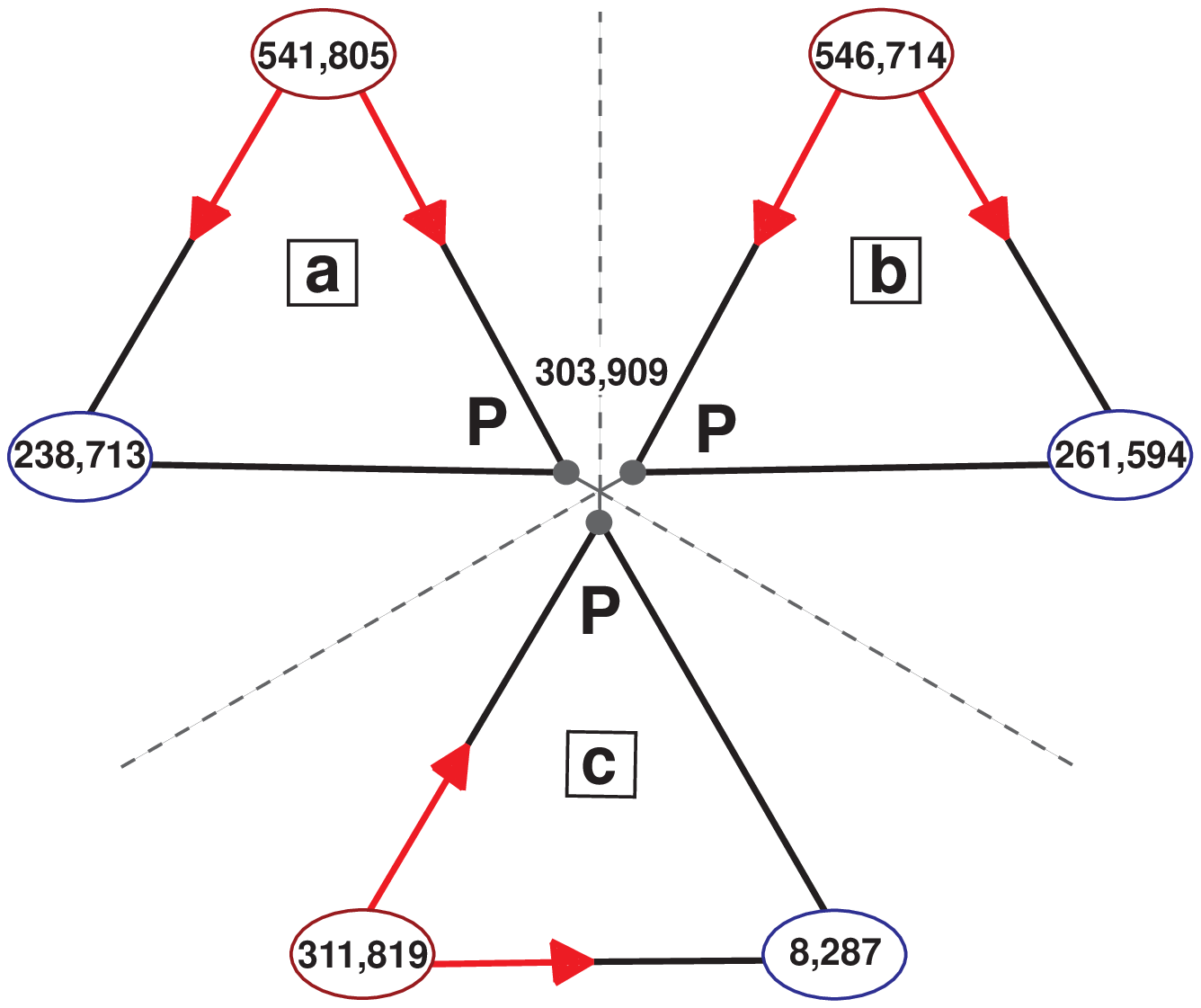} ~~~~~~ \\ 
        PP-reducible butterfly & PP-reducible  
    \end{tabular}
  \end{center}
  \caption{\label{f:AAP-star} 
    Color online. Examples of PP-reduction of pair connection, panel $\C A$ and   of triple connection, panel $\C B$.  Common (PP- and PPP-) modes are split stressing the difficulty of energy exchange between the corresponding triads. Dashed lines separate PP-irreducible sub-clusters, discussed in \Sec{ss:PPred}. The rest of the notations  are as in \Fig{f:1}.
  }
\end{figure*}

Large clusters can be divided into "almost separated" subclusters
connected by PP-connections (e.g. \Figs{f:ext-cat}, \ref{f:PPP-reduction} and \ref{f:AAP-star}). If such a sub-cluster cannot be divided further into smaller clusters connected by PP-connection we refer to
it as a PP-irreducible cluster. For example,  triangle cluster in \Fig{f:3-lin}~$\C I$  can be PP-reduced to a triad and a PA-butterfly, while  clusters in \Fig{f:AAP-star}$\,\C A$ and $\C B$  can be PP-reduced to  two and three  individual triads respectively.  The cluster in \Fig{f:PPP-reduction} is PP-reduced into three triads and PA-butterfly, while extended caterpillar   in \Fig{f:ext-cat} is   PP-reduced into  three triads, PA-,  AA-butterfly  and   AAA-star. One can see that the 14-triad cluster, shown in \Fig{14,16}~$\C A$ can be PP-reduced into 8 triads and 6-triad cluster. The 16-triad cluster (\Fig{14,16}~$\C B$)  can be PP-reduced into 5 triads,
 AA- and   PA-butterfly,  AAP-star and   4-triad cluster, consisting of AAA-star with one AP-connected triad.

\begin{figure*}
  \begin{center}
    \begin{tabular}{|c||c|}\hline
      $\C A$ & $\C B$ \\
      ~~~\includegraphics[width= 8.5 cm ]{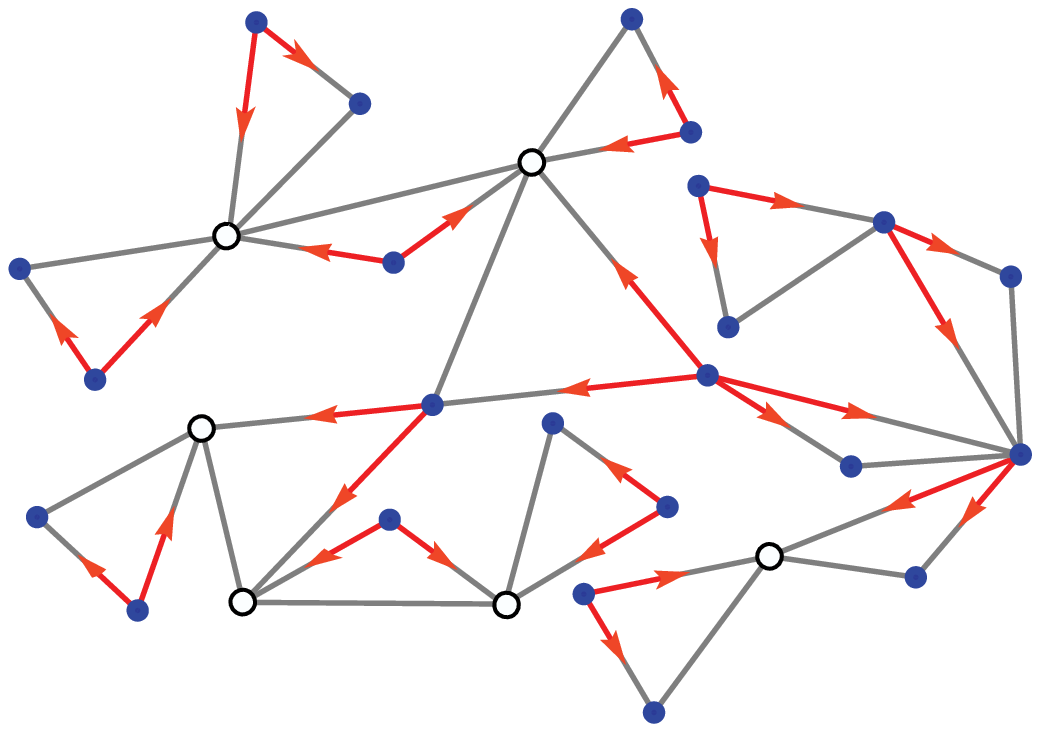}~~~ &    
      ~~~ \includegraphics[width= 7.7 cm ]{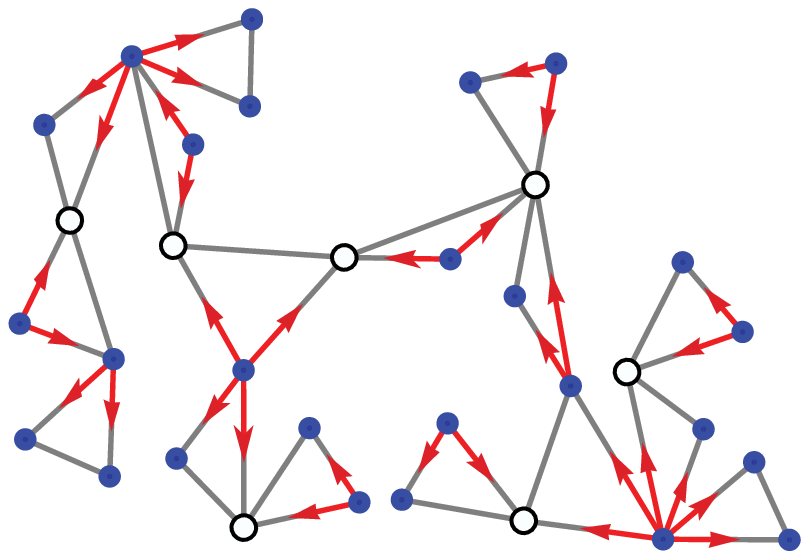}~~~  \\ 
      14-triad PP-reducible cluster   & 16-triad PP-reducible cluster  \\
      \hline
    \end{tabular}
  \end{center}
  \caption{
    Large  PP-reducible clusters. Common PP- and PPP-modes are shown by empty circles. The rest of modes are denoted by full (blue) circles. As in previous figures, outgoing (red) arrows indicate  A-modes.
  }
  \label{14,16}
\end{figure*}

  \begin{figure*}
  \begin{center}
  \begin{tabular}{|c |}\hline \\
    ~~\includegraphics[width= 17.5cm ]{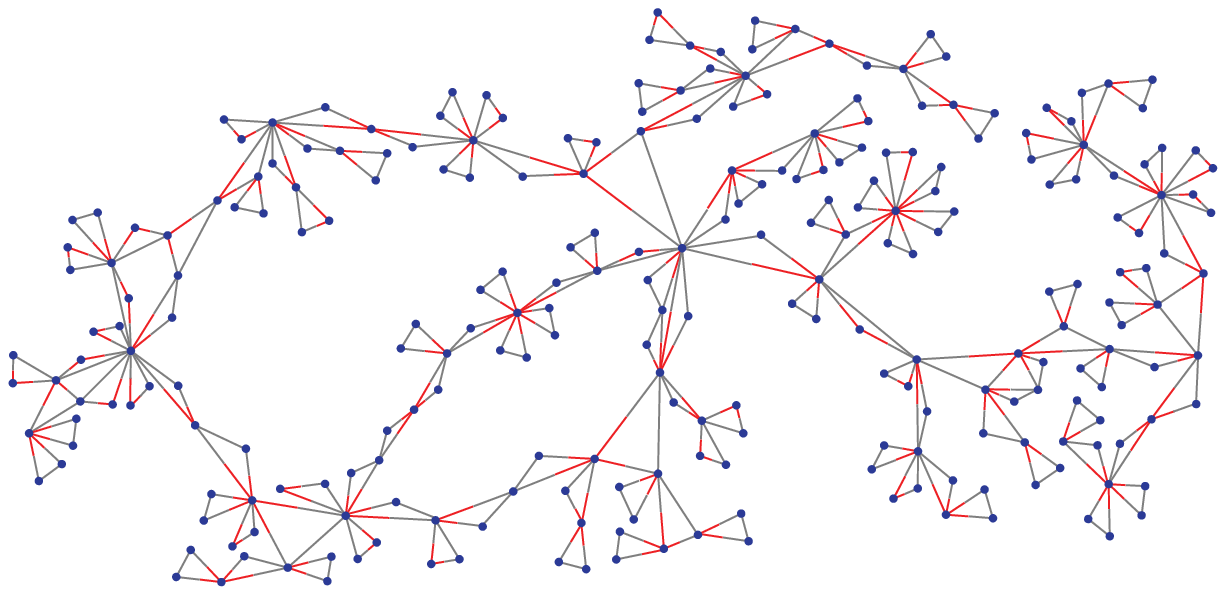}~~\\  
    The 130-triad PP-irreducible sub-cluster belonging to the monster \\
   \hline
  \end{tabular}
\end{center}
\caption{
    \label{largestPP}Color online.  Largest PP-irreducible cluster in the spectral domain $\ell, |m|\le 1000$.
}
\end{figure*}

In the  spectral domain $\ell, |m|\le 1000$ the largest
PP-irreducible sub-cluster belonging to the monster consists of 130
triads and is shown in \Fig{largestPP}. The statistics of
PP-irreducible sub-clusters are presented in \Fig{f:histogram}~$\C B$.

We learn  from this analysis that many clusters cannot
carry energy flux through PP connections; their dynamics is
naturally reduced  to the dynamics of
PP-irreducible sub-clusters. Therefore it is sufficient to study carefully the
dynamics of these PP-irreducible clusters  to understand the properties
 of any cluster.


\section{ \label{s:free} Numerical analysis of  free evolution of typical sub-clusters}
Our goal is to describe the energy flux through resonant triads in the
 regime of finite-dimensional wave turbulence.  In the first
 Subsec.~\ref{ss:butF} we consider free evolution of the smallest
 clusters -- butterflies -- from asymmetrical initial conditions, in
 which only one triad is excited to high amplitudes, exceeding by
 orders of magnitudes the initial amplitudes in the other triad. The
 questions are how the energy flux from the energetic ``leading" triad
 depends on the type of connection, on the ratios of the interaction
 coefficients etc.  Subsec.~\ref{ss:3-6} is devoted to the free
 evolution of the triple-triad clusters: stars and chains from initial
 conditions in which only the leading $a$-triad is substantially
 excited, the levels of excitation of the two other triads are much
 smaller.  \emph{All these examples, and PA-PA-..PA chains, studied in
 the next Section, can serve as building blocks of bigger clusters and
 the knowledge about the energy flux through them allows to
 qualitatively predict efficiency of the energy transfer through
 bigger clusters.}

\subsection{\label{ss:num} Methodology and numerical procedure}
The equations of motion for all clusters were prepared for numerical analysis
using a specially designed algorithm that allowed an automatic implementation
for any cluster, given the number of triads and
the connectivity table. This served to avoid human errors in
implementing large sets of equations.

The equations of motion for the free evolution of butterflies and triple
clusters are stiff and were integrated using a multistep adaptive method
based on numerical differentiation formulas~\cite{num}.

For each system of equations the accuracy of integration was controlled by
testing the conservation of the relevant integrals of motion. The integration
parameters were adjusted to keep the standard deviation $\sigma (I)$ below
a given threshold for the duration of the numerical runs. Since the main
parameter that affects the accuracy in terms of the conservation of
the integrals of motion is the maximal allowed time step, the preliminary calculations
were carried out with a requirement $\sigma( I)\le 10^{-5}$; actually
in most cases $\sigma (I)\le 10^{-8}$ was achieved.

The evolution starting from several hundred initial conditions was
analyzed and several representative conditions were chosen for the  study of
energy transfer in the clusters. All the conclusions regarding the discovered
dependencies were verified by control calculations with stricter
accuracy requirements.

The equation of motion for forced chain clusters,  studied in the next \Sec{s:stat}, were integrated by both
adaptive methods~\cite{num}  and by 4th order constant time-step Runge-Kutta for better
control of accuracy.  By construction of the model, these equations
required accurate description of the last triad in the cluster to
ensure proper energy dissipation. The integration parameters were
adjusted to reproduce this fastest evolution, and therefore were
automatically suitable for all other triads. We verified the
convergence of the resulting statistics with respect to all relevant
parameters.

\subsection{\label{ss:butF} Free evolution in  butterflies}

The simplest topology that allows consideration of the energy flux between
resonant triads is the double-triad clusters -- butterflies.

\subsubsection{\label{sss:ic} Initial conditions, choice of the interaction coefficients and data representation}

In this Subsection we show that details of the time evolution in
butterflies are very sensitive to the initial conditions, which define
the values of the dynamical invariants.  Therefore a reasonable
choice of initial conditions, allowing to shed light on a ``typical" time evolution in a
relatively compact form is not obvious. At initial time $t=0$ we assign most of the energy to two individual
(not common) modes of one (leading) triad, denoted below for concreteness as
$a$-triad. The initial amplitudes of these two modes we denote as $B_0$ and $\~B_0$  (\Fig{f:but}).
To study the influence of the energy distribution between these modes we
will use two types of initial conditions:
    \begin{subequations}\label{i-cond}\begin{eqnarray}
  \label{act1}
\mbox{Type I}:&& B_0=3.9+0.50~i\,,  \  \~B_0=3.7+0.93~i\,,
~~~~~~ \\
    \label{act}
\mbox{Type II}:&&    B_0=5.3+0.50~i\,,   \  \~B_0=0.9+0.93~i\ .  ~~~~~~
    \end{eqnarray}\end{subequations}
Both distributions~\eq{i-cond} are complex and normalized such that
$|B_0|^2+|\~B_0|^2\approx 30$. The difference between \Eq{act1} and
\eq{act} is that in \Eq{act1} both amplitudes are similar, while in
\Eq{act} they are quite different.

For different butterflies  we choose in the leading triad:
\begin{subequations}\label{ic-a}
\begin{eqnarray}\nn
\mbox{PP-butterfly with}&&
B_{1|a}(0)=B_{1|b}(0):  \\ \label{ic-PP}
B_{2|a}(0)&=&\~B_0 \,, \
B_{3|a}(0)=B_0\, ; \\ \nn
\mbox{AA-butterfly with}&&
B_{3|a}=B_{3|b}: \\ \label{ic-AA}
B_{1|a}(0)&=&B_0\,, \
B_{2|a}(0)=\~B_0\ ;\\ \nn
\mbox{PA-butterfly with}&&
B_{1|a}=B_{3|b}: \\
\label{ic-AP2}
 B_{2|a}(0)&=&\~B_0\,,\
B_{3|a}(0)=B_0\,,
\end{eqnarray} \end{subequations}
as it is shown in \Fig{f:but}.

The initial conditions in the driven $b$-triad we choose the same for
all types of butterflies. They have much smaller initial amplitudes,
for example:
\begin{eqnarray}\label{ic-b}\begin{cases}
B_{1|b}(0)=B_{1,0}\,, \quad B_{1,0}\= C\, (0.05+0.02~i)\,,\\
B_{2|b}(0)=B_{2,0}\,, \quad B_{2,0}\= C\, (0.02+0.05 ~i)\,,\\
B_{3|b}(0)=B_{3,0}\,, \quad B_{3,0}\= C\, (0.10-0.02~i) \ .\end{cases}
\end{eqnarray}
To study the dependence of the energy flow between triads on the
initial level of excitation of the driven triad we vary the energy
content in the driven triad changing the coefficient $C$ in \Eq{ic-b},
taking in addition to $C=1$ also $C=0.1$ and $C=0.01$. In all the further simulations $C=1$ if else is not mentioned.

In this way the initial conditions for all butterflies are as similar
as possible and we can study the difference in time evolutions, caused
by different types of connections.

 Last but not least are the interaction coefficients. For concreteness
 we chose interaction coefficients corresponding to $\Delta_{14}$ and
 $\Delta_{16}$ ($Z_{14}\approx 75 $, and $Z_{16}\approx 15$) as
 prototypes and use either $\{Z_a=75\,, \ Z_b=15\}$ or vise versa and
 sometimes $\{Z_a= Z_b=15\}$. Since the change in the interaction
 coefficient re-normalizes the corresponding time scale, only their
 ratio is important for the dynamics. In our case these ratios are $=
 5\,, \ 1/5$ or $1$; this allows us to study the butterfly dynamics
 with very different values of the interaction amplitudes, which is
 typically the case. Having in mind that special choices of the ratios
 of interaction coefficients may lead to integrability of clusters of
 resonant triads~\cite{ver} (and see also~\cite{08-BK}) we verified
 that small variations of these ratios does not changed our
 conclusions concerning energy transfer in clusters.

Recall that any butterfly has three quadratic integrals of motion,
that involve only squares of five amplitudes of modes  and
therefore only 2 combinations of them  are independent. For the presentation we chose such combinations that are orthogonal
to the corresponding invariants. Namely, for PP-butterflies, connected via
$B_{1|a}=B_{1|b}$ modes:
  \begin{subequations}\label{bat}\begin{equation}\label{batPP}
J_{2,3|a}\= |B_{2|a}|^2-|B_{3|a}|^2\,,\quad J_{ 2,3|b}\= |B_{2|b}|^2-|B_{3|b}|^2\,;
\end{equation}
for AA-butterflies, connected via $B_{3|a}=B_{3|b}$ modes:
\begin{equation}\label{batAA}
J_{1,2|a}\= |B_{1|a}|^2+|B_{2|a}|^2\,,\quad J_{1,2|b}\= |B_{1|a}|^2+|B_{2|b}|^2\,,
\end{equation}\end{subequations}
and for AP-butterflies, connected via $B_{1|a}=B_{3|b}$ modes,
$J_{2,3|a}$  and  $J_{1,2|b}$.

\begin{figure}
\begin{center}
\begin{tabular}{|c|}
  \hline  $\C A $ \\
  ~~\includegraphics[width=8.3cm]{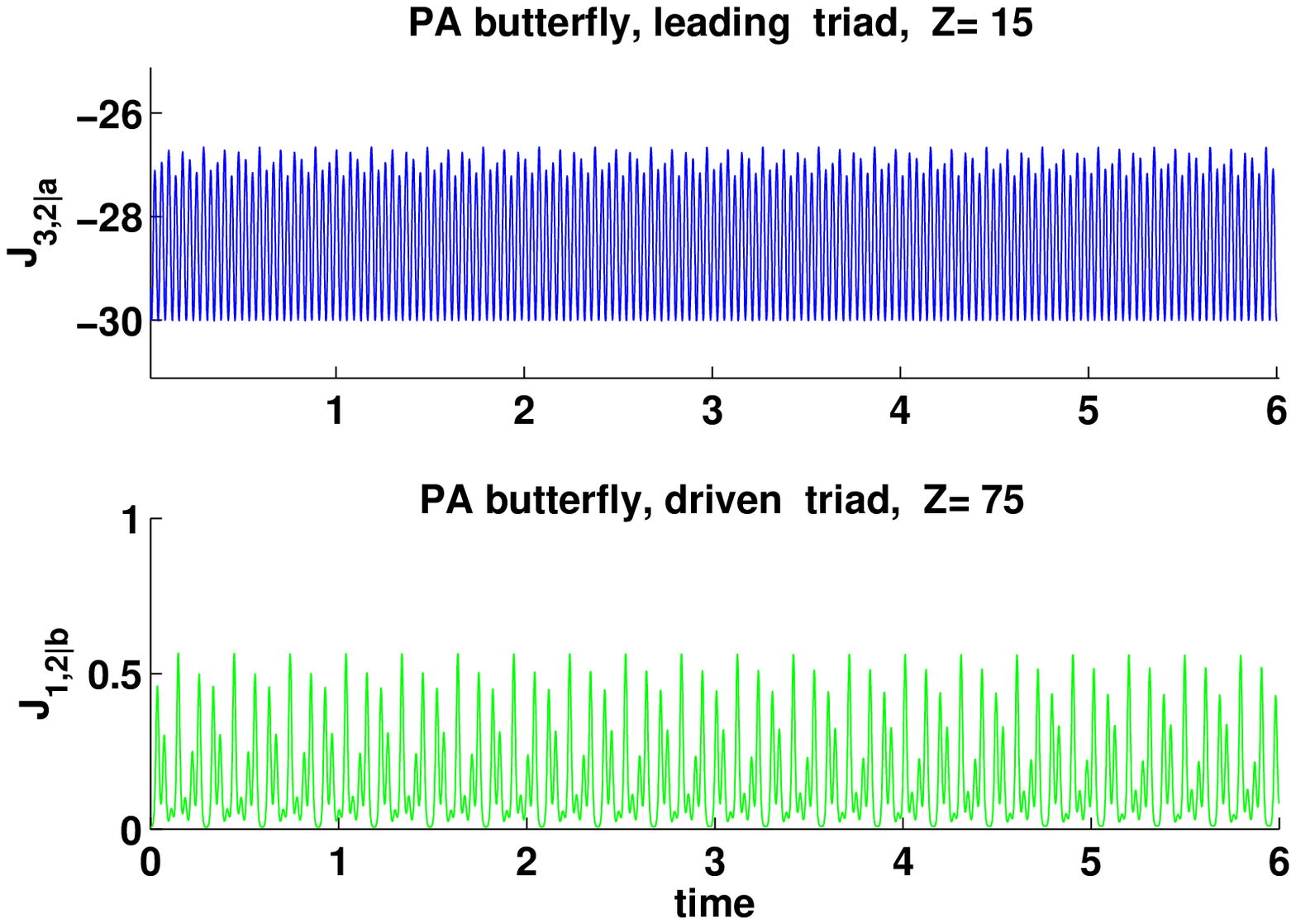}\\ 
  \hline \hline
   $\C B $\\
  ~~\includegraphics[width=8.3cm]{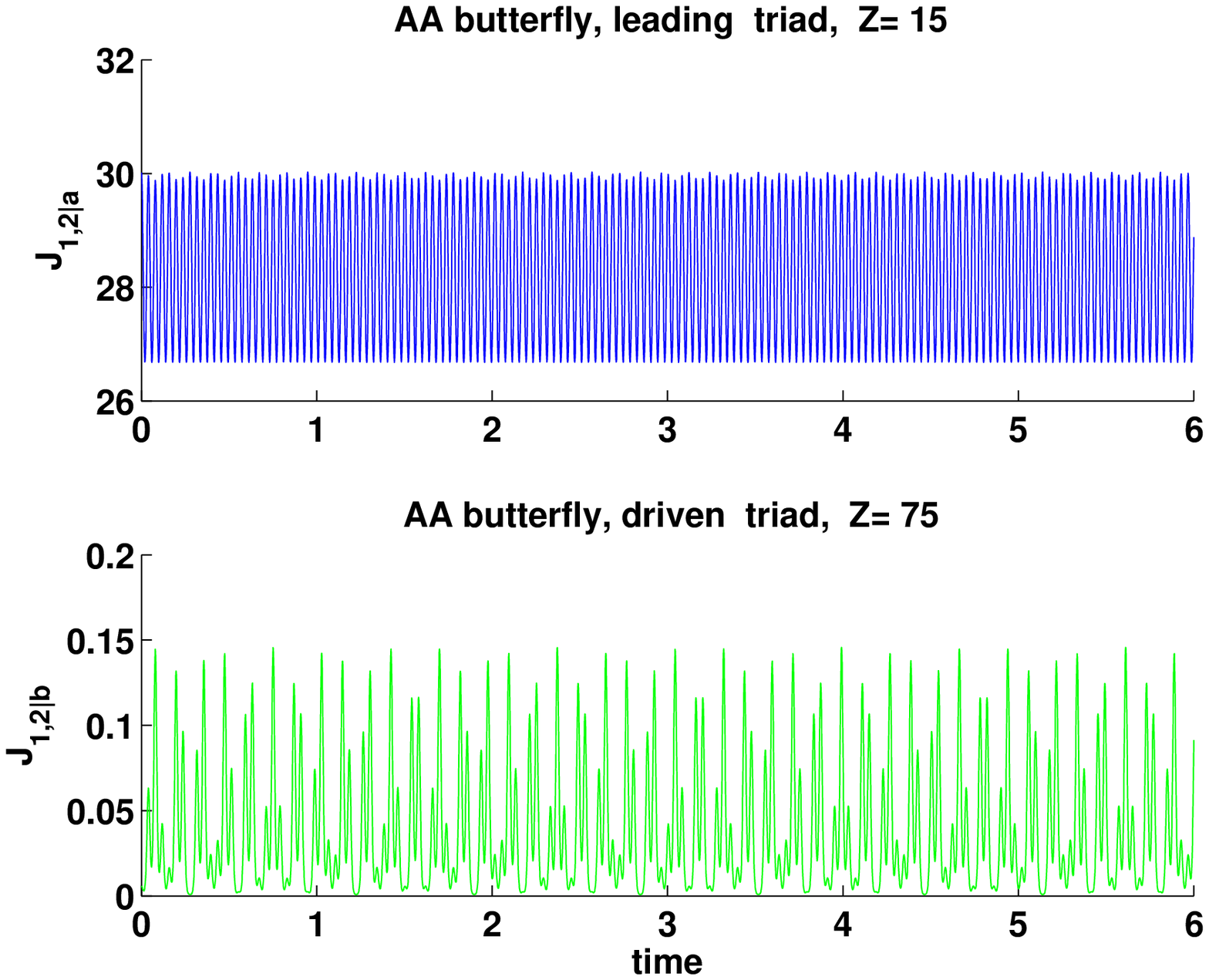}~~\\   
\end{tabular}
\end{center}
\caption{\label{f:PA-AA} Color online. Suppression of the energy transfer by initial conditions in PA- and AA butterflies.   }
\end{figure}

\begin{figure*}
\begin{center}
\begin{tabular}{ |c||c| }
  \hline $\C A $& $\C B $\\
  \includegraphics[width=8.5cm ]{fig11f.eps}~~ &
  \includegraphics[width=8.5cm]{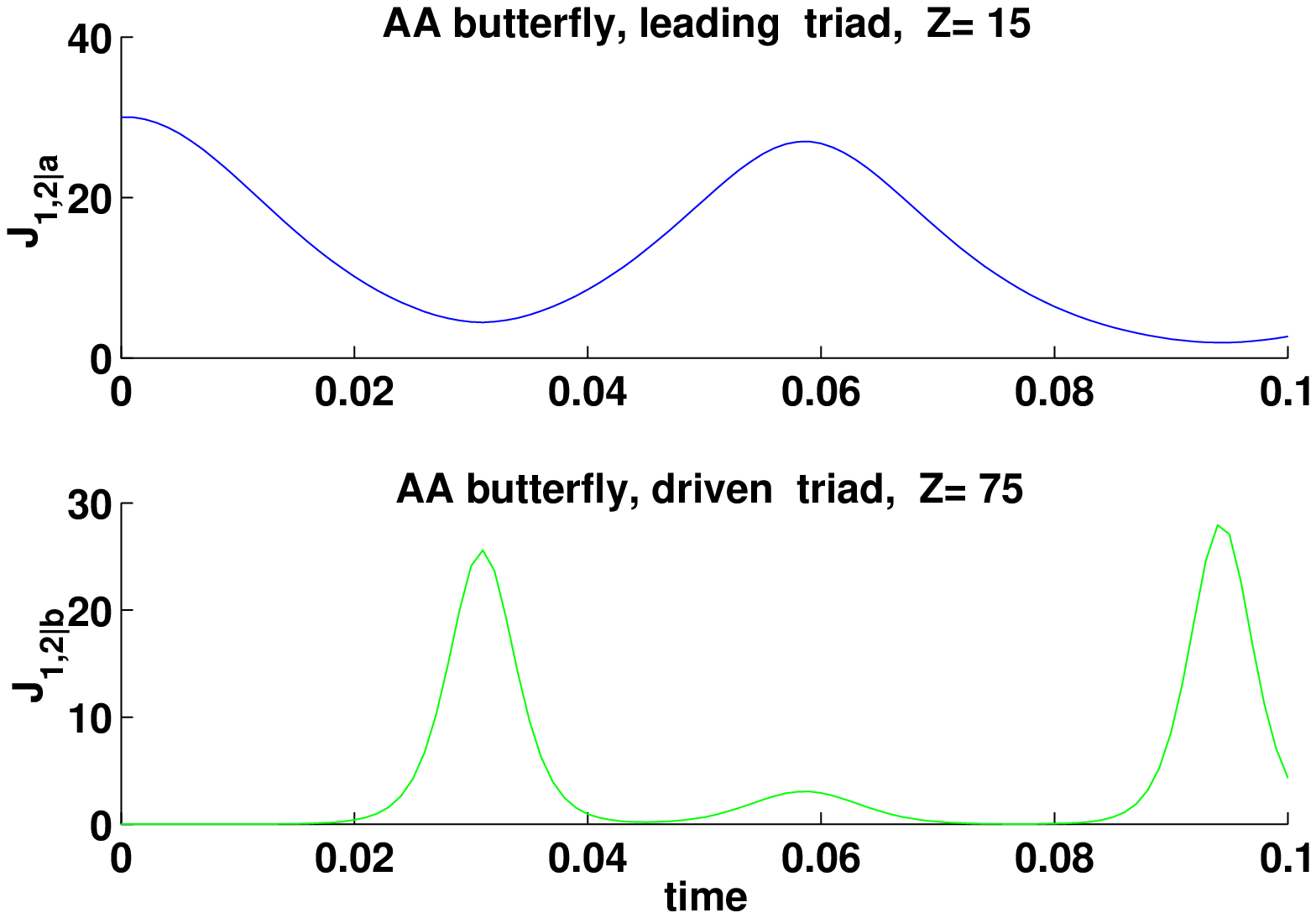}  \\ 
Long time evolution  for $C=1\,, 0.1$ and  $0.01$ & Initial stage of the evolution with $C=1$. \\
  \hline \hline
 $\C C $& $\C D $\\
 \includegraphics[width=8.5cm]{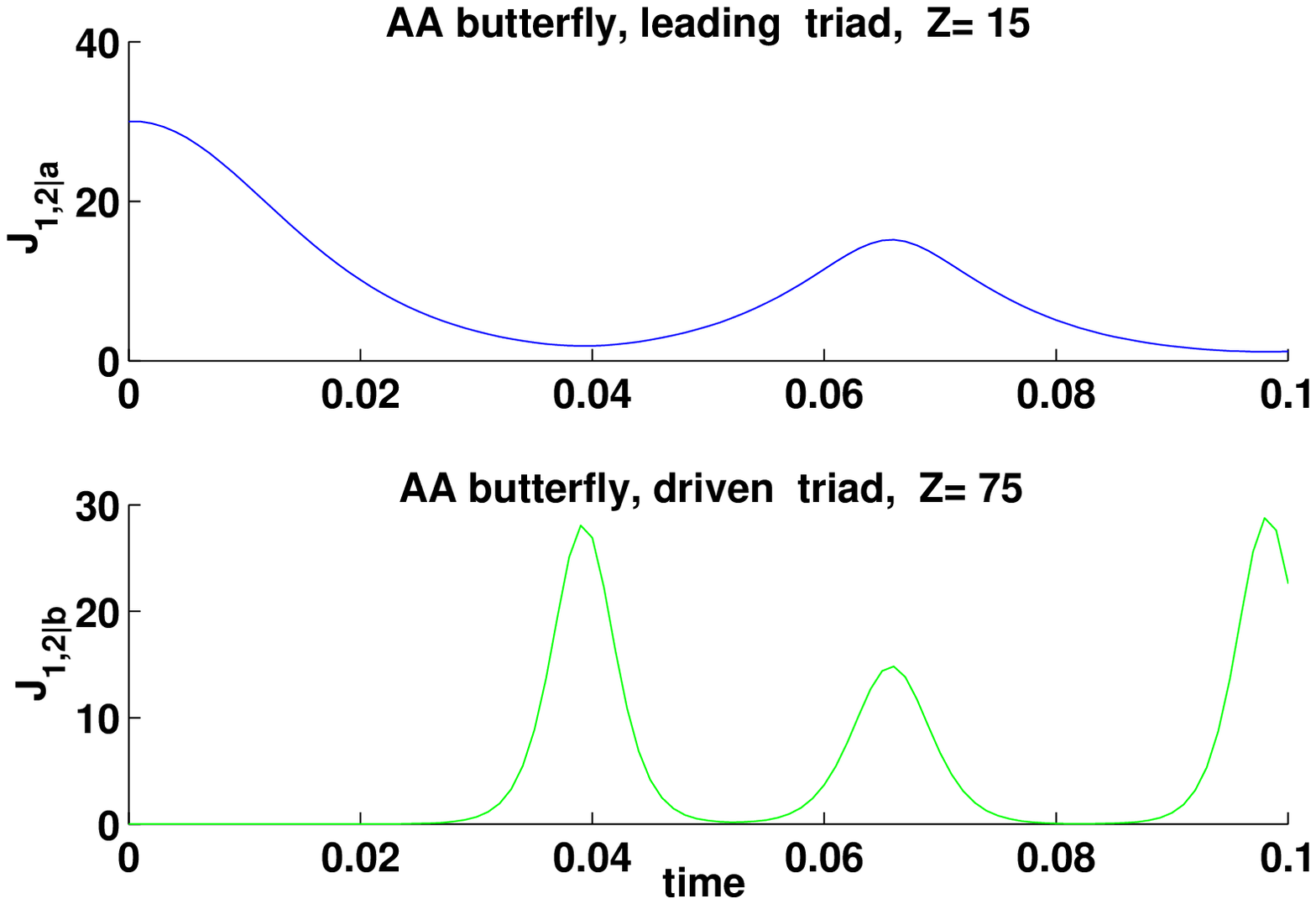}~~ & 
  \includegraphics[width=8.5cm]{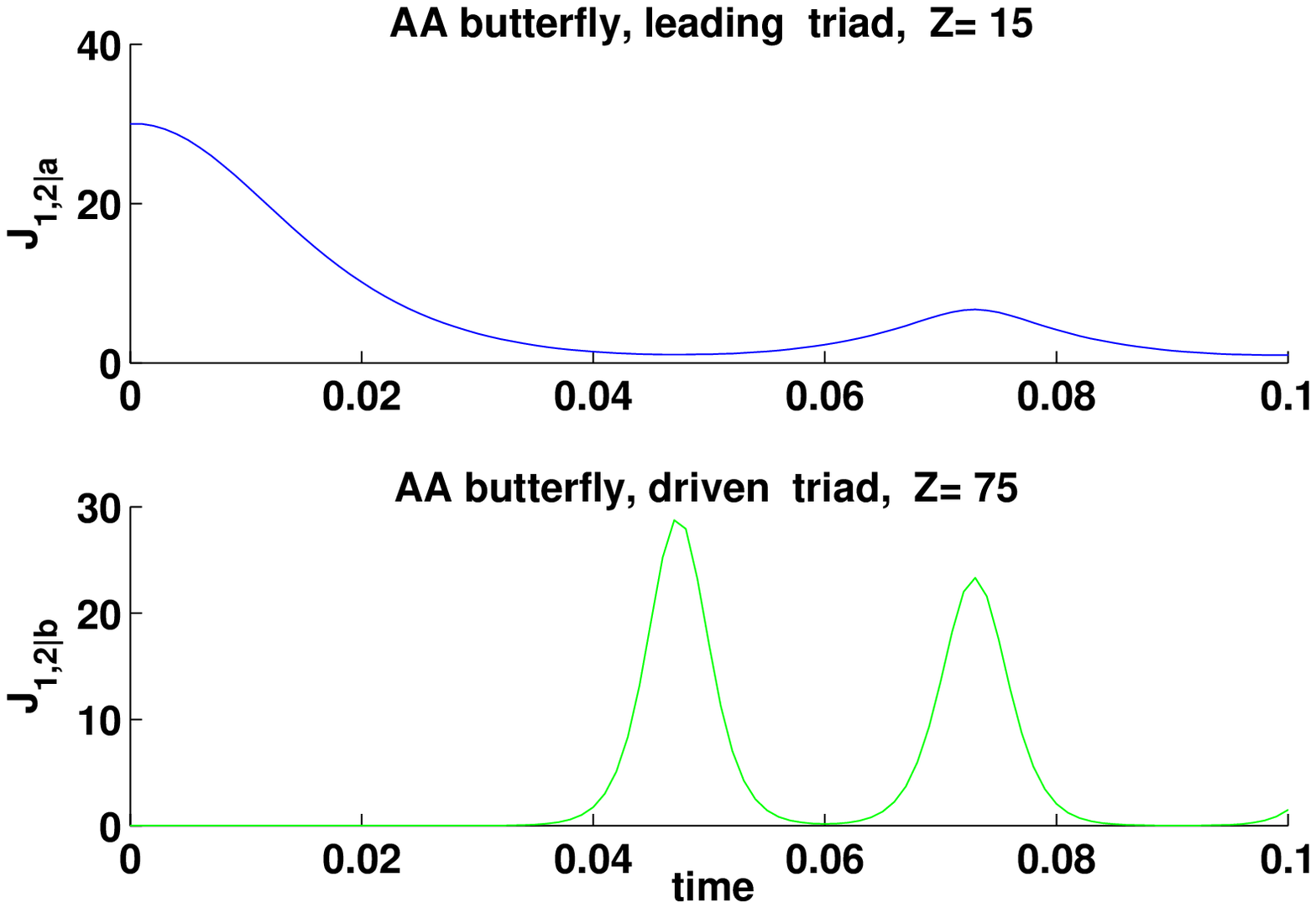} \\ 
  Initial  stage of the evolution with $C=0.1$ & Initial  stage of the evolution with $C=0.01$\\
  \hline
\end{tabular}

  \end{center}
\caption{\label{f:AA-ev2} Color online. Time evolution of AA-butterfly with different level of excitation in the driven triad, governed by parameter $C$ in \Eq{ic-b}.   }
\end{figure*}

 Time evolutions for these three types of butterflies with various
 initial conditions and choices of the ratio $Z_a/Z_b$ (5 or 1/5) are
 shown in \Figs{f:PP-PA-AA} -- \ref{f:AA-ev2}. Following Subsection~\ref{sss:dep}
 is devoted to discussion of these numerical results.

\subsubsection{\label{sss:dep} Effect of the type of connections and of the ratio $Z_a\big / Z_b$}

\begin{figure*}
\begin{center}
\begin{tabular}{|c || c|}
 \hline   $\C A$ & $\C B$ \\
  ~~~~\includegraphics[width=8.3cm ]{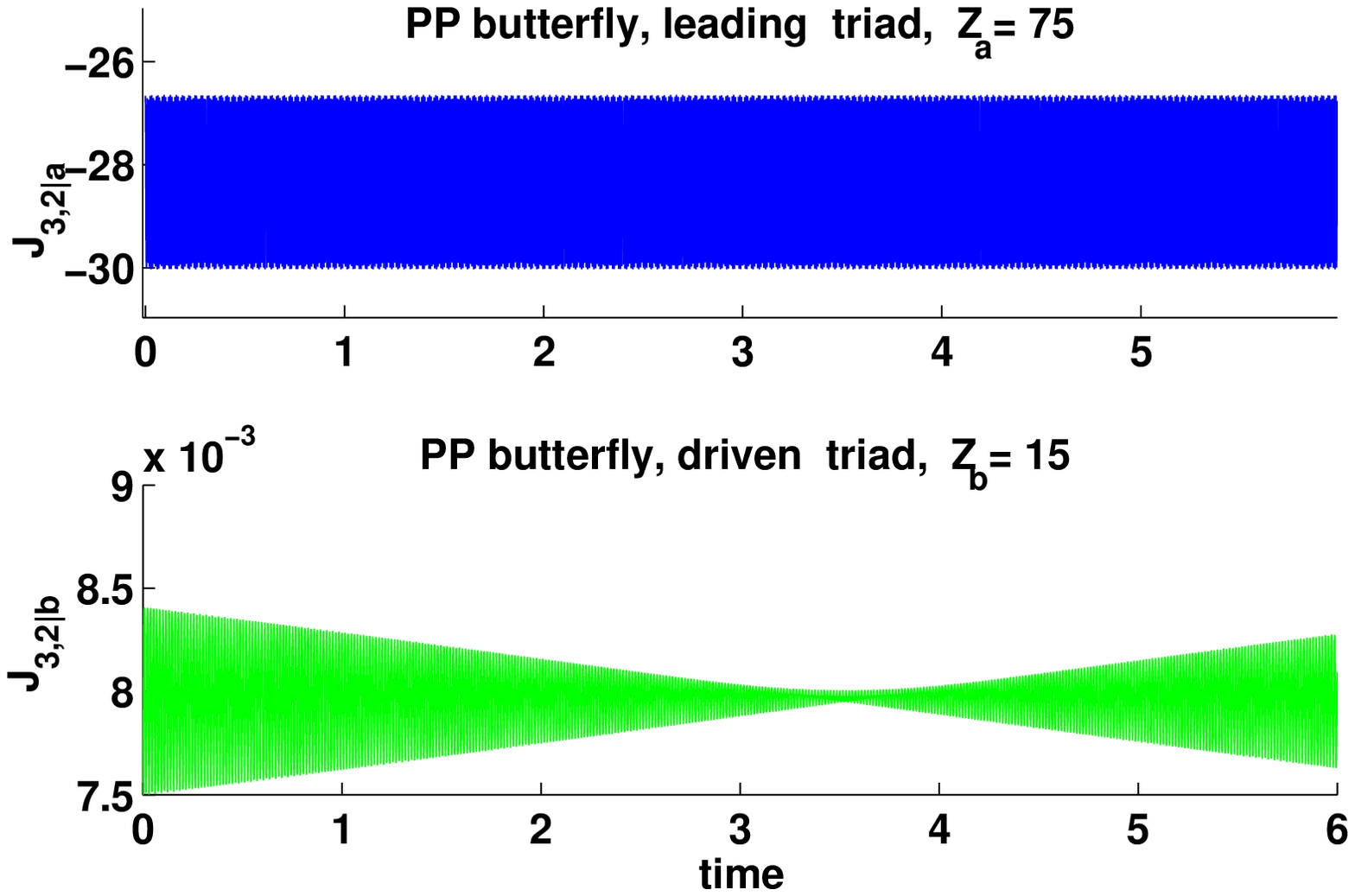}~~  & 
  ~~\includegraphics[width=8.3cm ]{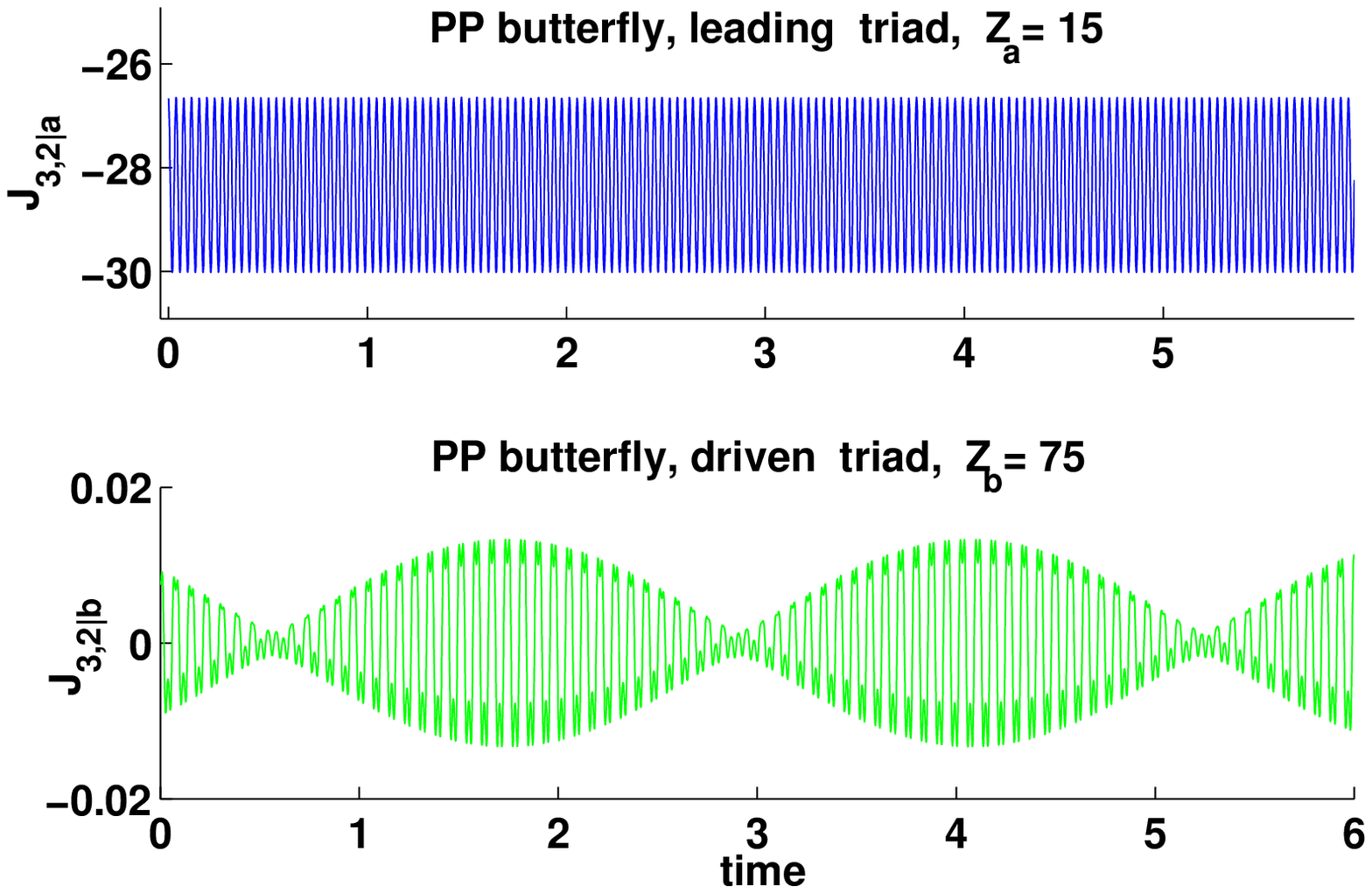}~~ \\   
  Blockade of the energy transfer in PP-butterfly&  Blockade of the energy transfer in  PP-butterfly \\  \hline \hline
 $\C C$ & $\C D$ \\
 ~~\includegraphics[width=8.3cm ]{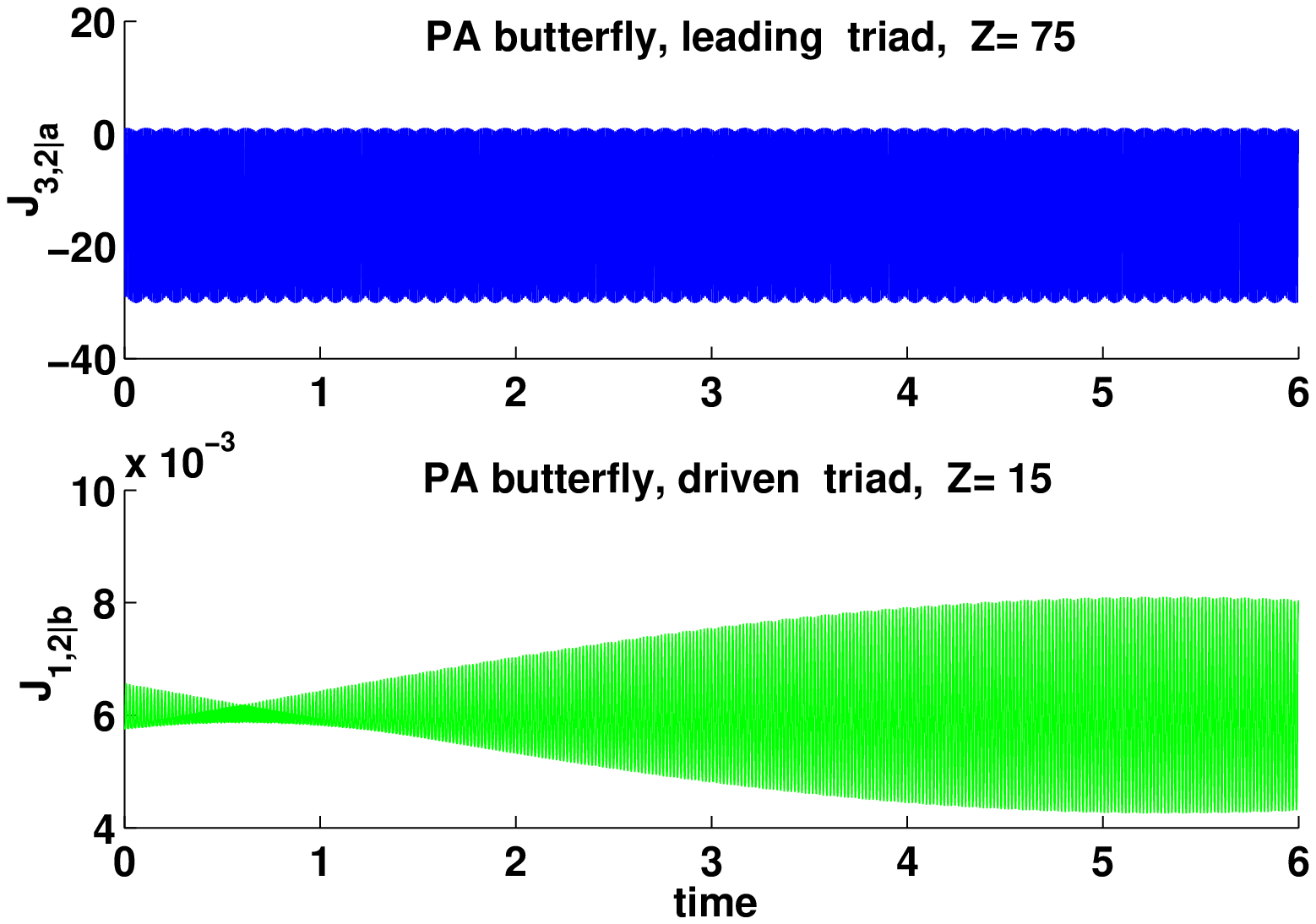}~~  &  
  ~~\includegraphics[width=8.3cm] {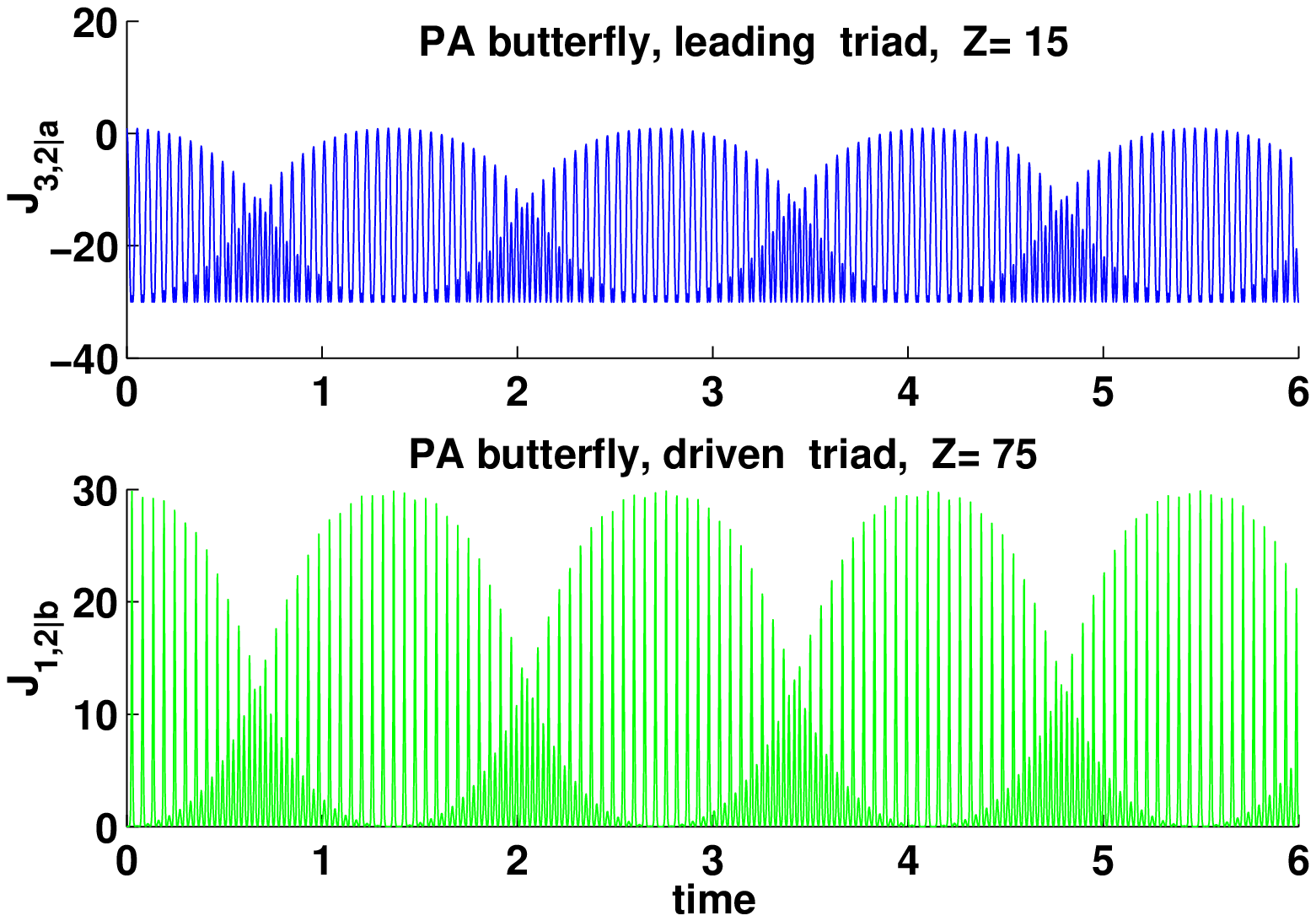}~~ \\ 
  Strong suppression of the energy transfer in  PA-butterfly& Efficient   energy transfer  in  PA-butterfly \\ \hline \hline
  $\C E$ & $\C F$ \\
  ~~\includegraphics[width=8.3cm ]{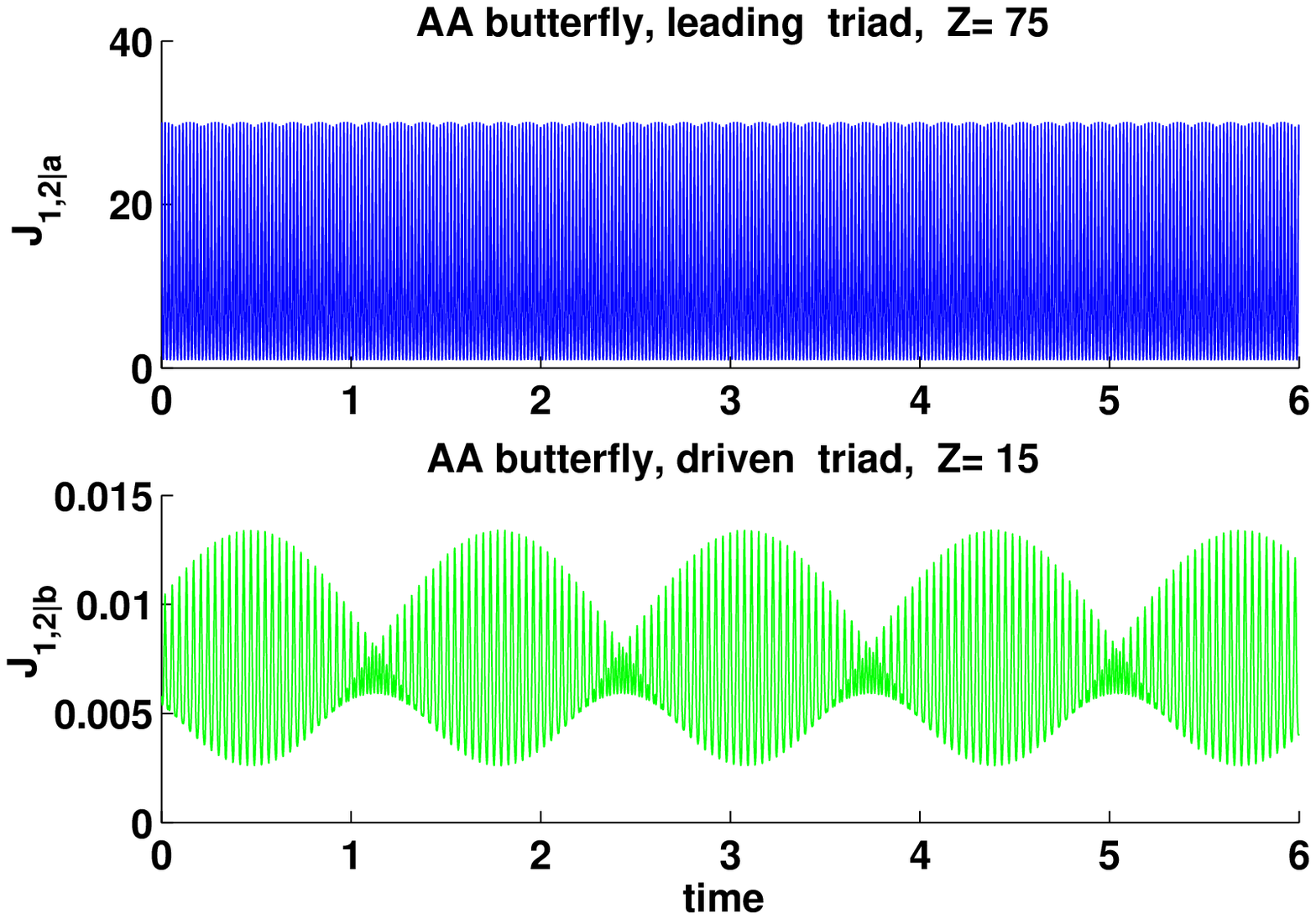}~~& 
 \includegraphics[width=8.3 cm]{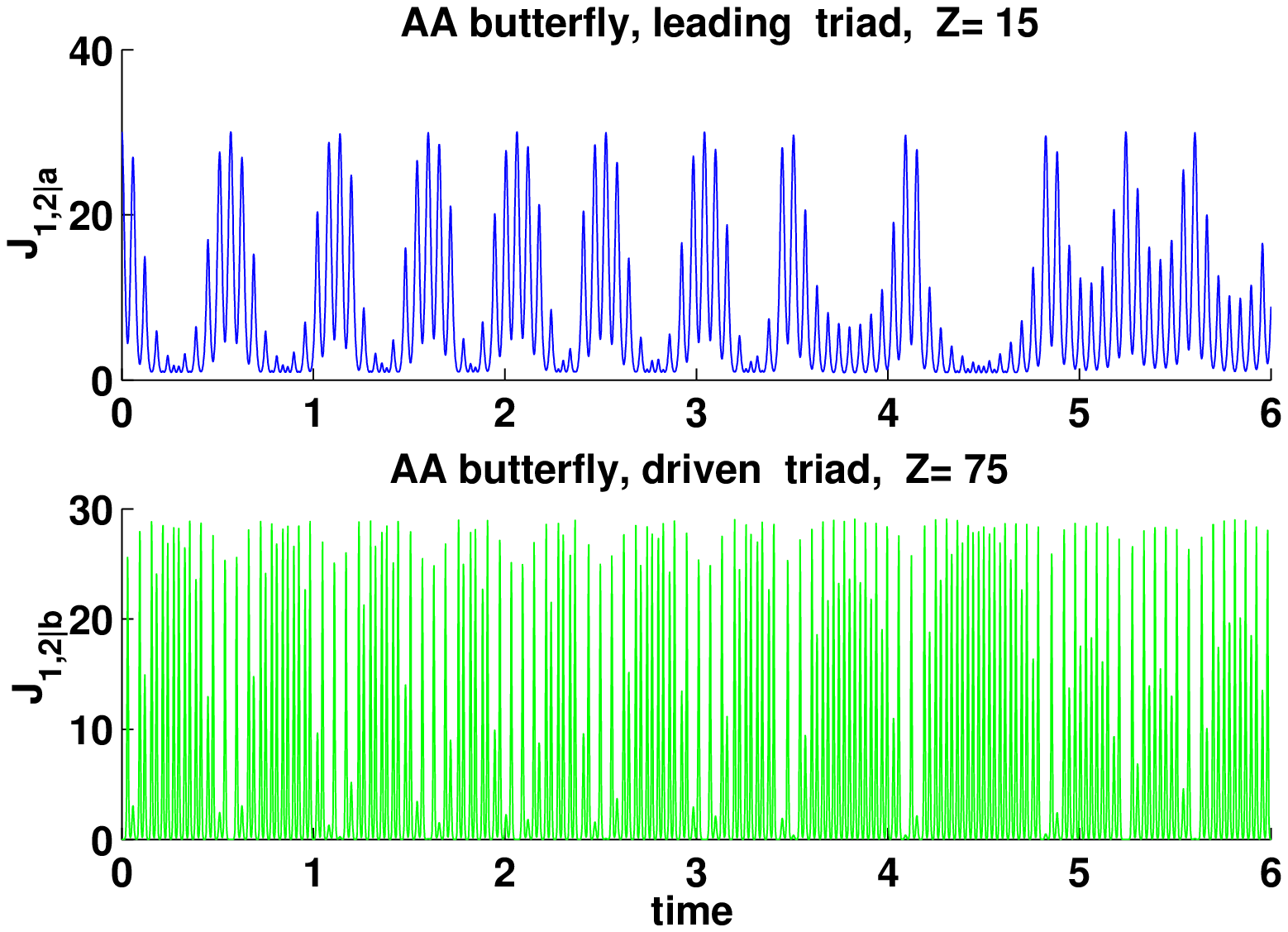} \\    
  Strong suppression of the energy transfer in  AA-butterfly& Efficient   energy transfer  in  AA-butterfly \\
  \hline
\end{tabular}
\end{center}
\caption{\label{f:PP-PA-AA} Color online. Time evolution of PP-, PA- and AA-butterflies with $Z_a=75$, $Z_b=15$, left panels and $Z_a=15$, $Z_b=75$, right panels.  Initial conditions are given by \Eqs{act1}, \eq{ic-a} and \eq{ic-b}. In \Figs{f:PP-PA-AA} -- \ref{f:PAA}   time is measured in arbitrary units}
\end{figure*}

\begin{figure*}
\begin{center}
\begin{tabular}{ c c }
    $\C A $& $\C B $  \\
 \includegraphics[width=9cm ]{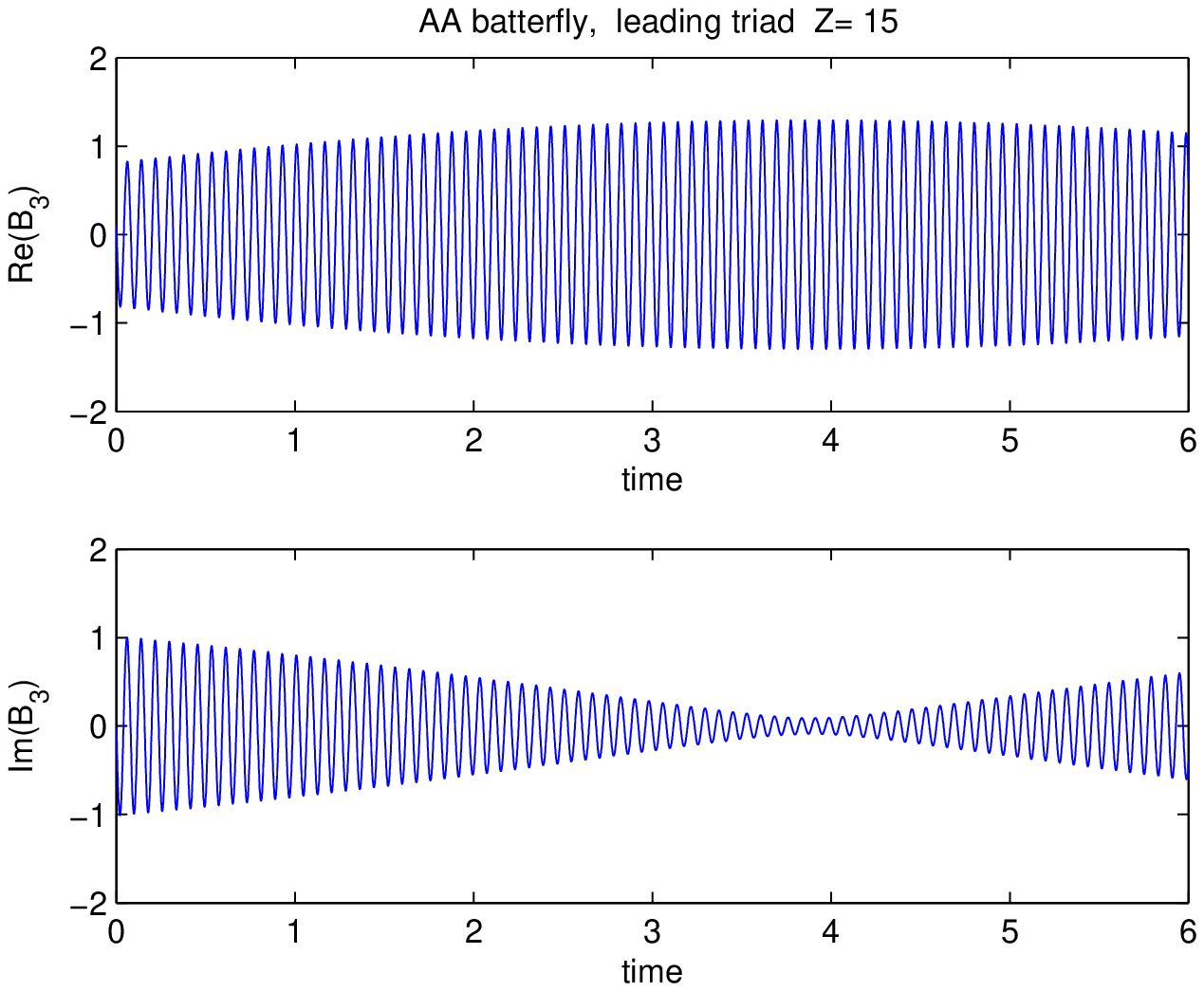} &        
 \includegraphics[width=9 cm] {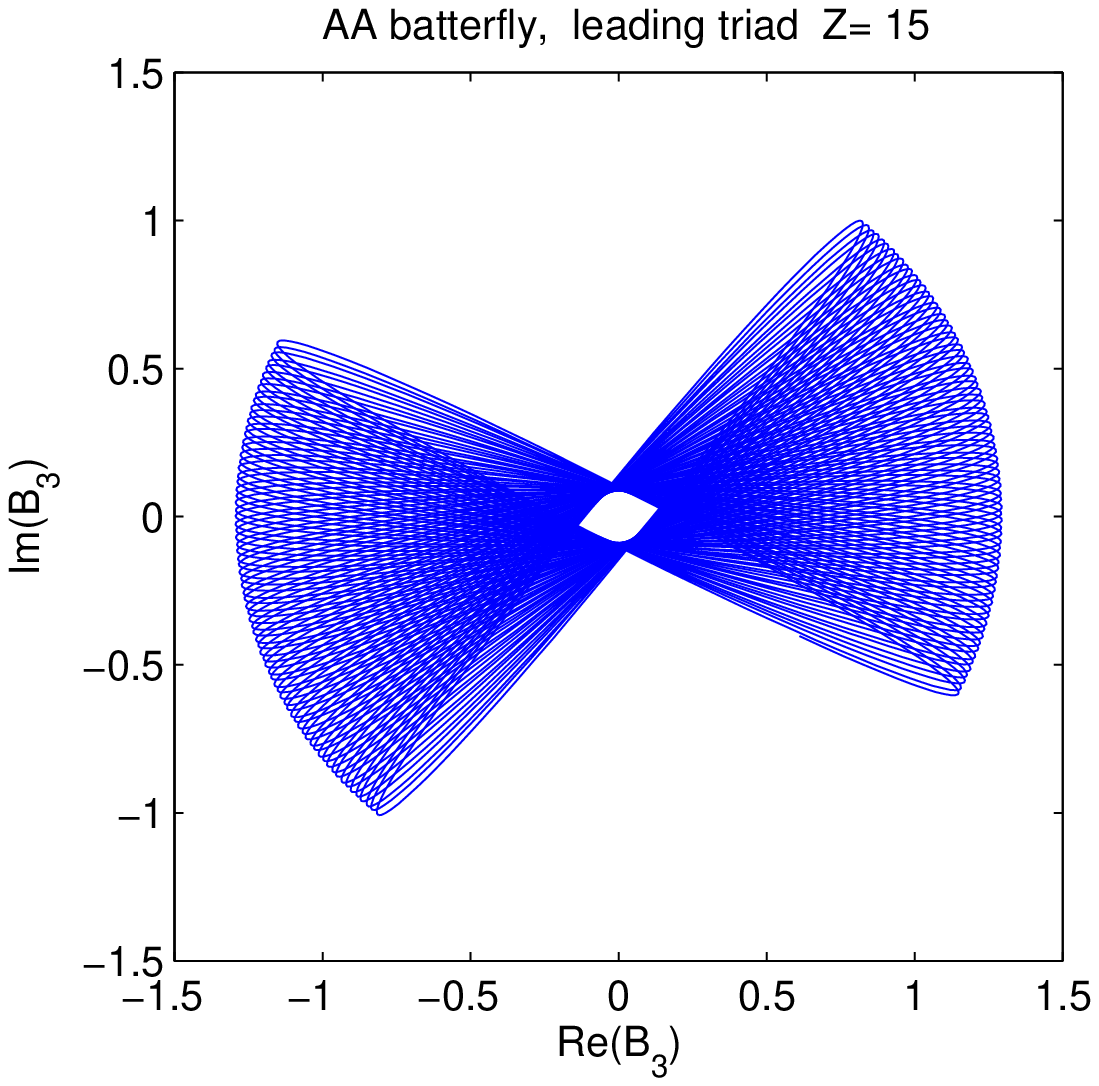} \\ 

\end{tabular}

  \end{center}
\caption{\label{f:AA-ev3} Time evolution of the common mode
  Re$B_{3|a}(t)$ and Im $ B_{3|a}(t)$, panel $\C A$, and the parametric
  representation Re$B_{3|a}(t)$ vs.  Im $B_{3|a}(t)$, panel $\C B$, for
  AA-butterfly with $ Z_a=15\,, Z_b=75 $ and the same initial
  conditions~\eq{act}, \eq{ic-AA} and \eq{ic-b} as in
  \Fig{f:PP-PA-AA}. }
\end{figure*}%
 \paragraph{PP-butterfly} has the   most trivial time-evolution, see    \Fig{f:PP-PA-AA}  for $Z_a=75, \ Z_b=15$, (panel  $\C A$) and $Z_a=15, \ Z_b=75$, (panel  $\C B$).  As expected, there is practically no energy exchange between triads: amplitudes $J_{2,3|a}\,, J_{2,3|b}$, defined by \Eq{batPP}, oscillate within the domains, that are determined by initial conditions. We show these evolutions   just to illustrate our analytical result that PP-connection is non-penetrative  for energy in both directions at any time.

\paragraph{AA-butterfly} is a promising candidate for the energy transfer. Time evolutions of $J_{1,2|a}(t)$ and $J_{1,2|a}(t)$ for AA-butterfly [defined by \Eqs{batAA}]  are shown    in \Fig{f:PP-PA-AA}$\,\C E\,, \C F$. Indeed, as one sees in   panel $\C F$  the peaks of the  amplitudes $J_{1,2|a}(t)$ in the leading triad and $J_{1,2|b}(t)$ in the driven triad  are close to 30 for  $Z_a=15, \ Z_b=75$.
 Quite unexpectedly, the energy transfer may be not efficient even
 with AA connections. Indeed, as one sees in \Fig{f:PP-PA-AA}~$\C E$,
 when the leading $a$-triad has larger $Z_a=75$, the sum
 $|B_1|^2+|B_2|^2$ in the leading triad oscillates between $\approx
 30$ and $\approx 20$ being close to its initial value $\approx
 30$. At the same time, this sum in the driven triad oscillates close
 to its initial value $\approx 0.01$. Therefore the energy transfer is
 strongly suppressed if $Z_a\gg Z_b$.

 To understand the difference between these two cases, consider time
dependence of the common mode $B_3(t)$, for the case $ Z_a=15\,,
Z_b=75 $ shown in \Fig{f:AA-ev3}~$\C A$. One sees fast oscillations of the
common mode $B_3(t)$ with (almost) zero mean and some frequency $\O_a$
that can be estimated as $Z_a B_a $, where $ B_a \=\sqrt{J_{1,2|a}}$
is a characteristic value of mode magnitudes in $a-$triad.  An
illustration of this estimate one sees in \Fig{f:PP-PA-AA}~$\C E$,
 where (for the same value of $|B_a|$) the oscillation frequency of the triads with $Z_a\approx 75$ is much higher than that for triads with $Z_a\approx 15$. As one can see from the equations of
motion (\ref{EMbuts}) for butterflies,  the mean  energy flux from the
leading, $a-$, to the driven $b$-triad (with A-connection), $\varepsilon_{a\to
b}$, can be written as
\begin{equation}\label{flux}
\varepsilon_{a\to b}=  2 Z_b  \mbox{Re} \big [ B_{1|b} B_{2|b} B_{3|b}^* \big ] \propto \cos \phi _b\,,
 \end{equation}
  where $\varphi_{b}=\varphi_{1|b}+\varphi_{2|b}-\varphi_{3|b}$ is the
triad phase, introduces by \Eq{dc}. Fast oscillations of $ B_a (t)$,
equivalent to fast growth of $\varphi_{3|b}(t)$ with the speed $d
{\varphi}_{3|a}(t)/d t = d {\varphi}_{3|b}(t)/d t\approx \O_a(t)$ ,
lead to self-averaging of $\cos \varphi_{b}$ almost to zero, if phases
$\varphi_{1|b}$ and $\varphi_{2|b}$ cannot react fast enough to
variations of $\varphi_{3|b}(t)$. This is a qualitative explanation of
the observed fact that the energy flux from $a-$ to $b-$triad is
strongly suppressed if $Z_a\gg Z_b$ and is observable (but still
suppressed) if $Z_a \lesssim Z_b$. One can hope that the $a-b$ energy
flux can be significant if $ B_{3|a} (t)$ has a nonzero mean and
therefore $\< \varphi_{3|b}(t)\>$ also does not vanish. But this is
impossible under the initial conditions of interest: we do not want to
put initially energy to $b-$triad, taking large value of $ B_{3|a}
(t)=B_{3|b} (t) $ at time zero. Therefore at $t=0$ $|B_{3|a}|\ll
|B_{1|a}|, |B_{2|a}|$. If so, according to the equation of motion, the
time derivative $\dot{B}_{3|a}=-Z_a B_{1|a}B_{2|a}$ is large enough to
allow $B_{3|a}$ to cross quickly zero and to reach significant value
with different sign. Having in mind periodical character of evolution
of isolated triad it practically means that $\<B_{3|a}\>\ll
\sqrt{\<|B_{3|a}|^2\>}$.

\paragraph{PA- and AP-butterflies.} Considering free evolution from asymmetrical initial conditions we will distinguish PA-butterfly,  in which the leading triad has P-connection and  the driven triad has A-connection,  from AP-butterfly,  in which the leading triad has A-connection and  the driven triad has P-connection.

We found that in AP-butterflies the energy transfer to the driven triad
 is blocked by the second of the conservation laws~\eq{APintC} and
its evolution is similar to that of PP-butterfly.

 On the contrary, PA-butterflies demonstrate time evolution similar to that
 of AA-butterflies: compare in \Fig{f:PP-PA-AA} panels $\C C$ and $\C D$ for
 PA-butterfly with   panels $\C E$ and $\C F$ for AA-butterflies.  One sees that
 corresponding plots in both panels are quantitatively the same: in
 the left panels (with $Z_a=75\,, \ Z_b=15$) the energy transfer is
 strongly suppressed (with $J\sim 0.01$ in the driven triad), while in
 the right panels (with $Z_a=15\,, \ Z_b=75$) the energy transfer is
 efficient (with $J\sim 10\div 30$ in the driven triad).

\begin{figure}
\begin{center}
 ~~\includegraphics[width=9 cm ]{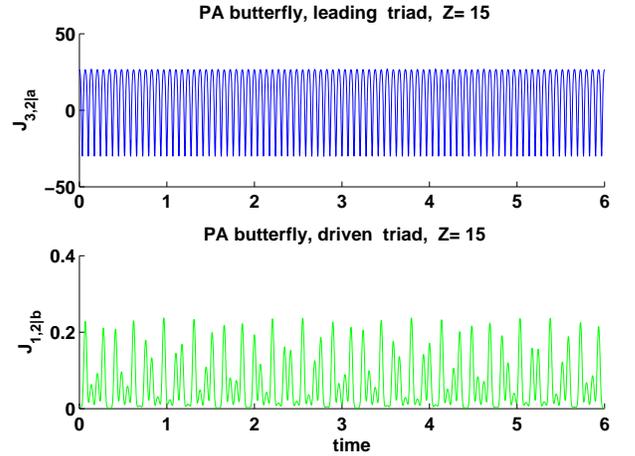}~~ 
  \end{center}
\caption{\label{f:PA-ev1}    Color online. Suppression of the energy transfer in PA-butterfly with   equal values of the interaction coefficients: $Z_a=Z_b =15 $. Initial conditions are the same as in \Fig{f:PP-PA-AA} for PA-butterfly: \Eqs{act1}, \eq{ic-AP2} and  \eq{ic-b}. }
\end{figure}

To complete discussion on how the energy transfer depend on the ratio
$Z_a\big / Z_b$ we present in \Fig{f:PA-ev1} time evolution from the
same as in \Fig{f:PP-PA-AA} initial conditions, but with equal values
of the interaction coefficients, taking for concreteness
$Z_a=Z_b=15$. The excitation level of the driven triad
$J_{1,2|b}\simeq 0.2$. As expected, this level is larger than $\simeq
0.01$ for $Z_a\big / Z_b=5$ and smaller than $\simeq 20$ for $Z_a\big
/ Z_b=1/5$.

\subsubsection{\label{sss:dep} Effect of initial conditions}
As we mentioned, the time evolution and efficiency of the energy transfer between triads crucially depend on the initial conditions. To demonstrate this we present in \Fig{f:PA-AA}  the evolution of PA-butterflies ( panel~$\C A$) and AA-butterflies (panel $\C B$) for the same ``efficient" values of the interaction coefficients ($Z_a=15, \ Z_b=75$), as in \Fig{f:PP-PA-AA}$\, \C D$, $\C E$ and $\C F$. The only difference is in initial conditions,  that on the first glance are very similar to that used above in \Fig{f:PP-PA-AA}.   Namely, for PA-butterflies we simply interchange  initial amplitudes of individual modes in the leading $a$-triad, replacing $B_0 \Leftrightarrow \~ B_0$ in \Eq{ic-AP2}.  For AA-butterflies  we replace almost equal values of $B_0$ and $\~ B_0$  in \Eq{act1} by different in an order of magnitude  values~\eq{act} with the same sum $|B_0|^2+|\~B_0|^2\approx 30$. The result of this ``minor" change is crucial: in PA-butterfly the excitation level $J_{1,2|b}$  decreases from $\simeq 20$ to $\simeq 0.4$ and in AA-butterfly from from $\simeq 20$ to $\simeq 0.1$.

 In order to rationalize this effect notice that  the energy flux from the $a$- to $b$-triad is proportional to the level of excitation of the common mode,  $\<|B_{3|b}|^2\>$ for both butterflies under consideration. To estimate the upper bound for $\<|B_{3|b}|^2\>$ we can approximate $a$-triad as an isolated one, neglecting the feedback effect of the driven triad on the leading one. In this approximation we can use the invariants~\eq{MR} for isolated triad:
\begin{eqnarray}\label{AAints2}
I_{1,3|a}= |B_{1|a}|^2+ |B_{3|a}|^2\!\! , \  I_{2,3|a}= |B_{2|a}|^2+ |B_{3|a}|^2.~~
\end{eqnarray}
\paragraph{AA-butterfly.} In AA-butterfly initially   the amplitude $|B_{3|a}|\ll |B_{1|a}|, \ |B_{2|a}|$. Therefore $I_{1,3|a}= |B_{1|a}(t=0)|^2\=|B_{1|a}^{\,(0)}|^2$ and $I_{2,3|a}= |B_{2|a}^{\,(0)}|^2$ and thus
\begin{eqnarray} \label{res1}
\begin{cases}
  |B_{3|a}|^2\le |B_{1|a}^{\,(0)}|^2  \\
  |B_{3|a}|^2\le |B_{2|a}^{\,(0)}|^2   \\
  \end{cases}\!\!\!\!\!\!\!\!
&\Rightarrow&  \!\!\!  |B_{3|a}|^2\le \min\{ |B_{1|a}^{\,(0)}|^2 , |B_{2|a}^{\,(0)}|^2   \}  .~~~~~~
\end{eqnarray}
This means that the efficiency of the energy transfer is determined by the smallest initial magnitudes of the individual modes; fixing sum of their squares one has the most efficient transfer at (almost) equal initial magnitudes. This was realized in the first  version of the initial conditions~\eq{act1},  that leads to the best energy transfer, in which, according to \Fig{f:PP-PA-AA}$\,\C F$ $\max{J_{1,2|b}}\approx 30$.

Let us show, that this is indeed the  maximal possible value of the  level of excitation of the driven triad $J_{1,2|b}$. To this goal consider the invariants~\eq{AAint} of AA-butterfly, that can be combined as follows:
\begin{subequations}
\label{AAres1}
\begin{equation}
\label{AAints1}
    I\Sb{AA}= |B_{1|a}|^2+|B_{2|a}|^2+2|B_{3|a}|^2+|B_{1|b}|^2+|B_{2|b}|^2\ .
\end{equation}
At $t=0$ $ B_{1|a}= B_0 $,  $ B_{2|a}=\~B_0 $ (or vise versa) and much larger than the rest of the amplitudes. Therefore $I\Sb{AA}=
 B_{1|a}|^2+|B_{2|a}|^2=J_{1,2|a}=|B_0|^2+|\~B_0|^2$ and thus
\begin{equation}
\label{AAres}
     J_{1,2|b} \simeq |B_{1|b}|^2+|B_{2|b}|^2\le I\Sb{AA}\approx 30\,, \quad \mbox{AA.}
\end{equation}\end{subequations}

Finally notice, that in view of the restriction~\eq{res1} it is clear, that with the second choice of the initial conditions~\eq{act} the energy transfer should be much less efficient. This is exactly what one sees  in  \Fig{f:PA-AA}$\,\C B$.

\paragraph{PA-butterfly.} The situation is a bit different for PA-butterfly with P- and A-modes being individual in the leading $a$-triad. In this case
 combining invariants~\eq{APintC} for PA-butterfly one finds new
 (dependent) invariants:
 \begin{subequations}\label{PAres1}
 \begin{eqnarray}\label{PAints1}
I\Sb{PA}&=&|B_{3|a}|^2-|B_{2|a}|^2 +2 |B_{3|b}|^2+|B_{1|b}|^2~~~~\\ \nn
&& +|B_{2|b}|^2=J_{3,2|a}+J_{1,2|b}+2|B_{3|b}|^2\ .
\end{eqnarray}
At $t=0$, $ B_{2|a}= \~B_0 $, $ B_{3|a}=B_0 $ (or vise versa) and much
larger than the rest of the amplitudes. Therefore $I\Sb{PA}=
J_{2,3|a}(0)$ which is equal to $-(|B_0|^2-|\~B_0|^2)$ for the old
initial conditions and to $I\Sb{PA}= |B_0|^2-|\~B_0|^2 $ for the new
ones. This allows one to write:
 \begin{equation}\label{PAres}
 I\Sb{PA}=  J_{2,3|a}(0)\approx \pm 26.7\,, \  \mbox{PA.}
 \end{equation}\end{subequations}
 Unfortunately, in this case invariant $I\Sb{PA}$ is not positive
 definite and therefore one cannot get rigorous restriction on
 $J_{1,2|b}$, similar to~\eq{AAres}.  Nevertheless, more detailed
 analysis allows us to think that the upper bound for $J_{1,2|b}$ for
 PA-butterfly should be the same as for AA-butterfly, i.e. about 30
 for our initial conditions.

 In order to clarify the dependence of the energy transfer on initial
 conditions we should estimate
 the amplitude of common mode similarly to the case of AA-butterfly, neglecting the excitation of the driven
 triad. In this case invariants~\eq{APintC} can be written as:
 \begin{eqnarray}\label{APints2}
I_{1,3|a}= |B_{1|a}|^2+ |B_{3|a}|^2\!\! , \  I_{1,2|a}=
|B_{1|a}|^2- |B_{2|a}|^2.~~
\end{eqnarray}
At initial moment of time the amplitude $|B_{1|a}|\ll |B_{2|a}|, \
|B_{3|a}|$. Therefore $I_{1,3|a}\simeq
|B_{3|a}(t=0)|^2\=|B_{3|a}^{\,(0)}|^2$ and $I_{3,2|a}=
|B_{3|a}^{\,(0)}|^2- |B_{2|a}^{\,(0)}|^2 $. Thus the first of
\Eqs{APints2} gives
\begin{eqnarray} \label{APres1}
   |B_{1|a}|^2=|B_{3|b}|^2\le | B_{3|a}^{\,(0)}|^2\,,
\end{eqnarray}
while the second one leads to the trivial restriction $|B_{1|a}|^2\ge
- | B_{3|a}^{\,(0)}|^2$, that is satisfied automatically.  The
conclusion is that the efficiency of the energy transfer is determined
by the initial magnitudes of the individual A-mode. This was realized
in the first version of the initial conditions in which we assigned
more energy to $B_{3|a}$ mode; that leads to the best energy transfer.

The next question is how to rationalize why the energy transfer into the
driven triad can exceed 50\%.  Intuitively the answer is rather obvious: we understood already that the energy transfer is much
more effective, if the accepting triad has larger interaction
coefficient $Z$. Therefore in the considered case, when $Z_b\gg Z_a$,
during long evolution with various values of the triad phase (that
determines the direction of the energy flux) and with similar value of
the triad excitations, the energy flux from  $a$- to $b$-triad
is more favorable than the flux in the opposite direction. This leads to the
asymmetry of the mean energy content between triads in favor of
the $b$-triad with larger value of the interaction coefficient:
$Z_b\gg Z_a$.

\paragraph{What depends on the excitation level of the driven triad?}
Up to now we have considered initial conditions in the leading triad, just
by mentioning that the driven triads have much smaller values of the
initial excitations. The reason for this neglect is that the time
evolution is insensitive to the initial excitations in the driven
triads, their level just have to be very small. To demonstrate this we
compare   the time evolution of AA-triad with $Z_a=15$, $Z_b=75$ from
the initial conditions \eq{act1}, \eq{ic-AA} and   with ``standard"
level of initial excitation in $b$-triad [$C=1$ in \Eqs{ic-b}] with
that for the much smaller initial energy contents in the driven triad,
see \Fig{f:AA-ev2}.  The long-time evolution is
practically independent of $C$ as shown in \Fig{f:AA-ev2}$\, \C A$. The $C$-dependence is visible only on the
initial stages of the evolution:   panel-$\C B$ with
$C=1$, -$\C C$  with $C=0.1$  and  -$\C D$ with
$C=0.01$. One sees that the position of the first maximum for $C=1$ is at
$t\sb{max}\approx 0.03$, for $C=0.1$ at $t\sb{max}\approx 0.04$, while
for $C=0.01$ at $t\sb{max}\approx 0.05$. This dependence is quite
understandable: according to \Eqs{AP} initial small perturbations in
the driven triad grow exponentially in time due to the parametric
instability of the common $B_{3|b}$ mode with respect of decay into
small individual modes: $B_{1|b}\,, \ B_{2|b}\propto \exp (Z_b
|B_{3|b}|t) $ modes. Therefore
\begin{equation}\label{t-est}
t\sb{max}\propto - \ln Z_b C\,, \end{equation}
 as observed.

\begin{figure*}
\begin{center}
\begin{tabular}{| c || c |}\hline
 $\C A $& $\C B $\\
    \begin{tabular}{ l }  \\
 ~\includegraphics[width= 8.3cm ]{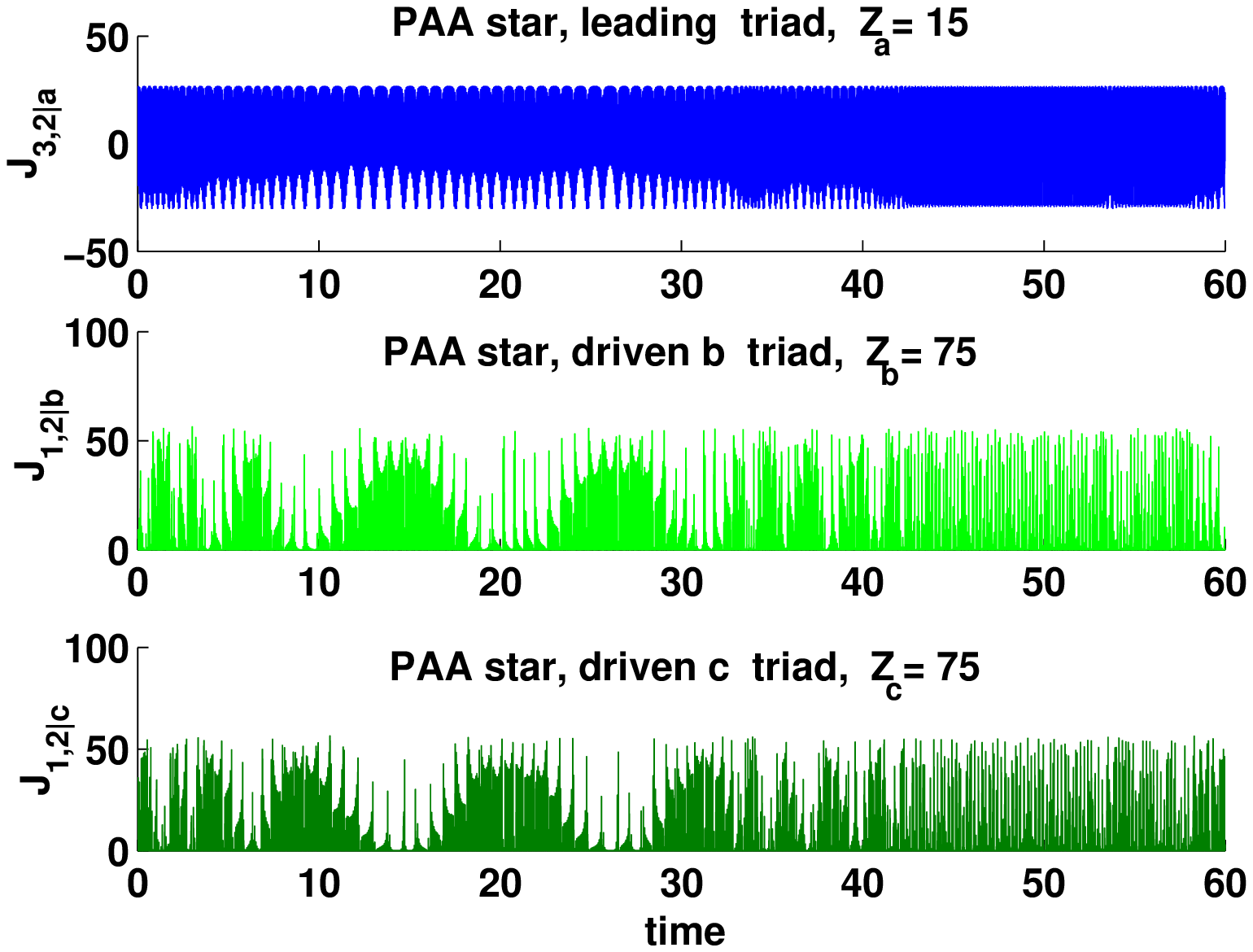}~~ \\ 
 ~~~~~Efficient energy transfer from $a$- to $b$-triad ($Z_b/Z_a=5$)\\ ~~~~~
and  from $a$- to $c$-triad ($Z_c/Z_a=5$)\\ \\
  \end{tabular} &

    \begin{tabular}{ l }  \\
 ~\includegraphics[width= 8.3cm ]{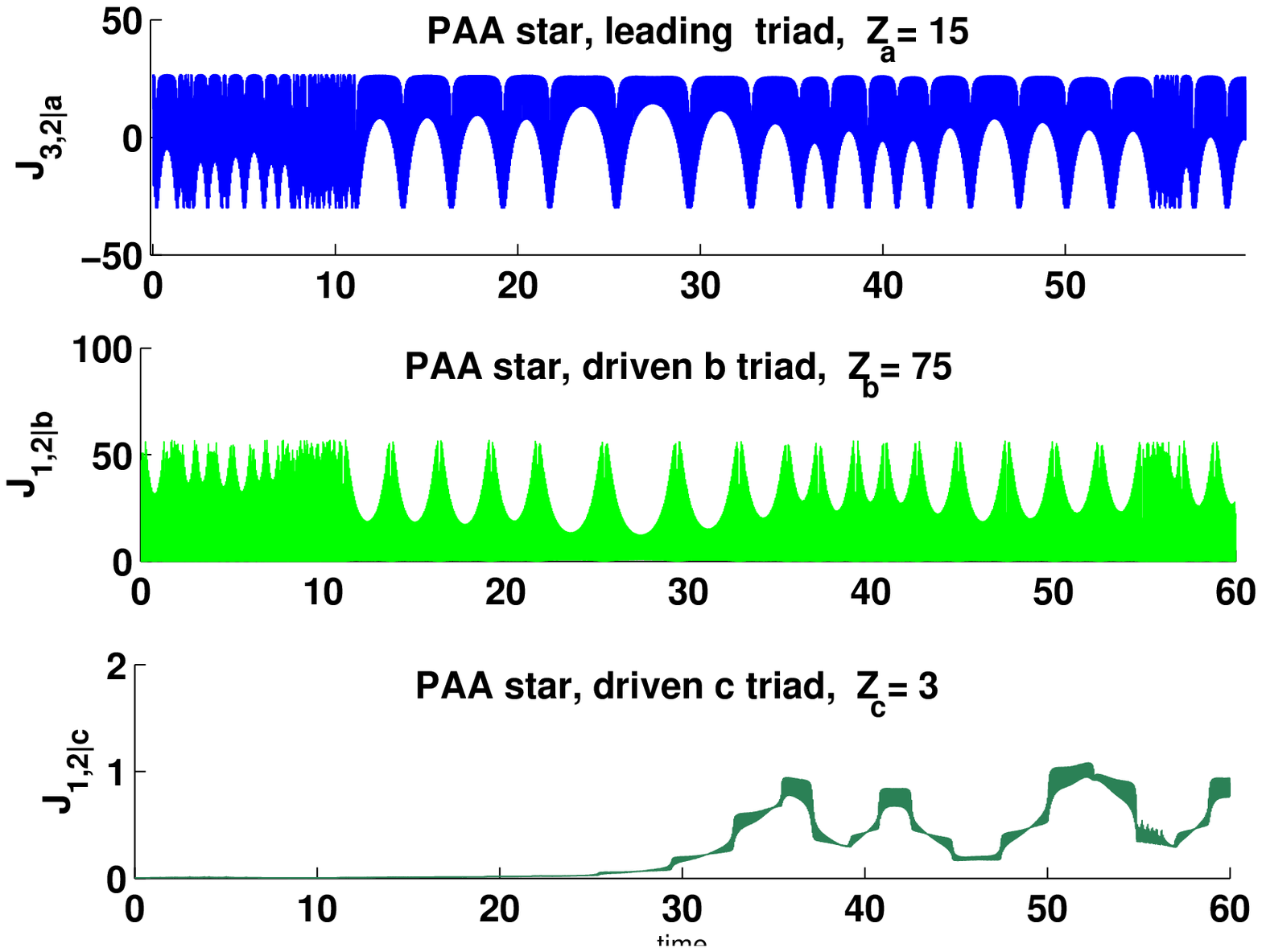}  \\ 
~~~~~ Efficiency of the  energy transfer from $a$-triad  to  \\
~~~~~$b$-triad ($Z_b/Z_a=1/5$)   is  a bit  smaller than that   \\   ~~~~~to $c$-triad ($Z_c/Z_a=5$)\\
  \end{tabular}\\  \hline\hline
 $\C C $& $\C D $\\
  \begin{tabular}{ l }    \\
 ~\includegraphics[width= 8.3 cm ]{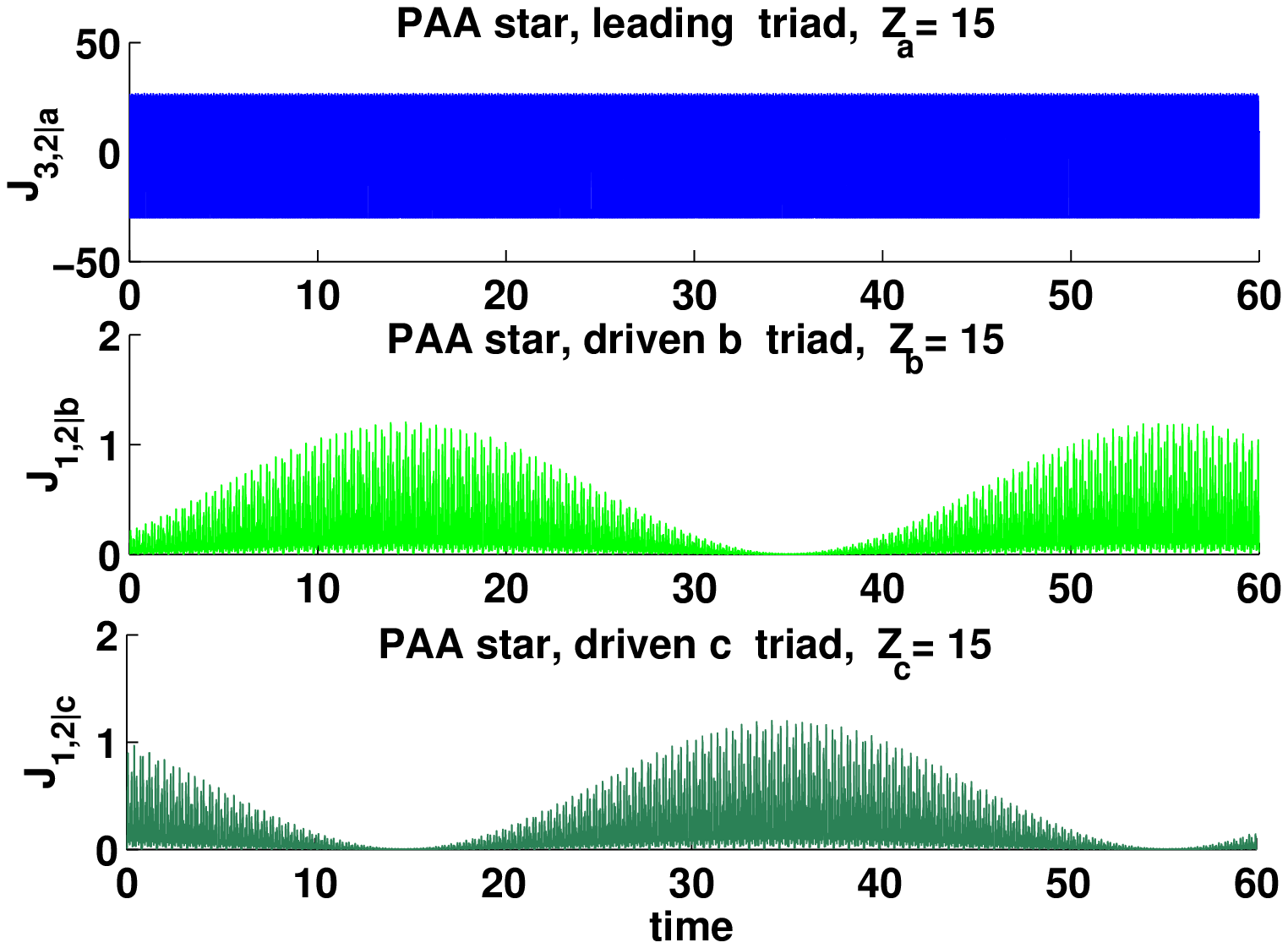}~~ \\ \\ 
 ~~~~~ Suppression of the energy transfer in PAA-star \\  ~~~~~junction with $Z_a=Z_b=Z_c$ \\ \\
 \end{tabular} &

    \begin{tabular}{ l }
 ~\includegraphics[width= 8.3 cm ]{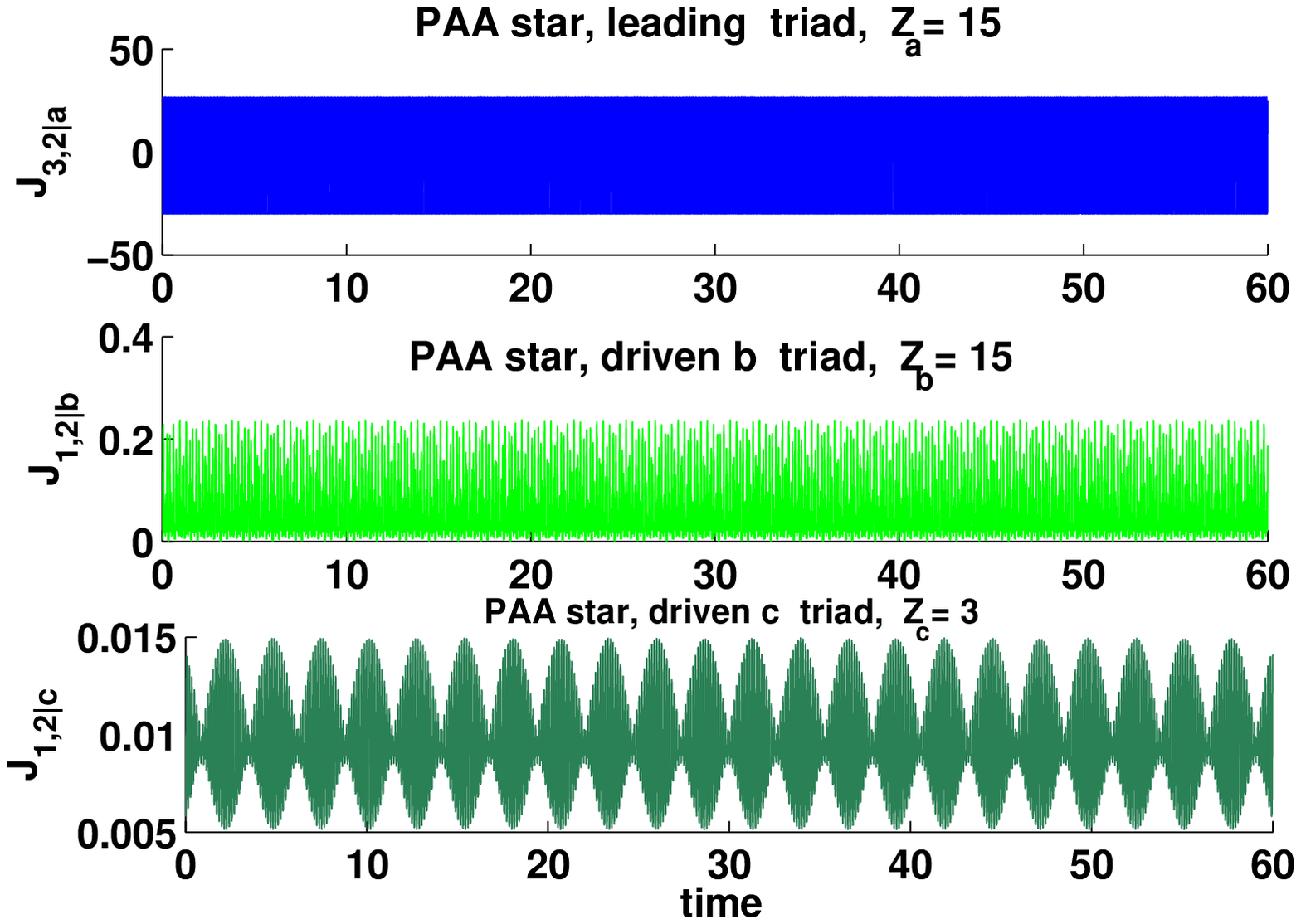}~~  \\ 
 ~~~~~ Strong suppression of the energy transfer in PAA-star \\  ~~~~~junction with $Z_a=Z_b>Z_c=3$ \\
  \end{tabular}\\  \hline
\end{tabular}
 \end{center}
\caption{\label{f:PAA} Color online. Efficiency of  PAA-star energy junction with  different  $Z_b/Z_a$ and $Z_c/Z_a$ [initial conditions~\eq{ic-PAA}].}
\end{figure*}

\begin{figure*}
\begin{center}
\begin{tabular}{|l||l|}
  \hline  ~~~~~$\C A $& ~~~~~$\C B $\\
  \includegraphics[width= 8.5cm ]{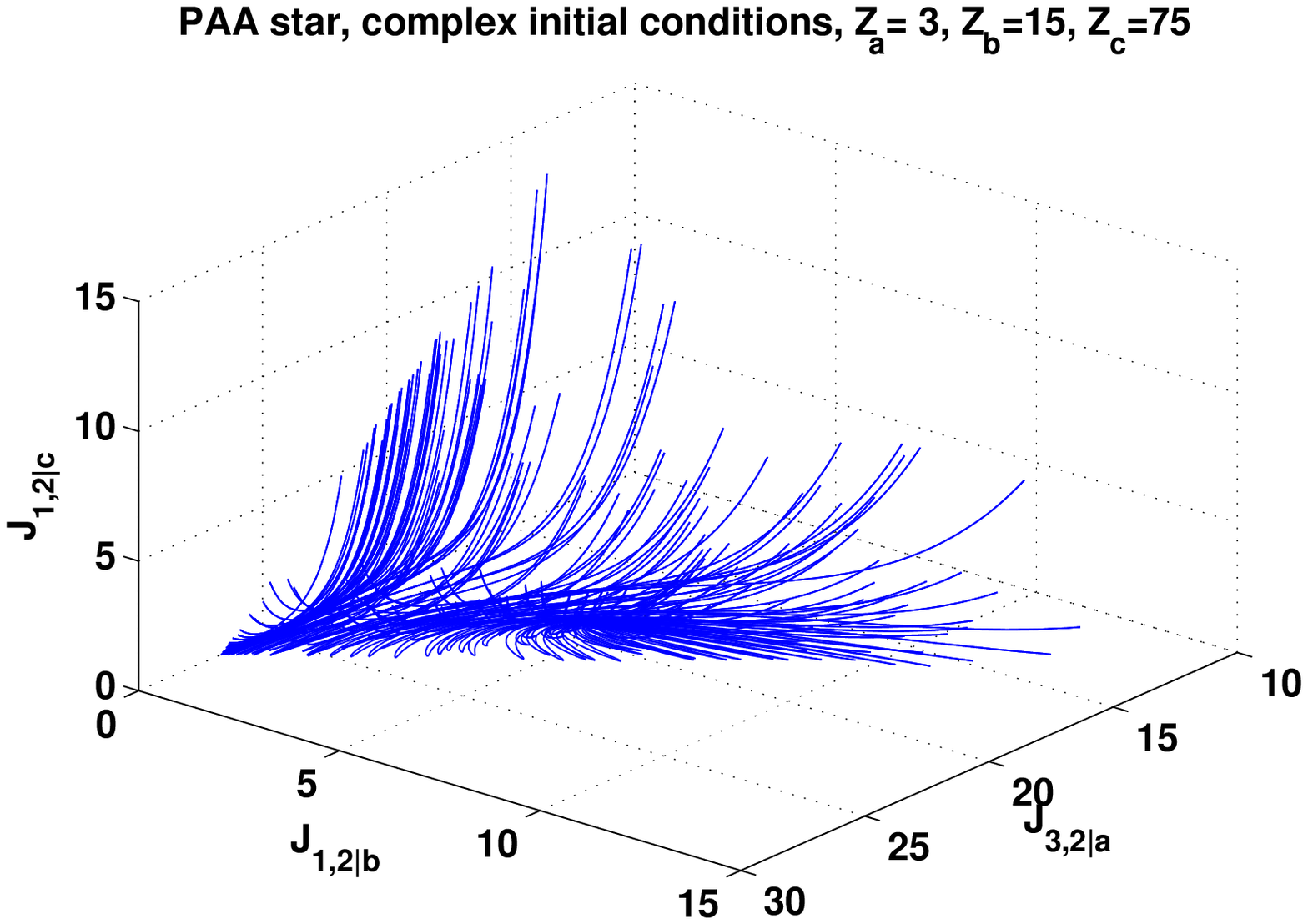}~~ & 
   \includegraphics[width= 8.5cm ]{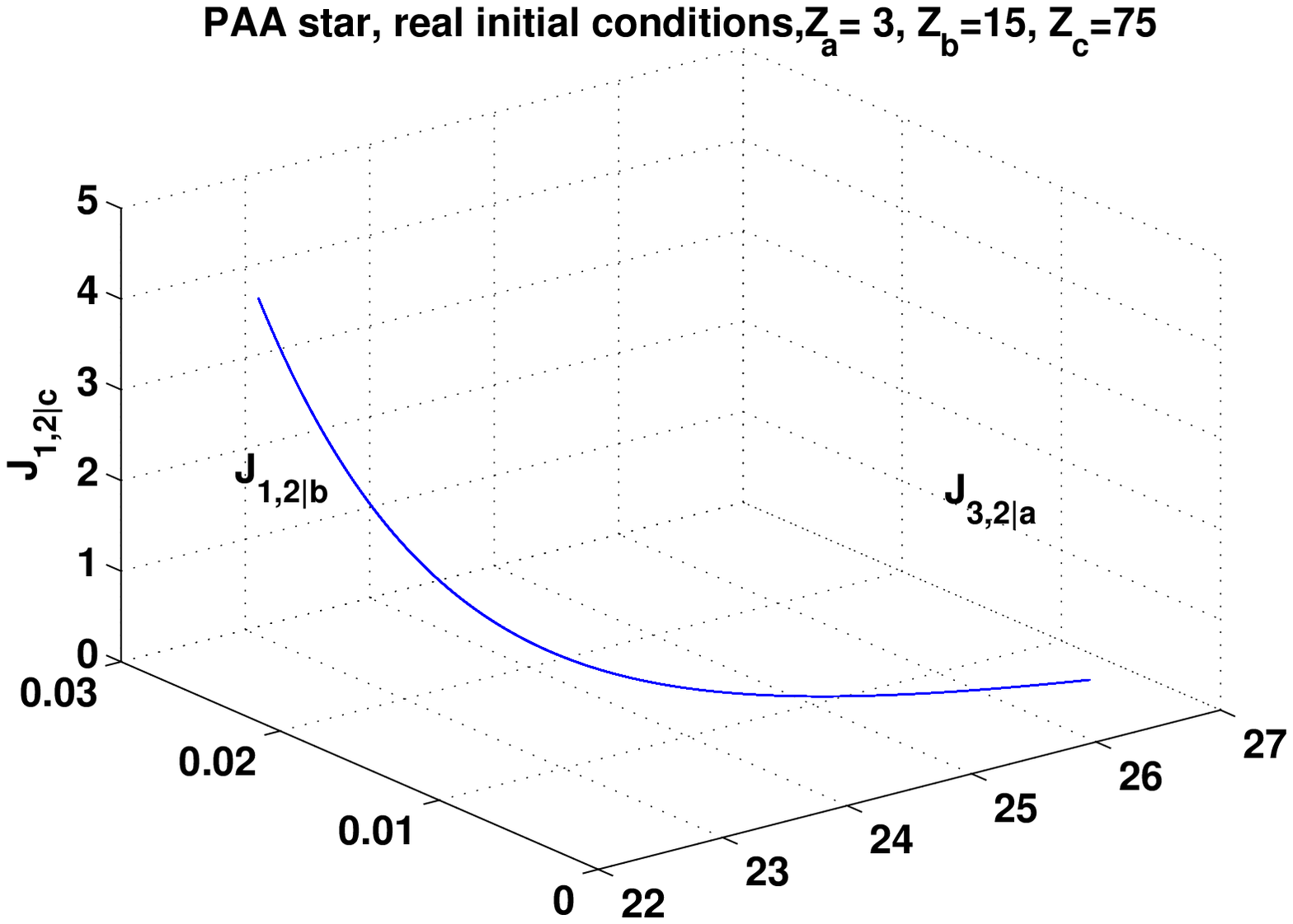}~~ \\ 
     \hline \hline  ~~~~~$\C C $& ~~~~~$\C D $\\
  \includegraphics[width= 8.5cm ]{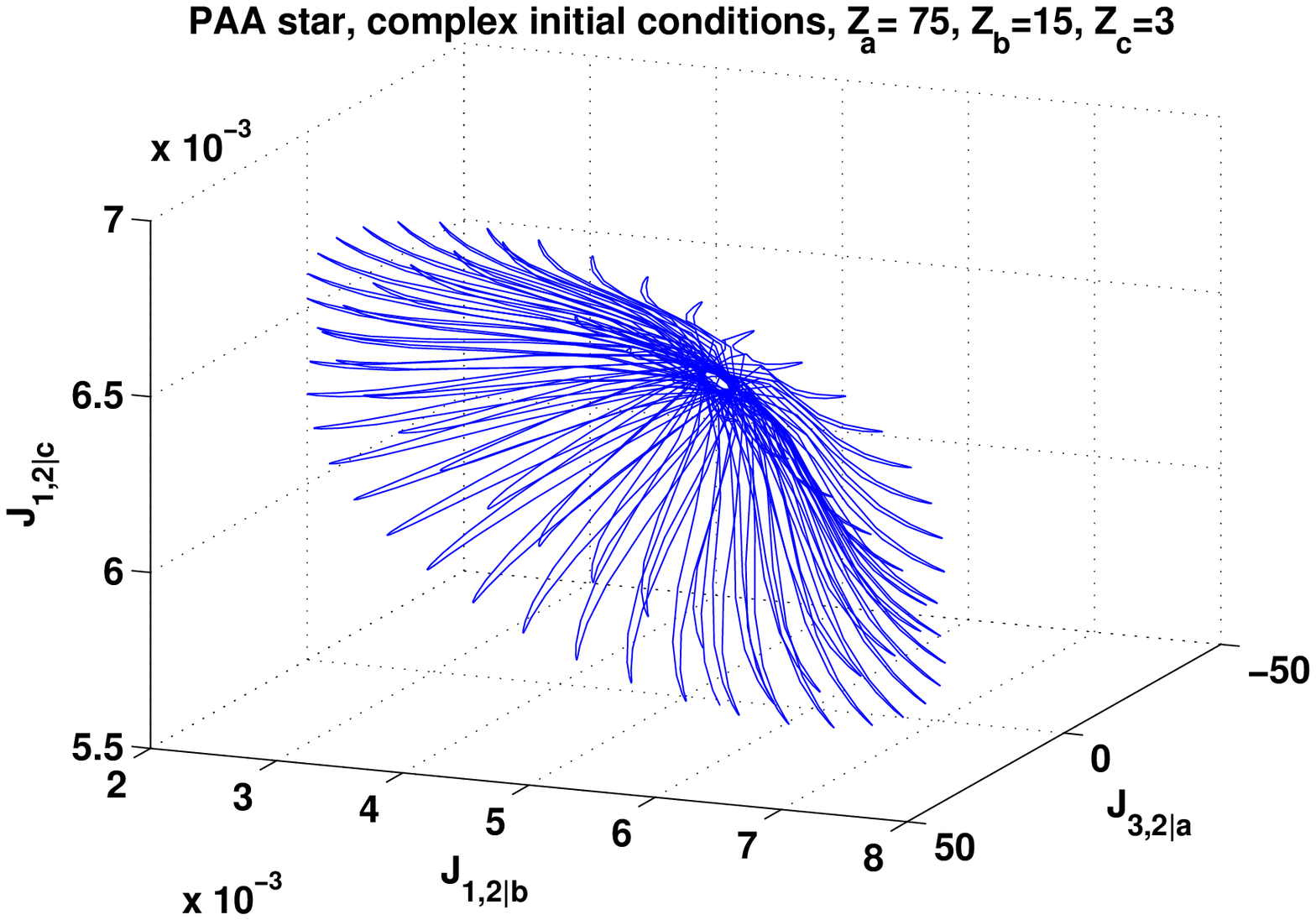}~~& 
  \includegraphics[width= 8.5cm ]{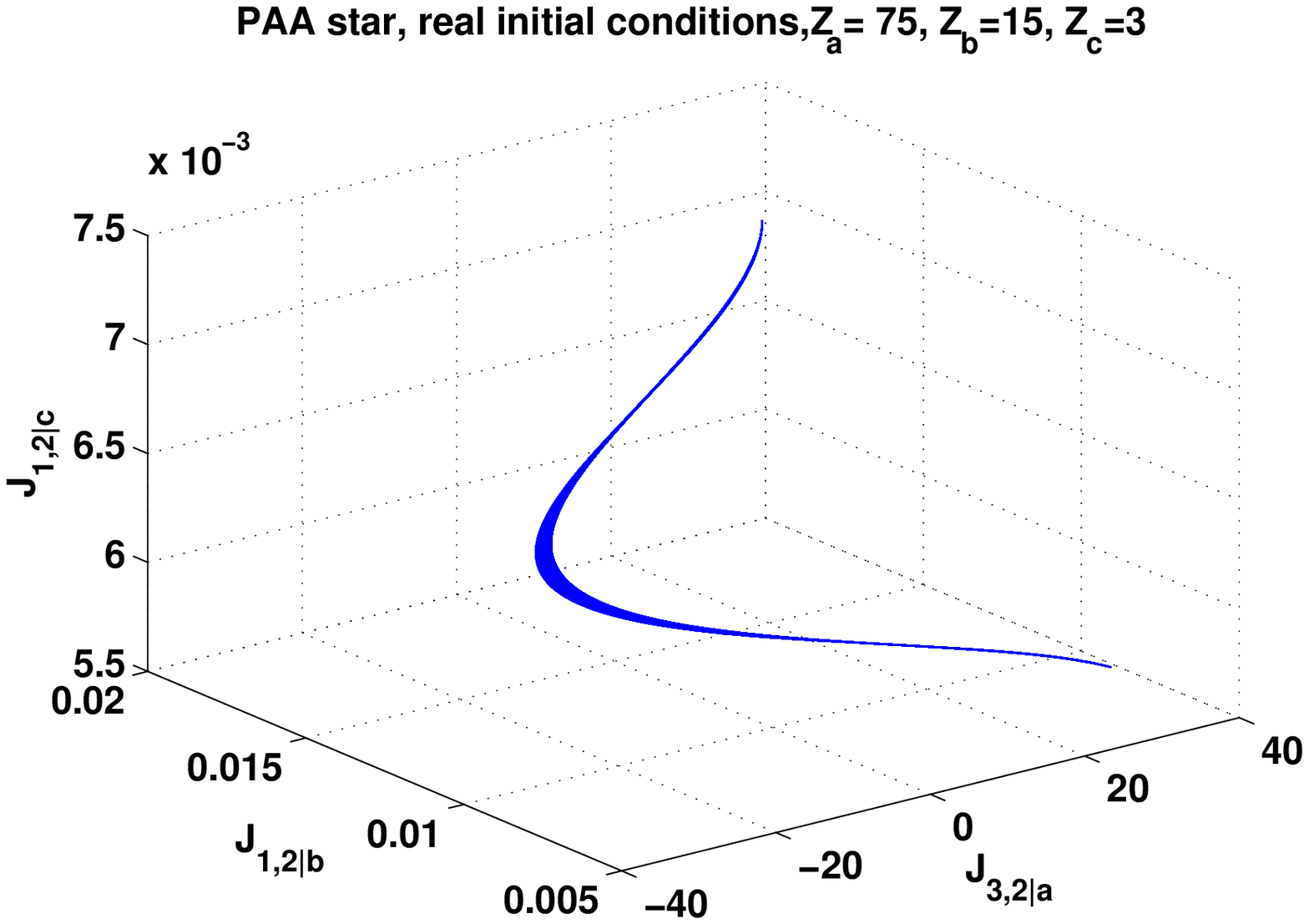}~~ \\  
  \hline
\end{tabular}

 \end{center}
\caption{\label{f:PAA-par} Color online. Three dimensional parametric
representation of trajectories of PAA-star with $Z_a=3$, $Z_b=15$,
$Z_c=75$, panels $\C A$ and $\C B$ and with $Z_a=3$, $Z_b=15$,
$Z_c=75$, panels $\C C$ and $\C D$. Left panels: Complex initial
conditions~\eq{ic-PAA}; Right panels: real initial conditions in which
complex number in \Eq{ic-PAA} are replaced by their absolute values.
}
\end{figure*}%


 \begin{figure}
\begin{center}
\begin{tabular}{|l||l|}
  \hline
  ~~~$\C A $& ~~~$\C B $\\
  ~\includegraphics[width= 3.2 cm ]{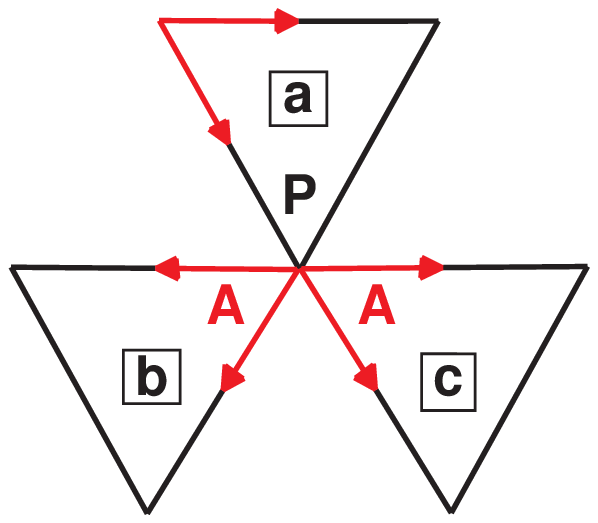}~ &     
  ~\includegraphics[width= 4.7 cm ]{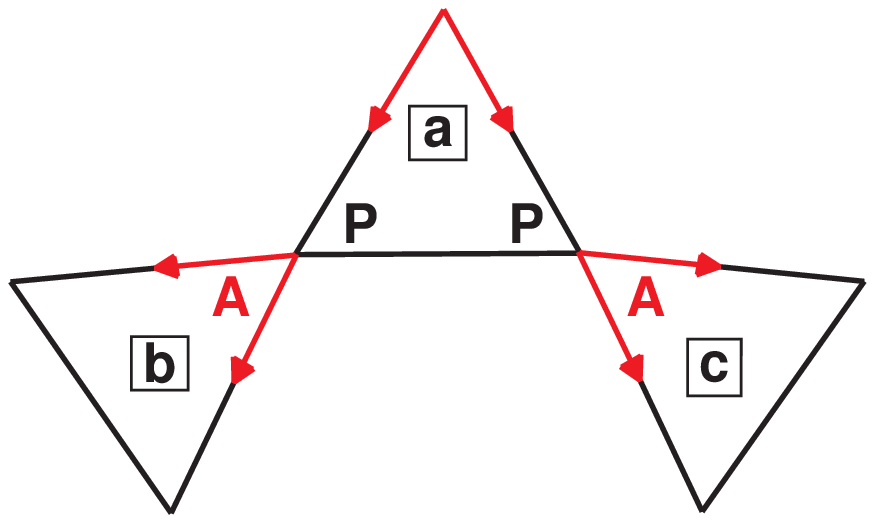}~ \\ 
\end{tabular}

  \end{center}
  \caption{Triple-star ($\C A $) and triple-chain ($\C B $) energy
  junctions. Initial energy is localized in the leading $a$-triad and
  goes during free evolution in two driven $b$- and $c$-triads. }
  \label{f:proto}
  \end{figure}

\subsection{\label{ss:3-6} Energy-junctions   in the triple-triad clusters}
In this Subsection we study free evolution of the triple-triad
clusters, that allow to shed light on the energy transfer from one triad to
two other triads. To this aim we consider here free evolution of
various triple-triad clusters, in which the leading $a$-triad will be
initially highly exited, while the level of excitation of two other
$b$- and $c$-triads will be relatively  small. One topological
option for effective energy flux from $a$- to $b$- and $c$-triad,
considered in the next Sec.~\ref{ss:3tr} is PAA-star. One more option is the triple-triangle
configuration, in which energy goes from one triad to two other,
PP-connected triads. This structure is very rare and will not be
discussed here.  Next option,
considered in Sec.~\ref{sss:AP-PA}, is AP-PA chain with the leading
$a$-triad in the middle.

 %

\begin{figure*}
\begin{center}
\begin{tabular}{| c || c |}\hline
$\C A$ & $\C B$ \\
    \begin{tabular}{ l }
 ~\includegraphics[width= 8.5cm ]{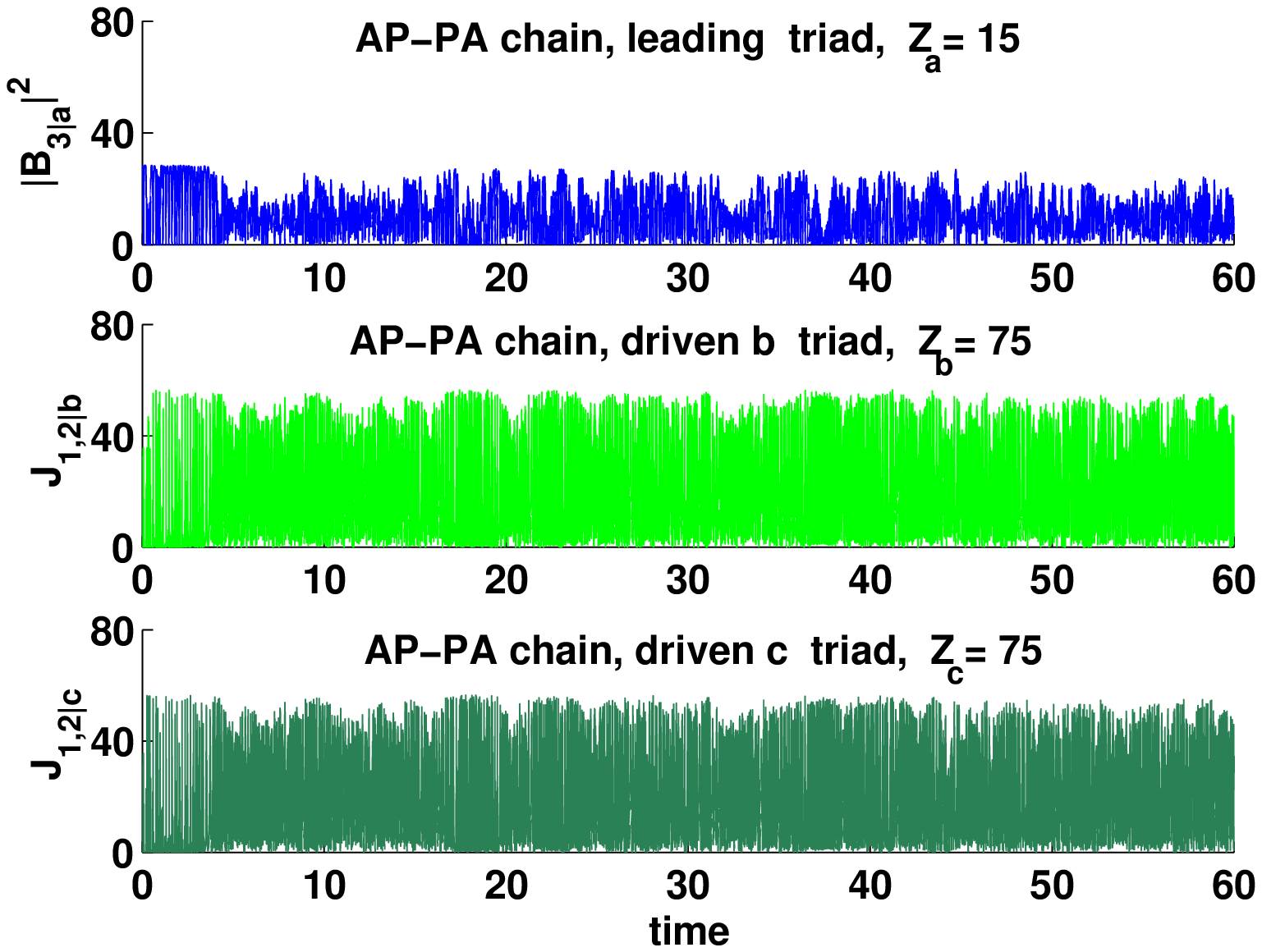}~~ \\ 
 ~~~~~ Efficient energy transfer from $a$- to $b$-triad ($Z_b/Z_a=5$)\\
  ~~~~~and  from $a$- to $c$-triad ($Z_c/Z_a=5$), similar to that \\ ~~~~~in PAA-star in the same panel in \Fig{f:PAA}.  \\
  \end{tabular} &

    \begin{tabular}{ l }
 \includegraphics[width= 8.6cm ]{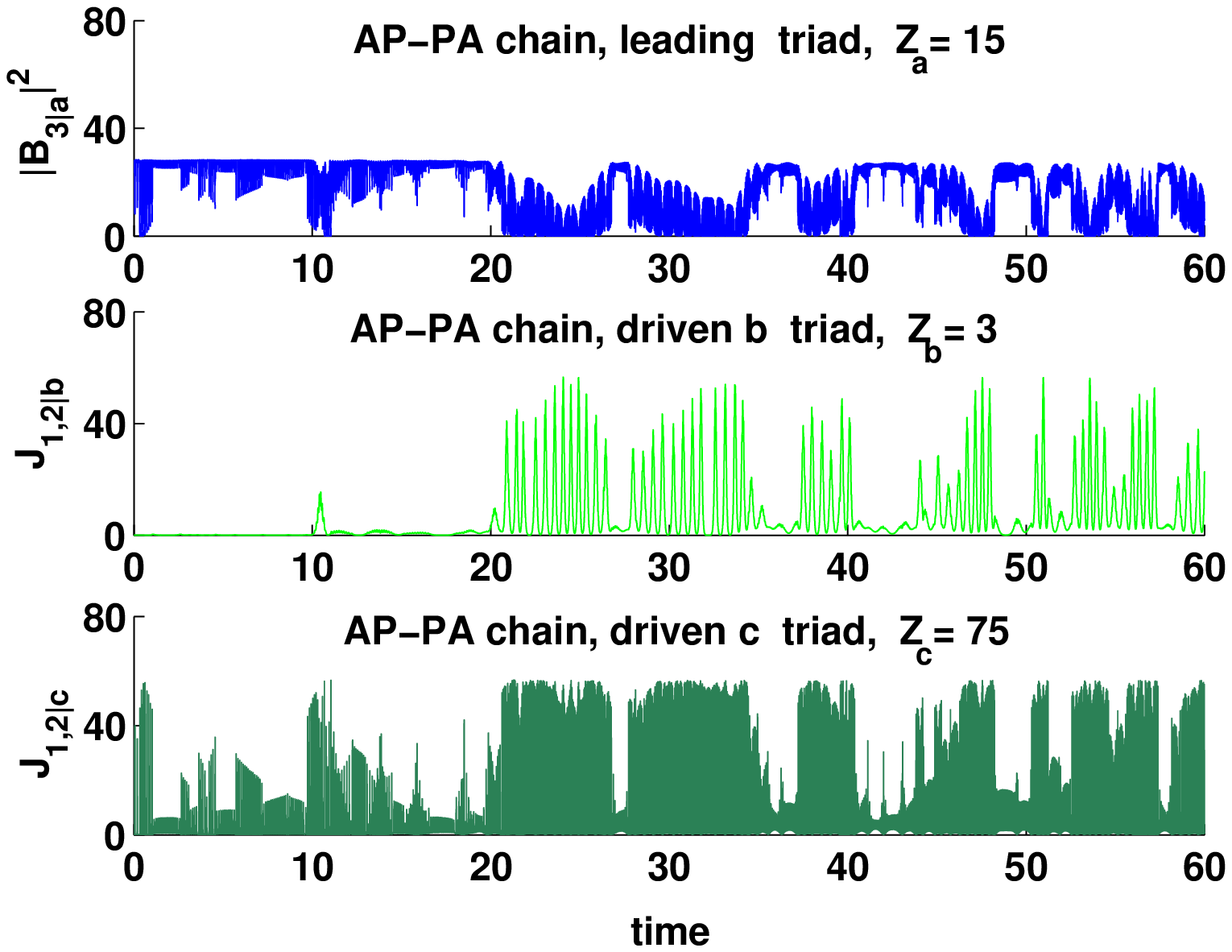}~~ \\ 
~~~~~ The  energy transfer from $a$-triad  to $b$-triad  is very \\
 ~~~~~intermittent, but much more efficient  than that  \\ ~~~~~in PAA-star 
  \end{tabular}\\  \hline\hline
$\C C$ & $\C D$ \\
  \begin{tabular}{ l }    \\
 \includegraphics[width= 8.5cm ]{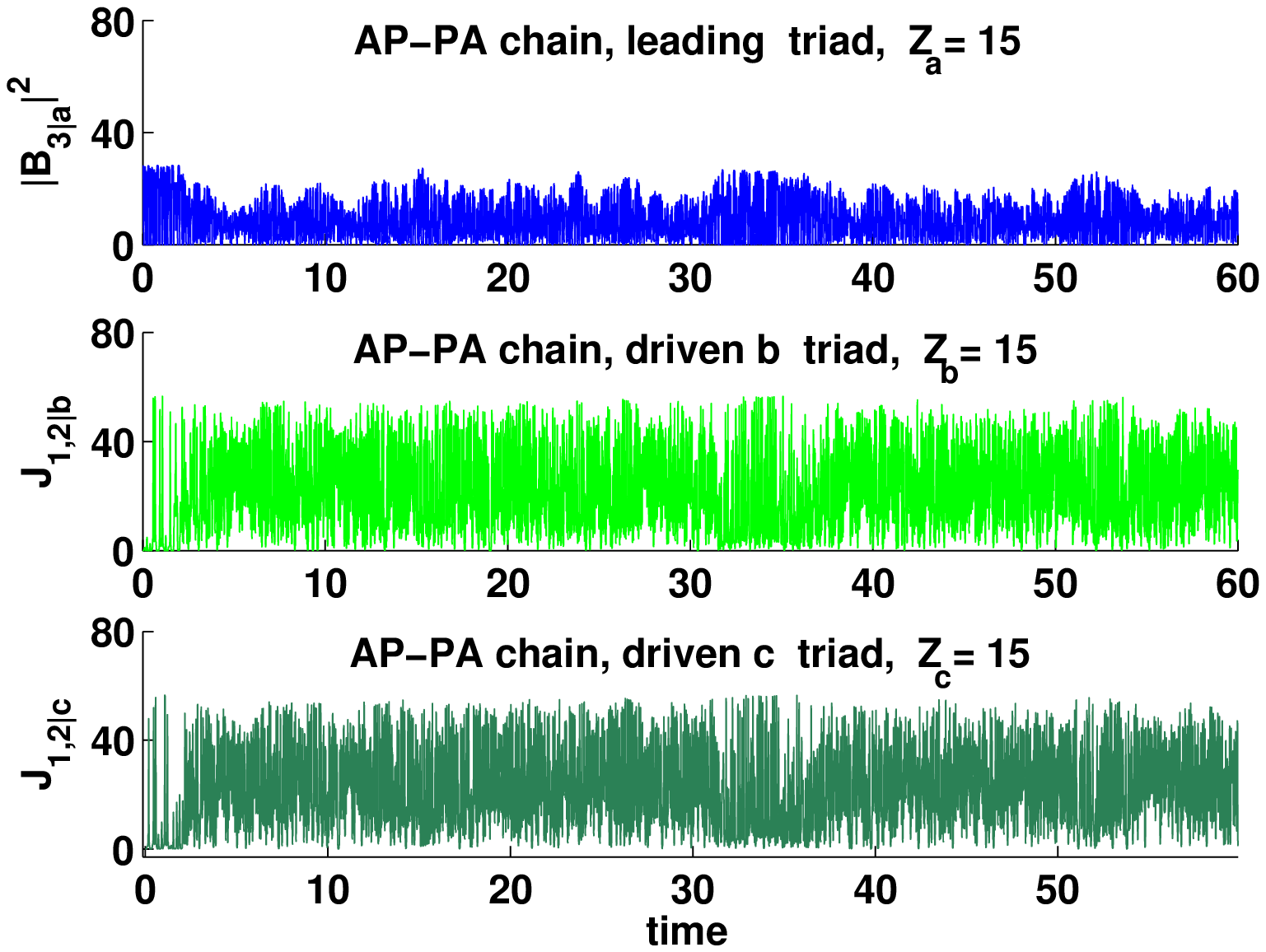} \\ 
 ~~~~~ Efficient energy transfer for AP-PA chain \\
  ~~~~~with $Z_a=Z_b=Z_c$, unlike PAA-star\\
 \\
 \end{tabular} &

    \begin{tabular}{ l }
 ~\includegraphics[width= 8.3 cm ]{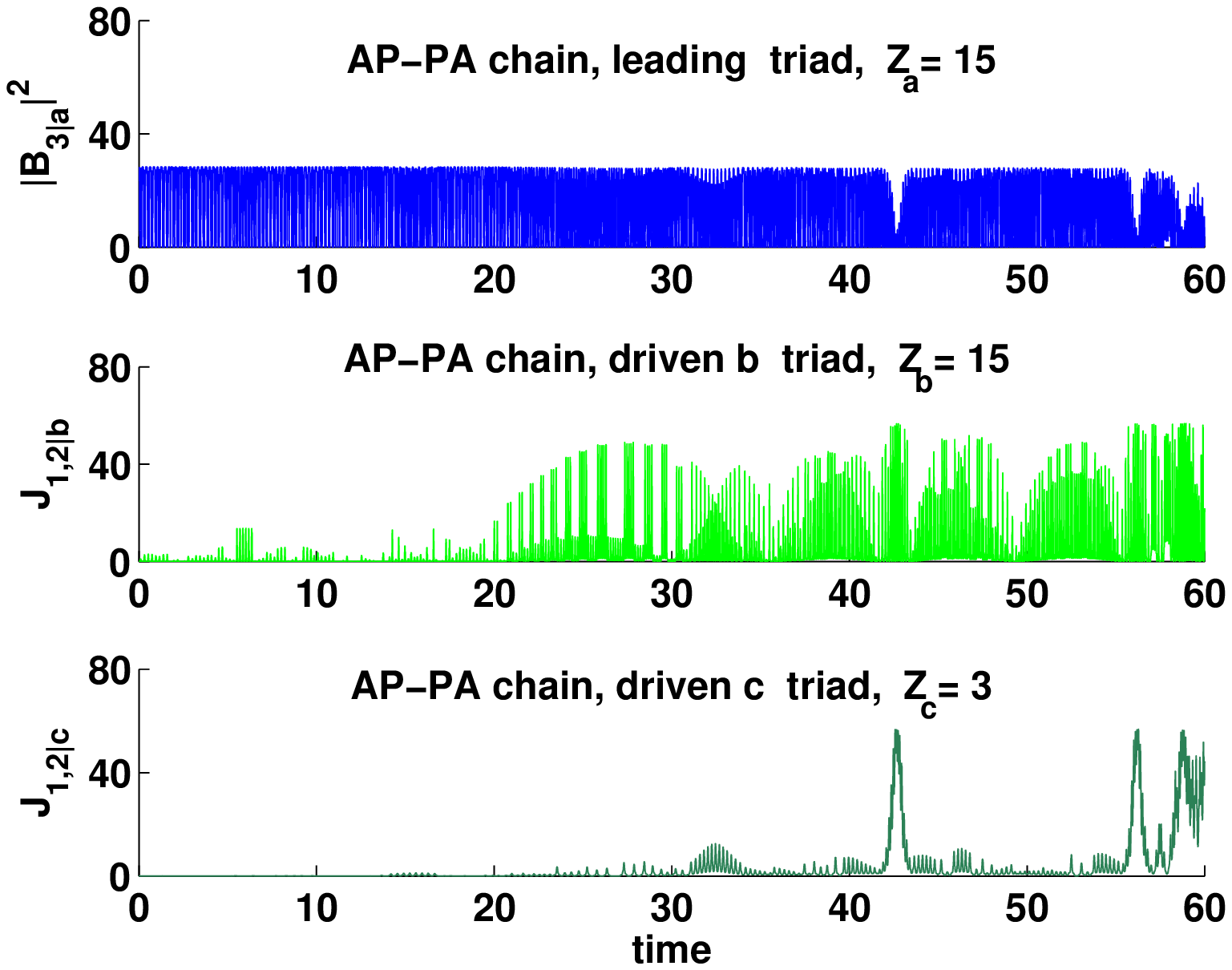}~~  \\ 
~~~~~ Efficient energy transfer for AP-PA chain \\
  ~~~~~with $Z_a=Z_b>Z_c$, unlike PAA-star\\
  \end{tabular}\\  \hline
\end{tabular}
 \end{center}
\caption{\label{f:AP-PA} Color online. Efficiency of  AP-PA-chain energy junction with  different  $Z_b/Z_a$ and $Z_c/Z_a$ [initial conditions~\eq{ic-AP-PA}]. }
\end{figure*}

\subsubsection{\label{ss:3tr} Triple-star junction}
 An effective component in the energy transfer through big clusters
is the PAA-star. PAA-star can accept energy via A-mode of $a$-triad and
transfer it to two other $b$- and $c$-triads, via their A-modes,
$B_{3|b}$ and $B_{3|c}$, connected to the (same) passive $B_{1|a}$ mode of the
$a$-triad (see \Fig{f:proto}, $\C A$). In order to clarify how an additional $c$-triad affects   the
energy in PA-butterfly studied above, we present in \Fig{f:PAA}
an evolution of PAA-star with the choice of the interaction coefficients
similar to that of above and with ``energy-transfer effective"
initial conditions, given by \Eq{ic-AP2} in $a$-triad, \Eq{ic-b} in
$b$-triad. For $c$-triad we took initial conditions similar to that
for $b$-triad \Eq{ic-b},   but with the complex conjugated $B_{1,0}$
and $B_{2,0}$.  Hence, the initial conditions are as follows:
\begin{subequations}\label{ic-PAA}
\begin{eqnarray}\label{ic-PAA_a}&& \begin{cases}
B_{1|a}(0)=B_{3|b}(0) =B_{3|c}(0)= B_{3,0} \,,\\
B_{2|a}(0)=\~B_0\,,\qquad\
B_{3|a}(0)=B_0\,;\end{cases}\\
 \label{ic-PAA_b}
&& ~~B_{1|b}(0)=B_{1,0}\,, \qquad
B_{2|b}(0)=B_{2,0}\, ; \\ \label{ic-PAA_c}
 && ~~B_{1|c}(0)=B_{1,0}^*\,,    \qquad
B_{2|c}(0)=B_{2,0}^* \ .
\end{eqnarray}
\end{subequations}

 Quite expectedly, when the leading triad has much smaller interaction
 coefficient than the driven ones (see   \Fig{f:PAA}$\, A$
 with $Z_a\ll Z_b=Z_c=75$), the energy transfer to both driven triads
 is fully effective. The new element with respect to the butterfly
 case is the energy oscillation between driven triads.

 Panel $\C B$ in \Fig{f:PAA} demonstrates an evolution of the
 PAA-star, in which driven $c$-triad has the interaction coefficient
 $Z_c=3$, i.e. five times smaller  than in the leading triad. As expected,
 the energy transfer to very slow $c$-triad is strongly suppressed,
 below the level $J_{1,2|c} \lesssim 1$.

The same level of excitation, $J_{1,2|b}\approx J_{1,2|c} \lesssim 1$
demonstrates the PAA-star with all equal interaction coefficients, see \Fig{f:PAA}$\, \C C$. The energy alternations between equivalent triads is
even more pronounced, than in case $Z_a\ll Z_b=Z_c=15$, that
demonstrates a  more stochastic behavior.

The comparison of \Fig{f:PAA}$\, \C B$ and $\, \C D$ shows that the decrease in $Z_b$ (from 75 to 15) suppresses energy transfer to the $c$-triad even further up to level $0.015 \ll 1$.

The overall conclusion is that qualitatively the difference in the
interaction coefficients affects the energy transfer in triple stars
similarly to that in butterflies, but on the quantitative level the
interplay of two driven triads cannot be neglected.

Another expected conclusion is that the triple clusters demonstrate
``more random" evolution than butterflies.  However, the level of
randomness strongly depends on the initial conditions. One sees this
from three-dimensional parametric
representation of PAA-star trajectories in  \Fig{f:PAA-par}   with different choice of the
interaction coefficients in the upper and lower panels. Left panels
shows randomization of trajectories, starting from complex initial
conditions~\eq{ic-PAA}, right panels -- much more regular behavior of
the trajectories, starting from similar, but real initial conditions
(in which complex number in \Eq{ic-PAA} are replaced by their absolute
values).  The qualitative explanations of this difference is very
simple. Equations of motion for triple stars~\eq{star-eqs} are
real. Therefore the mode amplitudes $B_{\dots}$, that start from the
real initial conditions, remain real during the evolution and the
dimensionality of the phase space in that case is smaller than that
during evolution from general (complex) initial conditions. Note also
that the real trajectories, although close to periodical ones, have small
random components that lead to the finite ``width" of the attractors
in \Eq{ic-PAA}, right panels.

\subsubsection{\label{sss:AP-PA} AP-PA-chain  energy junction}
Another example of an effective energy junction is the triple
 AP-PA-chains, in which energy can be accepted via A-mode of the
 middle $a$-triad and transferred to  other two $b$- and $c$-triads via
 their A-modes, $B_{3|b}$ and $B_{3|c}$, connected to different
 P-modes of the $a$-triad from the left and the right:
 $B_{3|b}=B_{1|a}$ and $B_{3|c}=B_{2|a}$ (see \Fig{f:proto}, $\C B$). A choice of the initial
 conditions that guarantees efficiency of the energy transfer is
 similar to \Eqs{ic-PAA} with the differences that are dictated by
 different types of the connections:
\begin{subequations}\label{ic-AP-PA}
\begin{eqnarray}\label{ic-AP-PA-a}&& \begin{cases}
B_{1|a}(0)=B_{3|b}(0) = B_{3,0} \,,\\
B_{2|a}(0)=B_{3|c}(0) = B_{3,0}^*,\quad
B_{3|a}(0)=B_0\,;~~~\end{cases}\\
 \label{ic-AP-PA-b}
&& ~~B_{1|b}(0)=B_{1,0}\,, \qquad
B_{2|b}(0)=B_{2,0}\, ; \\ \label{ic-AP-PA-c}
 && ~~B_{1|c}(0)=B_{1,0}^*\,,    \qquad
B_{2|c}(0)=B_{2,0}^* \ .
\end{eqnarray}
\end{subequations}

  In \Fig{f:AP-PA} we show the time evolution of the triple
  AP-PA-chain with the same four sets of the interaction coefficients,
  as in \Fig{f:PAA} for APP-stars.

   Comparing \Fig{f:AP-PA} with \Fig{f:PAA} panel by panel one
  concludes, that the triple AP-PA-chain,  in which the driven triads are
  connected to different modes of the leading triad, demonstrates much
  more efficient energy transfer, than the triple PAA-star, in which the
  driven triads are connected to the same mode of the leading
  triad. Another difference, is that the time evolution in the triple
  AP-PA-chain is much more intermittent than in the triple PAA-star.

\section{ \label{s:stat} Finite-dimensional wave turbulence  in the long-chain clusters}
In this Section we study stationary energy distributions of resonant
waves and energy exchange between triads in the long-chain clusters in
the regime of finite-dimensional wave turbulence with constant energy
flux and during free evolution.

\subsection{\label{sss:stas} Pumping and damping in chain clusters}
With this subsection we begin to consider the stationary dynamics of
the $N$-chain clusters, consisting of many resonant triads (with
$N=12, 16, 20$ and $24$). To do this numerically one should model the
energy pumping in the leading $a$-triad and energy dissipation in the
driven $N^{\rm th}$-triad.

The simplest and reasonable way to model the energy dissipation is quite obvious: one should introduce linear damping terms into the equations of motion for individual modes of the last driven triad. In our case these are P-modes and instead of \Eqs{AA}  we suggest  the following two equations for them:
\begin{equation}\label{Ab}\begin{cases}
\dot{B}_{1|N}= Z_N B_{2|N}^* B_{3|N}- \gamma_{1|N}{B}_{1|N}\,,
\\  \dot{B}_{2|N}= Z_N B_{1|N}^* B_{3|N}- \gamma_{2|N}{B}_{2|N}\,,
\end{cases}
\end{equation}
with some phenomenological damping frequencies $\gamma_{1|N}$ and
$\gamma_{2|N}$ that should be compared with the characteristic interaction frequency in these equations
\begin{equation}\label{O-Ab}
\Omega_{\rm int |N}\= Z_N\sqrt {\< |B_{3|N}|^2 \>}\ .
\end{equation}
Then one can distinguish the cases of symmetrical damping:
$\gamma_{1|N}=\gamma_{2|N}$, asymmetrical damping: $\gamma_{1|N}\ne
\gamma_{2|N}$, but $\gamma_{1|N}\simeq \gamma_{2|N}$ and strong
damping asymmetry: $\gamma_{1|N}\gg \gamma_{2|N}$. In addition
there are cases of weak damping, when $\displaystyle\sqrt
{\gamma_{1|N} \gamma_{2|N}}\ll \Omega_{\rm int |N}$, intermediate and
strong damping, when $\displaystyle \sqrt {\gamma_{1|N}
\gamma_{2|N}}\approx \Omega_{\rm int |N}$ and $\displaystyle\sqrt
{\gamma_{1|N} \gamma_{2|N}}\gg \Omega_{\rm int |N}$.
 To start with
in this paper we will restrict ourselves by the simplest case of weak
symmetrical damping.

In all cases we will include energy pumping term $P_{3|1}$ in the
equation of motion for the A-mode of the first (leading) triad, i.e. in the
equation for $B_{3|a}$. Now it reads:
\begin{equation}\label{pump-eq}
\dot{B}_{3|1}=-Z_1 B_{1|1} B_{2|a}+ P_{3|1}\ .
\end{equation}
How to mimic energy pumping is a much more delicate question.  Notice,
that in the problems of developed wave or hydrodynamic turbulence
researches usually fill free to model the energy source and sink in
the simplest possible manner~\cite{95LP}. For example, one introduces a random
force with Gaussian statistics (that has some justification) and
$\delta$-correlated in time, which is usually far from
reality, see e.g.~\cite{Fri}. The rationale is that for very high levels of turbulence excitation
the inertial interval of scales is large enough such that one expects
universal statistics of turbulence, independent of characteristics of
the energy pumping~\cite{95LP}. In our case of relatively low level of
excitation, when the number of excited degrees of freedom is not so
huge, the universality is, generally speaking, questionable:
statistics of the system dynamics can depend on the way the system is
excited.  One can imagine  few very different versions of the pumping
term.

 The first version mimics an instability of the $B_{3|1}$ mode with an inverse growth time $\nu_{(3|1)}$ :
\begin{subequations}\label{pump}\begin{equation}\label{pumpA}
P_{3|1}= \nu_{{3|1}}B_{3|1}\,,
\end{equation}

Next one can consider a periodic driving force:
\begin{equation}\label{pumpB}P_{3|1}= F_\omega \cos (\omega t)\,,
\end{equation}
that can mimic the effect of an ``external" triad, connected to $B_{3|1}$
mode. In this case $\omega$ is the leading frequency of energy
exchange between modes of the external triad.

 Also one can take a random white Gaussian noise
   \begin{equation}\label{pumpC} P_{3|1}=f(t)\,,  \quad
\< f(t)f(t') \> = f^2 \delta(t-t')\,,
 \end{equation}\end{subequations}
that   mimics the influence of numerous non-resonant triads.

In this paper we will utilize   a bit more realistic driving force $P_{3|a}(t)$  [periodic, but not $\propto \cos (\omega t)$, as in \Eq{pumpB}] that is produced by the equations of motion for an isolated triad. This mimics the effect of an ``external" triad, connected to $B_{3|a}$ mode in the approximation, when the feedback effect is neglected.

\subsection{\label{sss:ints} Condition of stationarity}
Consider first the dynamical invariants for isolated $N$-chain clusters (without pumping and damping terms).
Combining invariants~\eq{MR}, one has in this case a (dependent) invariant  for isolated triad in the form
\begin{subequations}\label{ints4}
\begin{equation}\label{ints4A}
I_{|1}=|B_{3|1}|^2+ \frac12 \big( |B_{1|1}|^2+ |B_{2|1}|^2  \big)\ .
\end{equation}\
Similarly, a combination of  invariants~\eq{APintC} yields the invariant  for PA-butterfly:
\begin{equation}\label{ints4B}
I_{|1,2}=|B_{3|1}|^2+ |B_{3|2}|^2+\frac12 \big( |B_{1|2}|^2+ |B_{2|2}|^2  \big)\ .
\end{equation}
And finally, from invariants~\eq{PA-PA-int} one derives new (dependent) invariant for PA-PA-chain:
\begin{equation}\label{ints4C}
I_{|1,2,3}=|B_{3|1}|^2+ |B_{3|2}|^2+ |B_{3|3}|^2+\frac12 \big( |B_{1|3}|^2+ |B_{2|3}|^2  \big)\ .
\end{equation}
Similarly, for isolated $N$-chain cluster one gets:
\begin{equation}\label{ints4D}
I_{|1,2,\dots N}=\sum _{j=1}^N
|B_{3|j}|^2+ \frac12 \big( |B_{1|N}|^2+ |B_{2|N}|^2  \big)\ .
\end{equation}\end{subequations}

All these invariants include $B_{3|1}$-mode, affected by the pumping, and $\frac12 \big( |B_{1|N}|^2+ |B_{2|N}|^2  \big)$ terms, affected by the damping. Therefore under the stationary conditions  the following equality must hold:
 \begin{subequations}\label{cond4}
\begin{equation}\label{cond4A}
\< P_{3|1} B_{3|1}^*\>= \frac12 \big( \gamma_{1|N} \<|B_{1|N}|^2\>+ \g_{2|N}\<|B_{2|N}|^2 \> \big)  \ .
\end{equation}
In particular for the pumping~\eq{pumpA} one has a simple relationship:
 \begin{equation}\label{cond4B}
\g_{3|1} \<  |B_{3|1}|^2\>= \frac12 \big( \gamma_{1|N} \<|B_{1|N}|^2\>+ \g_{2|N}\<|B_{2|N }|^2\>  \big)  \ .
\end{equation}
\end{subequations}
Equations~\eq{cond4} can be considered as the (nessesary) condition of
the energy balance: the left hand side describes the energy pumping,
while the right had side -- the energy damping. Each of these
quantities can be equated (under the stationary conditions) to the
energy flux through the clusters.

\subsection{\label{ss-chain-res}Universality of statistics of finite-dimensional wave turbulence in the long-chain clusters}
\subsubsection{Modeling the parameters  of the forced  chain clusters}
We present here our choice of the modeling parameters for the forced PA-PA-\dots-PA chain clusters. The cluster  consists of
$N$ triads and is constructed by connecting the first P-mode of $(n-1)$'s triad, $B_{1|n-1}$,  with the A-mode
 of the next, $n$'s triad, $B_{3|n}$.  The interaction coefficients  $Z_n$ were
chosen in geometric progression,
\begin{equation}\label{Zchain}
Z_n=\lambda Z_{n-1}=Z_1\lambda^{n-1}\, .
\end{equation}
  The chain is forced by an independent freely evolving triad with
$Z_0=0.125$ related to $Z_1$ as $Z_1=Z_0 \lambda$. The amplitude
$B_{1|0}$ of the forcing triad is added to the equation of $B_{3|1}$
of the first triad in the chain 200 times during the period $T_f$ of the
forcing triad.  We made special efforts to ensure the exact periodicity of the forcing.

The linear dissipation with the damping frequency $\gamma$ was added
to the equations of two P-modes of the last triad, $B_{1|N}$ and
$B_{2|N}$.  In most calculations $\gamma=10^{-4}$ was used.

We have studied the mean square amplitudes of ``free" modes,
$\<|B_{2|n}|^2\>$ and ``connected" modes
$\<|B_{3|n}|^2\>=\<|B_{1|n-1}|^2\> $ in the chains with different
number of triads, $N=16,\ 20,\ 24$, for the set of the frequency
scaling parameter $\lambda=1, \ \sqrt 2,\ 2,\ 2\sqrt 2 $ and using
different initial conditions in the pumping triad.

We found that initial conditions do not effect the stationary
 statistics in the cluster. For concreteness, for all presented
 simulations we chose for the pumping triad
\begin{eqnarray}\label{ic_first _triad}
     B_{1|0}&=&0.05+  0.02\, i\,, \\
     B_{2|0}&=&0.02+  0.05\, i\,,\\ \nonumber
     B_{3|0}&=&0.9-  0.93\, i\,;
 \end{eqnarray}
and for all  $N$ triads of the chain

\begin{eqnarray}\label{ic_all}
    B_{1|n}&=&0.05+  0.02\, i\,, \\
    B_{2|n}&=&0.02+  0.05\, i\,,\\\nonumber
    B_{3|n}&=&B_{1|n-1}\ .
\end{eqnarray}

The time of the chain evolution required to reach stationarity varied from
about 1000 $T_f$  of the pumping triad for $\lambda=1$ to
about 50 $T_f$ for $\lambda=2\sqrt 2$. For times later than  1000 $T_f$ we calculated
mean values in time, averaging values of $\langle |B_{j|n}|^2 \rangle
$ over $300 T_f$.  In the most fast converging case of
$\lambda=2\sqrt2$ we averaged over $50 T_f$. In all cases the
evolution was followed for much longer times and we verified that a
particular choice of the averaging interval does not affect results.

Our results are presented in  Figs.~\ref{forcedLdep}  as logarithm of $\langle
|B_{2|n}|^2 \rangle$ and  $\langle
|B_{3|n}|^2 \rangle$, normalized by its value in the first triad vs
triad number $n$ normalized by   $N$. To force all lines to start from the same point and not from $1/N$, we shifted triad numbers by $1$: $n \to n-1$ for presentation. Such a shift leads to better collapse of the data.

\subsubsection{\label{sss:force} $N$-,  $\lambda$- and $\gamma$-independence of the forced-chain statistics}
\begin{figure*}
\begin{center}
\begin{tabular}{| c || c |}\hline
$\C A$ & $\C B$ \\
    \begin{tabular}{ l }
 ~\includegraphics[width= 8.6cm ]{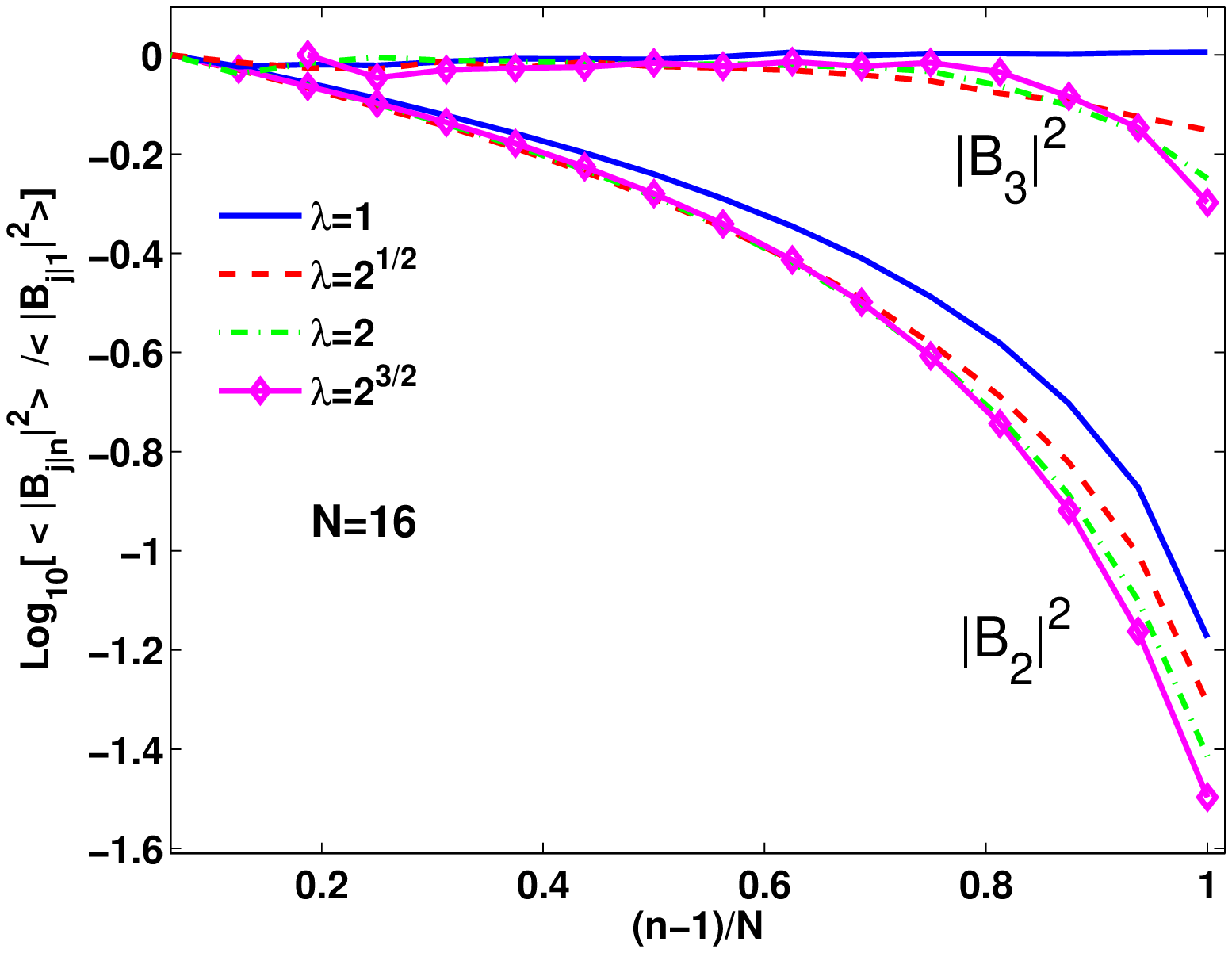}\\ 
 ~~~~~ Square amplitude distributions for for different \\
  ~~~~~frequency scaling factor $\lambda=1,\ \sqrt 2, \ 2,\ 2\sqrt 2$ with \\  ~~~~~$N=16$. For $\lambda \ge 2$ plots are practically converged.  \\
  \end{tabular} &

    \begin{tabular}{ l }
 \includegraphics[width= 8.2cm ]{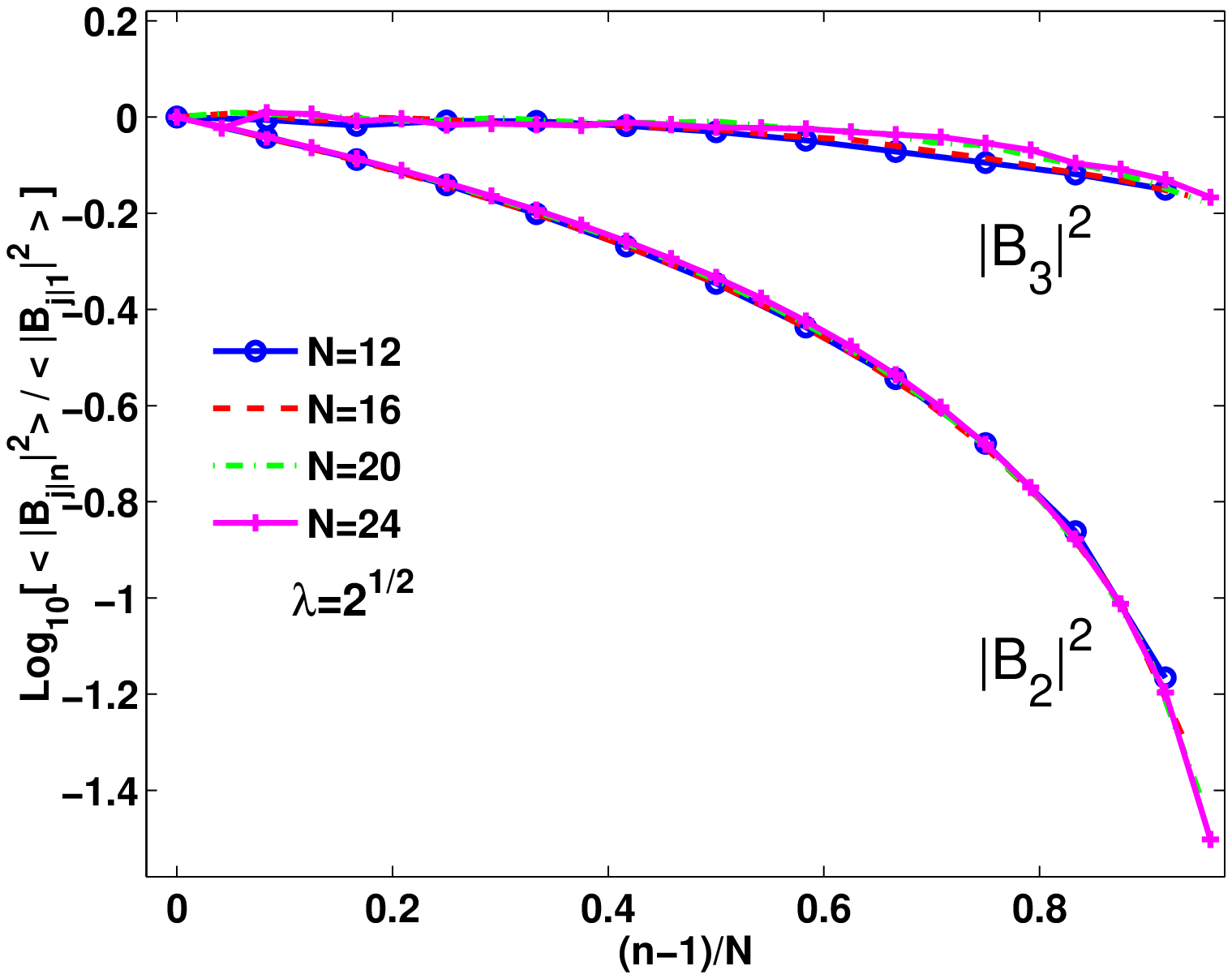} \\ 
~~~~~ Square amplitude distributions for  different \\
 ~~~~~ number of triads $N=16,\ 20,\ 24$ with $\lambda=\sqrt 2$.
  \\ ~~~~~All plots  practically coincide  \\
  \end{tabular}\\  \hline\hline
$\C C$ & $\C D$ \\
  \begin{tabular}{ l }
 \includegraphics[width= 8.5cm ]{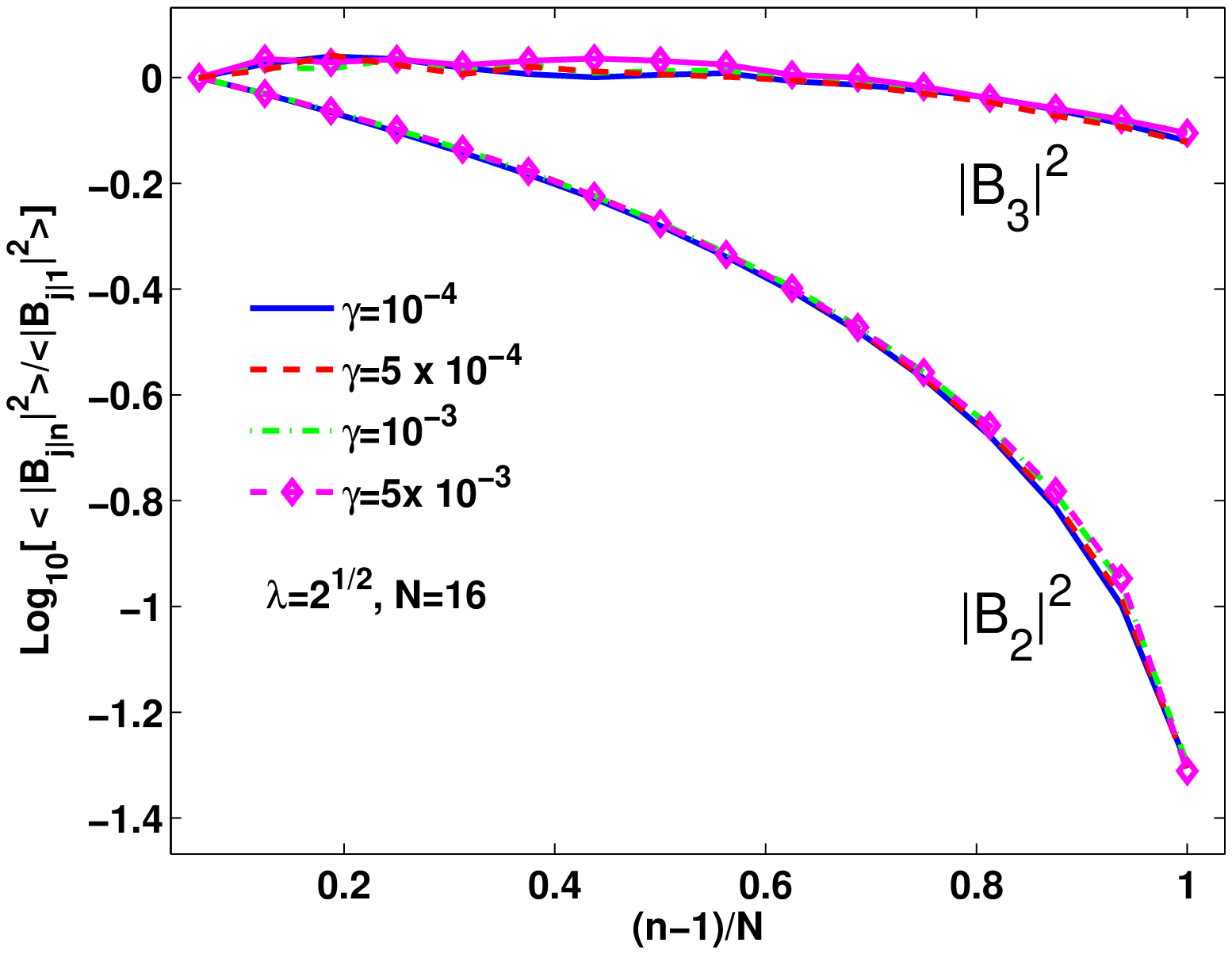} \\ 
 ~~~~~  Square amplitude distributions for  different  $\gamma$\\
 ~~~~~ for $\lambda=\sqrt 2$ and $N=16$.
 \\
 \\
 \end{tabular} &

    \begin{tabular}{ l }
 ~\includegraphics[width= 8.3 cm ]{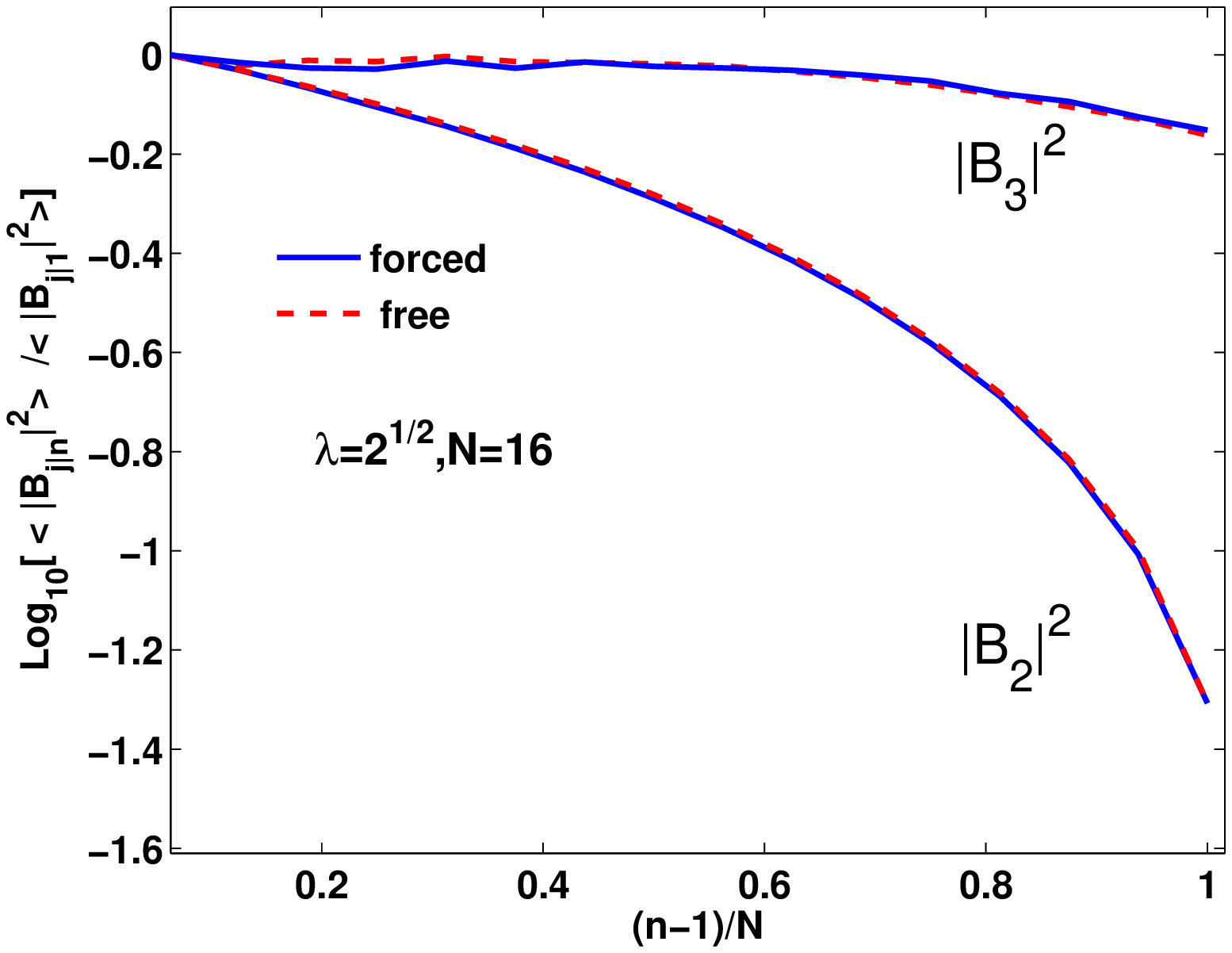}~~ \\ 
~~~~~ Comparison of the forced and free chain for $\lambda=\sqrt 2$\\
 ~~~~~and  $N=16$.  These two plots practically coincide \\

  \end{tabular}\\  \hline
\end{tabular}
 \end{center}
\caption{\label{forcedLdep} Comparison of the plots of $\ln \, \langle |B_{j|n}|^2 \rangle$ vs $n/(N-1)$ for different $\lambda$, panel $\C A$, for different $N$, panel $\C B$ and for  different $\gamma$,  panel $\C C$.  Panel $\C D$ compares distributions for the free and forced evolutions.}
\end{figure*}

Figure~\ref{forcedLdep}$\, \C A$  presents normalized mean-square amplitudes
$\langle |B_{2|n}|^2 \rangle$ and $\langle |B_{3|n}|^2 \rangle$ for
the chain with $N=16$ and a set of the scaling parameters $\lambda=1, \
\sqrt 2,\ 2$ and $ \lambda= 2\sqrt 2 $. One sees that the lines for
different $\lambda$ are quite close, quickly converging for larger
$\lambda$. We conclude that the distributions of $\langle
|B_{2|n}|^2 \rangle$ and $\langle |B_{3|n}|^2 \rangle$ are almost
$\lambda$-independent at least for $\lambda \ge 2 $.

Figure~\ref{forcedLdep}$\, \C B$ represents distributions of $\langle |B_{2|n}|^2
\rangle$ and $\langle |B_{3|n}|^2 \rangle$ for the chains with
$\lambda=\sqrt 2$ and different number of triads: $N=12, \ 16,\ 20$ and $N=24$. One
sees that the lines for different $N$ practically coincide. The conclusion is that the distributions of
$\langle |B_{2|n}|^2 \rangle$ and $\langle |B_{3|n}|^2 \rangle$ are
  $N$-independent in the limit of large $N$.

Panel $\C C$ in \Fig{forcedLdep} shows $\langle |B_{2|n}|^2 \rangle$
and $\langle |B_{3|n}|^2 \rangle$ for the chains with $\lambda =\sqrt
2$, $N=16$, and different, but small values of $\gamma= 10^{-4},\ 5
\times 10^{-4},\ 10^{-3},\ 5 \times 10^{-3}$. Collapse of these plots
is obvious. One can ask, what happens if one increases further the
damping parameter? Increasing $\gamma$, we found that in this case the
individual amplitudes in the last triad become very small. Its ability
to absorb the energy flux from the previous triad diminishes to the
stage at which it can not anymore serve as an energy sink.  This leads
to accumulation of the energy in the N-1'th triad and later in N-2nd
triad etc, thus creating a "bottleneck effect".

The reason for this effect is that the energy pumping of $B_{1|N}$ and
$B_{2|N}$ modes is $\propto \mbox {Re} \big [B_{3|N} ^*
B_{1|N}B_{2|N}\big ]$, while their energy damping (in the simple case
$\gamma_{1|N}=\gamma_{2,N}=\gamma $) is proportional to $\gamma
(|B_{1|N}|^2+|B_{2|N}|^2)$. In other words, both fluxes are quadratic
in the amplitudes $B_{1|N}$ and $B_{2|N}$. It means that for large
enough $\gamma$ the energy damping will be larger than the energy
pumping, whatever the values of $B_{1|N}$ and $B_{2|N}$ are. In this
case these amplitudes will decrease in time exponentially. In the
other case, when $\gamma$ is small enough, the energy damping will be
smaller than the energy pumping and amplitudes $B_{1|N}$ and $B_{2|N}$
will exponentially grow. In other words, there is a critical value of
$\gamma$, above which the last triad is not exited at all. A more
detailed analysis shows, that for $\gamma_{1|N}\ne\gamma_{2,N}$ the
critical parameter is $\displaystyle \sqrt{ \gamma_{1|N}\
\gamma_{2,N}}$.

\begin{figure*}
  \begin{tabular}{c c}

  $\C A$ & $\C B$\\
   \includegraphics[width=0.485\textwidth]{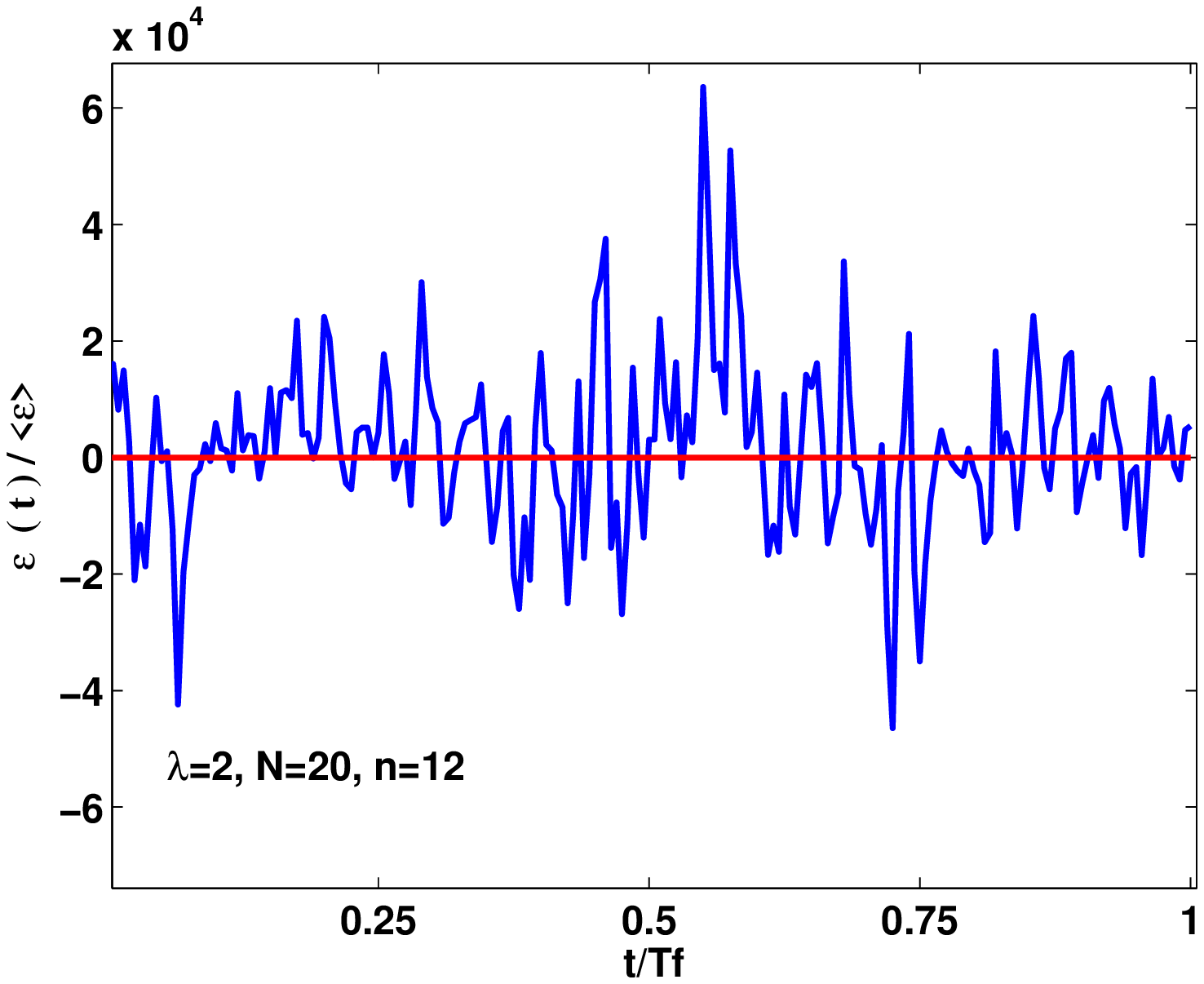} & 
   \includegraphics[width=0.52\textwidth]{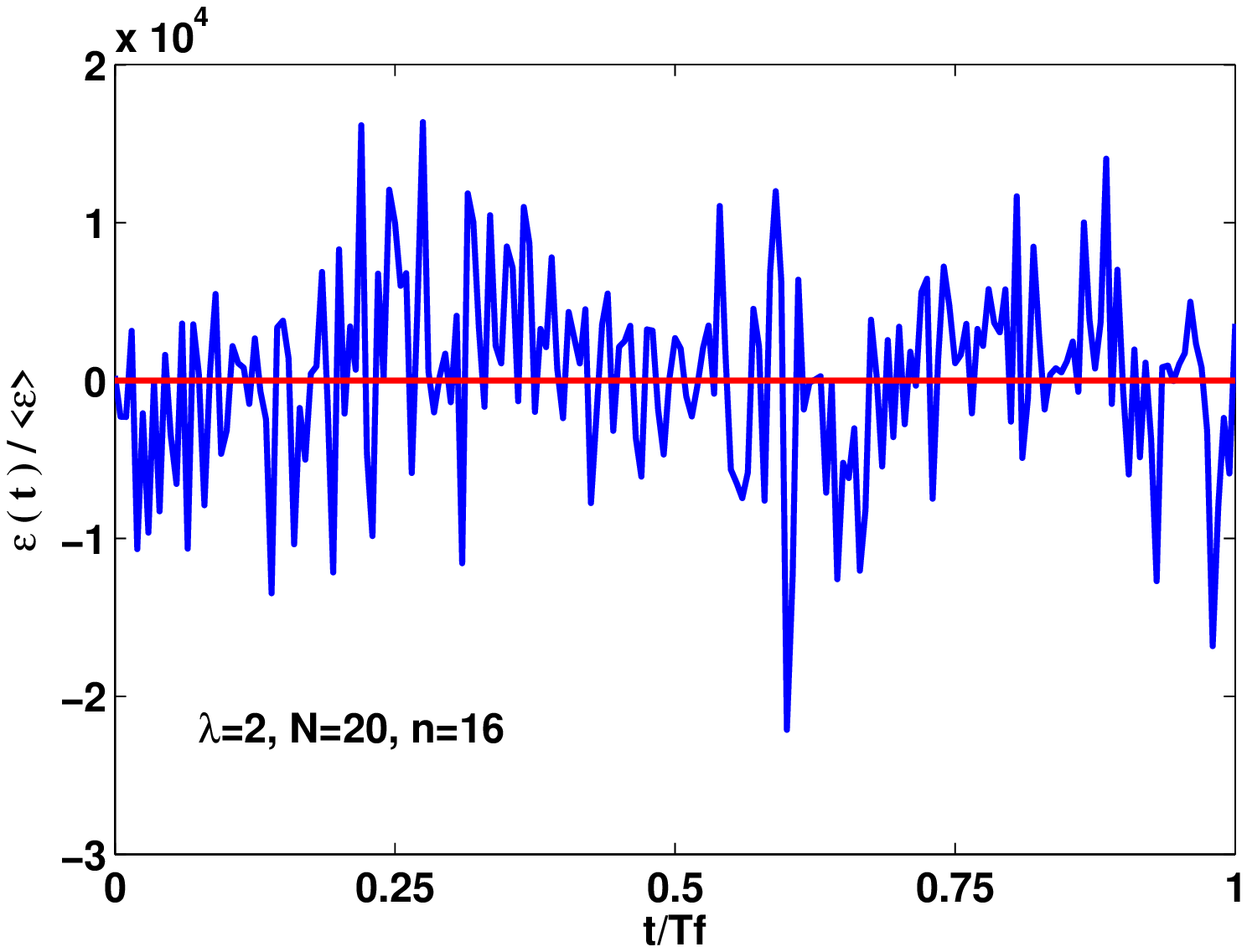}\\  

  \end{tabular}
\caption{
    The time evolution of the normalized flux over triad 12 ($\C A$) and over triad 16 ($\C B$) during one period of the forcing triad for $\lambda=2$ for $N=20$.
}
\label{f:flux}
\end{figure*}

\subsubsection{\label{sss:force} Forced- vs free-evolution statistics in
the chain clusters}

One more surprise is that in the conditions of the energy-flux
equilibrium in the scale-invariant situations (in our case,
$Z_n\propto \lambda^n$) we did not observe the expected power-like
behavior, when $\langle |B_{j|n}|^2 \rangle\propto \lambda ^{n\, \zeta
}$. Moreover, the ratio $\langle |B_{2|n}|^2 \rangle \big / \langle
|B_{3|n}|^2 \rangle$ depends on $n$ and the connected modes $\langle
|B_{3|n}|^2 \rangle$ weakly depend on $n$, as one expects in the
thermodynamic equilibrium, after a long free evolution.

That is why we decided to compare forced distributions of $\langle
|B_{2|n}|^2 \rangle$ and $\langle |B_{3|n}|^2 \rangle$ with the
distributions of $\langle |B_{2|n}|^2 \rangle $ and $ \langle
|B_{3|n}|^2 \rangle$ after long free evolutions (without forcing and
damping) in the same cluster. For concreteness we took cluster with
$\lambda=\sqrt 2$ and $N=16 $  after
forced evolution during time $t=1000 T_f$. To get the values of the forced amplitudes we averaged over last $500
T_f$. Then we took final complex amplitudes $ B_{2|n} $ and $
B_{3|n}$ as initial conditions for the free evolution during
another period $t=1000 T_f$, and again averaged over last $500 T_f$. Both
 distributions of  $\langle
|B_{2|n}|^2 \rangle$ and $\langle |B_{3|n}|^2 \rangle$ are presented in the  \Fig{forcedLdep}$\,\C D$. As one sees
these distributions practicably coincide.

  In other words, the distributions with constant energy flux and the
  distributions with zero energy flux are almost identical. To show,
  why this happens we presented in \Fig{f:flux} the time evolution of
  the instant energy flux via $n$-th triad
\begin{equation}\label{flux}
\varepsilon_{n}(t)=2 Z_{n} {\rm Re} [{B_{1|n}(t) B_{2|n}(t) B^*_{3|n}(t)}]\,,
\end{equation}
normalized by its mean value $\widetilde \varepsilon_{n}\= \<
 \varepsilon_{n}(t)\>$. One sees enourmous fluctuations of
 $\varepsilon_{n}(t)$ with peaks, reaching $\approx 6000\, \widetilde
 \varepsilon_{n}$. The computed mean value of the flux fluctuation
 $\delta \varepsilon_{n}\= \sqrt {\< \varepsilon_{n}^2(t)\>}$ is about
 100 times larger than $ \widetilde \varepsilon_{n}$. This means that
 there is a strong energy exchange between triads in the cluster, that
 essentially exceeds minor mean flux. Therefore one can switch off
 the mean energy flux without almost any effect on the cluster
 statistics. We think, that this general statement is valid for all
 kind on clusters, consisting of connected triads.

 The only difference between forced and free long-time evolution of
 the clusters is the restriction, that follows from the conservations
 laws, that are satisfied exactly in free evolutions and approximately in
 the forced case.

\section{\label{s:con}Conclusions}

\subsection{\label{ss:con1} Main points of understanding in finite-dimensional wave turbulence}

\textbullet~ In the first part of the paper  we studied the structure of finite clusters of resonant triads using the example of atmospheric planetary waves and showed that:

--~In  the physically relevant domain of atmospheric
planetary waves ($m\,,\ \ell \le 1000$, when the mode-scale is
larger than the height of the Earth atmosphere) we have found and
described the topology of all the clusters formed by resonantly interacting
planetary modes. The set od clusters contains
 isolated triads and subsets of 2-, 3-, $\dots$, 16 and  3691 connected triads, with 2- 3-, $\dots$, 9-mode and (maximum)
 10-mode connections;

--~Analyzing the integrals of motion we suggested a
classification i) of triad modes into two types  - active (A) and
passive (P),  and ii) of connection types between triads - AA, AP and
PP. We showed explicitly  that through  the AA-connection
 the energy can flow in both directions, through the AP-connection only from   P-connected triad to the A-connected one,
 but not vice-versa. The PP-connections are almost impenetrable for energy in both directions. Therefore from the viewpoint of the energy transfer, large clusters can be subdivided into smaller ones (by cutting PP-connections);

--~We introduced a notion of PP-irreducible sub-clusters that cannot be further divided by cutting PP-connections and studied their statistics for planetary waves in the spectral domain $m\,,\ \ell \le 1000$. There are  3005  triads, 400 PA- and AA-butterflies,  143 triple-triad PP-irreducible sub-clusters, etc.,  and only   two 130-triad ones in this domain;

--~To first approximation, almost all triads in the
methodologically significant domain, Table.~1,  can be considered as
completely or almost separated from the rest of the atmospheric
planetary waves and therefore the energy oscillations between them can
really lead to intra-seasonal oscillations in Earth's atmosphere as
suggested in~\cite{KL-07}.

\textbullet~ In the second part of the paper we presented a general analysis of the energy transfer in the PP-irreducible sub-clusters. Studying  free evolution from asymmetrical initial conditions, when almost all the initial energy is localized in one (leading) triad we showed:

 --~ the energy flux through the AA- and PA-connections are qualitatively the same and crucially depend on the initial conditions in the  leading triad and on the ratio of the interaction coefficients in the leading and  driven triads;

 --~ the energy flux from the leading to two driven triads in triple-triad clusters depends on the type of  energy-junction (triple AAA-,  PAA-stars or PA-AP chain) and on the relations between three interaction coefficients.

 \textbullet~ We also studied  forced stationary  energy transfer in a long chain consisting of $N$ PA-connected triads with the interaction coefficients $Z_n\propto \lambda^n$, $1\le n \le N$, with forcing in the first triad and damping in the last one.  We showed that stationary energy distributions between triads are universal in the following sense:

--~ For $\lambda \ge 1$ the distributions are almost $\lambda$-independent; For  $\lambda \ge 2$ the distributions practically collapse on each other.

--~ the distributions are practically independent of the damping parameter $\gamma$ for $\gamma< \gamma\sb {th}$. For $\gamma>\gamma\sb {th}$ the stationary state does not exist.


--~ Distributions for the forced case and for the free evolution (with the forced state as initial condition) coincide due to extremely large  fluctuations of the energy flux, exceeding its mean value in orders of magnitude.

  \textbullet~ Our analysis of finite-dimensional wave turbulence is based on the simple form~\eq{3} of the
   basic equations   and does not exploit the explicit form of the
   interaction coefficients $Z$ which vary substantially for
   different wave systems. This means that our results
    can be used directly for arbitrary 3-wave resonant systems governed by
   \Eqs{3}, e.g. drift waves, gravity-capillary waves, etc.

\subsection{\label{ss:con2}Remaining unclear issues in finite-dimensional and mesoscopic regimes of weak wave turbulence} %

\textbullet~ Having in mind the extremely rich variability of the structure of concrete finite clusters of resonance triads and the various important aspects of the cluster dynamics and statistics (some of them were not mentioned at all), this paper opens more questions than answers. Among them:

--~ What might be the peculiarities of free evolutions of finite
  clusters from other initial conditions, more general than those
  studied in this paper, especially for particular clusters which may
  be important in some applications?

--~ How energy goes through more complicated junctions, than triple
stars, studied in this paper, such as stars with four and five and
more triads, etc.?

--~ How integrability of clusters (with special choice of the ratios
    of interaction coefficients and/or initial
    conditions~\cite{ver,08-BK}) or closeness to the integrable cases
    affect the energy transfer and the statistics of the mode
    amplitudes?

--~ What is the dimensionality of the cluster trajectories, how does it depends on the details of the initial conditions and/or  the ratios of  interaction coefficients for various cluster structure?

--~ How different-mode, different-time correlation functions depend on the mode position in a cluster and on the time difference between them?

--~ What is the values of the flatness (ratio of the forth-order correlation functions to the square of the second-order ones) and how does it depend on the cluster structure, initial conditions, etc.?

--~ To what extent can the statistics of the amplitudes of individual modes  be considered close to Gaussian at least for very large (but finite) clusters?

--~  How does this closeness (if it exists) depend  on the mode position in a cluster and on initial conditions?

--~ If this closeness exists (we believe that it does), how can one formulate appropriate closures that will lead to an analytical statistical description of the finite cluster behavior?

\textbullet~ A few of the most important questions, at least from the theoretical viewpoint, are:

--~ What is the principal difference between finite size cluster behavior for three-wave resonances, discussed here, and that for four-wave resonance systems?

--~ How mode statistics and possible applicability of the closure procedures changes, if one accounts for small damping of the mode energy and external random noise, that can mimic interaction of a cluster with the ``rest of the
world"?

--~  How all the features of a cluster behavior (both for those studied and those left open) get modified  if one accounts for quasi-resonances that can become crucially important with increasing the level of the modes excitation?

--~ How to describe the statistical behavior of finite-size dynamical systems  (in the case of three- and four-wave interactions) with further  increases in the system size or in the level of system excitation? How this behavior approachs (step by step, probably via an intermediate kind of behavior) the limit of infinite system with quasi-Gaussian statistics, that can be successfully  described with the help of wave-kinetic equations?

\subsection{The road ahead}

Our feeling is that all these (and many similar and different, but related) questions are the subject of  a  new fields in nonlinear wave physics, \emph{finite-dimensional wave turbulence} and \emph{mesoscopic wave turbulence}. This subject offers very interesting issues both from the physical and the methodological viewpoints, with possible important applications in numerous physical examples.

\acknowledgements
We thank E. Kartashova for providing us with the data set of the exact solutions of resonance conditions for
planetary waves and for useful comments on a draft of this paper.
We acknowledge the supports of the Austrian Science Foundation (FWF)
under project P20164-N18 "Discrete resonances in nonlinear wave
systems" and  of the Transnational Access
Programme at RISC-Linz, funded by European Commission Framework 6
Programme for Integrated Infrastructures Initiatives under the
project SCIEnce (Contract No. 026133).

\end{document}